\documentclass[a4paper,11pt]{article}
\usepackage{jcappub}
\usepackage{multirow}
\usepackage{verbatim}
\usepackage{tabularx}
\usepackage{orcidlink}
\usepackage{amsmath}
\usepackage{appendix}
\usepackage {siunitx} 
\usepackage[utf8]{inputenc}
\usepackage{lineno}

\DeclareUnicodeCharacter{2212}{\textminus}
\DeclareRobustCommand{\ion}[2]{%
\relax\ifmmode
\ifx\testbx\f@series
{\mathbf{#1\,\mathsc{#2}}}\else
{\mathrm{#1\,\mathsc{#2}}}\fi
\else\textup{#1\,{\mdseries\textsc{#2}}}%
\fi}
\usepackage[capitalise]{cleveref}

\title{\boldmath 
Selection of high-redshift Lyman-Break Galaxies from broadband and wide photometric surveys}

\author[1]{{Constantin~Payerne}\orcidlink{0000-0002-1818-929X},}
\author[2]{{William~d'Assignies~Doumerg}\orcidlink{0000-0002-9719-1717},}
\author[1]{{Christophe~Yèche}\orcidlink{0000-0001-5146-8533},}
\author[1]{{Vanina~Ruhlmann-Kleider}\orcidlink{0009-0000-6063-6121},}
\author[3]{{Anand~Raichoor}\orcidlink{0000-0001-5999-7923},}
\author[4, 5]{{Dusting~Lang}\orcidlink{0000-0002-1172-0754},}
\author[3]{{Jessica~Nicole~Aguilar},}
\author[7]{{Steven~Ahlen}\orcidlink{0000-0001-6098-7247},}
\author[9]{{St\'ephane~Arnouts},}
\author[14]{{Davide~Bianchi}\orcidlink{0000-0001-9712-0006},}
\author[10]{{David~Brooks},}
\author[3]{{Todd~Claybaugh},}
\author[6]{{Shaun~Cole}\orcidlink{0000-0002-5954-7903},}
\author[11]{{Axel~de~la~Macorra}\orcidlink{0000-0002-1769-1640},}
\author[12]{{Arjun~Dey}\orcidlink{0000-0002-4928-4003},}
\author[13]{{Biprateep~Dey}\orcidlink{0000-0002-5665-7912},}
\author[10]{{Peter~Doel},}
\author[2, 10]{{Andreu~Font-Ribera}\orcidlink{0000-0002-3033-7312},}
\author[15, 16]{{Jaime~E.~Forero-Romero}\orcidlink{0000-0002-2890-3725},}
\author[3]{{Satya~Gontcho~A~Gontcho}\orcidlink{0000-0003-3142-233X},}
\author[17]{{Gaston~Gutierrez},}
\author[18]{{Stephen~Gwyn}\orcidlink{0000-0001-8221-8406},}
\author[19, 20, 21]{{Klaus~Honscheid},}
\author[12]{{Stephanie~Juneau},}
\author[3]{{Andrew~Lambert},}
\author[3]{{Martin~Landriau}\orcidlink{0000-0003-1838-8528},}
\author[22]{{Laurent~Le~Guillou},}
\author[3]{{Michael~E.~Levi}\orcidlink{0000-0003-1887-1018},}
\author[1]{{Christophe~Magneville},}
\author[23, 2]{{Marc~Manera}\orcidlink{0000-0003-4962-8934},}
\author[12]{{Aaron~Meisner}\orcidlink{0000-0002-1125-7384},}
\author[24, 2]{{Ramon~Miquel},}
\author[25]{{John~Moustakas}\orcidlink{0000-0002-2733-4559},}
\author[13]{{Jeffrey~A.~Newman}\orcidlink{0000-0001-8684-2222},}
\author[1, 3]{{Nathalie~Palanque-Delabrouille}\orcidlink{0000-0003-3188-784X},}
\author[5, 4]{{Will~Percival}\orcidlink{0000-0001-7145-8674},}
\author[27]{{Vincent~Picouet},}
\author[5, 4]{{Francisco~Prada},}
\author[8]{{Ignasi~P\'erez-R\`afols}\orcidlink{0000-0001-6979-0125},}
\author[28]{{Graziano~Rossi},}
\author[29]{{Eusebio~Sanchez}\orcidlink{0000-0002-9646-8198},}
\author[30]{{Marcin~Sawicki}\orcidlink{0000-0002-7712-7857},}
\author[3]{{David~Schlegel},}
\author[31, 32]{{Michael~Schubnell},}
\author[12]{{David~Sprayberry},}
\author[32]{{Gregory~Tarl\'{e}}\orcidlink{0000-0003-1704-0781},}
\author[12]{{Benjamin~A.~Weaver},}
\author[26]{{Hu~Zou}\orcidlink{0000-0002-6684-3997},}

\affiliation[1]{IRFU, CEA, Universit\'{e} Paris-Saclay, F-91191 Gif-sur-Yvette, France}
\affiliation[2]{IFAE, The Barcelona Institute of Science and Technology, Campus UAB, 08193 Bellaterra Barcelona, Spain}
\affiliation[3]{Lawrence Berkeley National Laboratory, 1 Cyclotron Road, Berkeley, CA 94720, USA}
\affiliation[4]{Perimeter Institute for Theoretical Physics, 31 Caroline St. North, Waterloo, ON N2L 2Y5, Canada}
\affiliation[5]{Department of Physics and Astronomy, University of Waterloo, 200 University Ave W, Waterloo, ON N2L 3G1, Canada}
\affiliation[6]{Institute for Computational Cosmology, Department of Physics, Durham University, South Road, Durham DH1 3LE, UK}
\affiliation[7]{Physics Dept., Boston University, 590 Commonwealth Avenue, Boston, MA 02215, USA}
\affiliation[8]{Departament de F\'isica, EEBE, Universitat Polit\`ecnica de Catalunya, c/Eduard Maristany 10, 08930 Barcelona, Spain}
\affiliation[9]{Aix Marseille Université, CNRS, CNES, LAM, 38 rue Frédéric Joliot-Curie, 13388 Marseille cedex 13, France}
\affiliation[10]{Department of Physics \& Astronomy, University College London, Gower Street, London, WC1E 6BT, UK}
\affiliation[11]{Instituto de F\'{\i}sica, Universidad Nacional Aut\'{o}noma de M\'{e}xico,  Cd. de M\'{e}xico  C.P. 04510,  M\'{e}xico}
\affiliation[12]{NSF NOIRLab, 950 N. Cherry Ave., Tucson, AZ 85719, USA}
\affiliation[13]{Department of Physics \& Astronomy and Pittsburgh Particle Physics, Astrophysics, and Cosmology Center (PITT PACC), University of Pittsburgh, 3941 O'Hara Street, Pittsburgh, PA 15260, USA}
\affiliation[14]{Dipartimento di Fisica Aldo Pontremoli, Universit\`a degli Studi di Milano, Via Celoria 16, I-20133 Milano, Italy}
\affiliation[15]{Departamento de F\'isica, Universidad de los Andes, Cra. 1 No. 18A-10, Edificio Ip, CP 111711, Bogot\'a, Colombia}
\affiliation[16]{Observatorio Astron\'omico, Universidad de los Andes, Cra. 1 No. 18A-10, Edificio H, CP 111711 Bogot\'a, Colombia}
\affiliation[17]{Fermi National Accelerator Laboratory, PO Box 500, Batavia, IL 60510, USA}
\affiliation[18]{NRC Herzberg Astronomy and Astrophysics, 5071 West Saanich Road, Victoria, BC V9E 2E7, Canada}
\affiliation[19]{Center for Cosmology and AstroParticle Physics, The Ohio State University, 191 West Woodruff Avenue, Columbus, OH 43210, USA}
\affiliation[20]{Department of Physics, The Ohio State University, 191 West Woodruff Avenue, Columbus, OH 43210, USA}
\affiliation[21]{The Ohio State University, Columbus, 43210 OH, USA}
\affiliation[22]{Sorbonne Universit\'{e}, CNRS/IN2P3, Laboratoire de Physique Nucl\'{e}aire et de Hautes Energies (LPNHE), FR-75005 Paris, France}
\affiliation[23]{Departament de F\'{i}sica, Serra H\'{u}nter, Universitat Aut\`{o}noma de Barcelona, 08193 Bellaterra (Barcelona), Spain}
\affiliation[24]{Instituci\'{o} Catalana de Recerca i Estudis Avan\c{c}ats, Passeig de Llu\'{\i}s Companys, 23, 08010 Barcelona, Spain}
\affiliation[25]{Department of Physics and Astronomy, Siena College, 515 Loudon Road, Loudonville, NY 12211, USA}
\affiliation[26]{National Astronomical Observatories, Chinese Academy of Sciences, A20 Datun Rd., Chaoyang District, Beijing, 100012, P.R. China}
\affiliation[27]{California Institute of Technology, Cahill Center for Astrophysics, Pasadena, CA, USA}
\affiliation[28]{Department of Physics and Astronomy, Sejong University, Seoul, 143-747, Korea}
\affiliation[29]{CIEMAT, Avenida Complutense 40, E-28040 Madrid, Spain}
\affiliation[30]{Department of Astronomy \& Physics and Institute for Computational Astrophysics, Saint Mary’s University, 923 Robie Street, Halifax, NS B3H 3C3, Canada}
\affiliation[31]{Department of Physics, University of Michigan, Ann Arbor, MI 48109, USA}
\affiliation[32]{University of Michigan, Ann Arbor, MI 48109, USA}
\emailAdd{constantin.payerne@cea.fr}

\abstract{In this paper, we investigate the possibility of selecting high-redshift Lyman-Break Galaxies (LBG) using current and future broadband wide photometric surveys, such as the  Ultraviolet Near Infrared Optical Northern Survey (UNIONS) or the Vera C. Rubin Legacy Survey of Space and Time (LSST), using a Random Forest algorithm. This work is conducted in the context of future large-scale structure spectroscopic surveys like DESI-II, the next phase of the Dark Energy Spectroscopic Instrument (DESI), which will start around 2029.
We use deep imaging data from the Hyper Suprime Camera (HSC) and the Canada-France-Hawaii Telescope Large Area U-band Deep Survey (CLAUDS) on the COSMOS and XMM-LSS fields. To predict the selection performance of LBGs with image quality similar to UNIONS, we degrade the $u, g, r, i$ and $z$ bands to UNIONS depth.
The Random Forest algorithm is trained with the $u,g,r,i$ and $z$ bands to classify LBGs in the $2.5 < z < 3.5$  range. 
We find that fixing a target density budget of $1,100$ deg$^{-2}$, the Random Forest approach gives a density of $z>2$ targets of $873$ deg$^{-2}$, and a density of $493$ deg$^{-2}$ of confirmed LBGs after spectroscopic confirmation with DESI. This UNIONS-like selection was tested in a dedicated spectroscopic observation campaign of 1,000 targets with DESI on the COSMOS field, providing a safe spectroscopic sample with a mean redshift of 3. This sample is used to derive forecasts for DESI-II, assuming a sky coverage of 5,000 deg$^2$. We predict uncertainties on Alcock-Paczynski parameters $\alpha_\perp$ and $\alpha_{\parallel}$ to be 0.7$\%$ and 1$\%$ for $2.6<z<3.2$, resulting in a potential 2$\%$ measurement of the dark energy fraction at high redshift. Additionally, we estimate the uncertainty in local non-Gaussianity and predict $\sigma_{f_{\rm NL}}\approx 7$, which would be comparable to the current best precision achieved by \textit{Planck}. The latter forecast suggests that achieving the precision required to place stringent constraints on inflationary models ($\sigma_{f_{\rm NL}} \approx 1$) using spectroscopic galaxy surveys necessitates the development of a next-generation (Stage V) spectroscopic survey.} 

\begin{document}
\maketitle
\newpage
\section{Introduction}
Lyman Break Galaxies (LBGs, \citep{Steidel1996LBG}) are young and actively star-forming galaxies at $z>1.5$. Due to absorption by neutral hydrogen present in the intergalactic medium (IGM) along the line-of-sight as well as within galaxies themselves, LBG rest-frame spectra display a decrement of flux bluewards the Lyman-alpha transition at 1216~\r{A}, down to the Lyman-limit at 912~\r{A}. The presence of these features enables the detection of these galaxies at redshifts $2.5 < z < 3.5$ using the $u$-dropout technique, which selects galaxies with a strong flux deficit in the $u$-band -- spanning approximately from 3300 to 4000~\r{A} -- compared to the flux measured in the $g$ or $r$ band flux \citep{RuhlmannKleider2024LBGCLAUDS}. Dropout techniques can be extended at higher redshifts (e.g. $g$- or $r$-dropouts \citep{malkan2017,ono2018,harikane2022,RuhlmannKleider2024LBGCLAUDS}) by measuring the flux decrement at wavelength above $4000$~\r{A}. LBGs are then good candidates to probe the early Universe using broadband photometry.

LBGs have historically played an important role in the study of galaxy formation and evolution at high redshifts \citep{Steidel1996LBG,Steidel1999LBG,Giavalisco2004LBG,Reddy2008LBG,Hildebrandt2009lbg,Finkelstein2022lbg,Harikane2023lbg}. More recently, they have emerged as promising cosmological probes, serving as highly biased tracers of the matter density field in the high redshift, matter-dominated Universe. LBGs enable the measurement of the growth of structure and the evolution of dark energy within the redshift range $2 < z < 6$ (e.g., via cross-correlation with CMB lensing; \citep{Miyatake2022lbgcmblensing}). Extensive high-redshift LBG samples covering large cosmic volumes offer opportunities to constrain primordial non-Gaussianities through large-scale dependent bias \citep{Schmittfull2018fnl,Chaussidon2024fnl} and to investigate the sum of neutrino masses \citep{Yu2018neutrinomass}. For clusters at $z > 1$, where shape measurement of background sources is challenging, the lensing magnification of high-redshift LBG samples can be used to estimate cluster masses \citep{Tudorica2017lbgmagnificationclusters} and constrain the low-redshift matter density field through inverse galaxy-galaxy lensing (IGGL; \citep{Cross2024IGGL}), which is largely independent of galaxy bias.

LBGs are currently under the scope of the Dark Energy Spectroscopic Instrument (DESI, \citep{DESI2022instrument}) science program. DESI is a multiplexed instrument currently installed on the Mayall telescope in Arizona and consists of a 4-meter focal plane equipped with 5,000 fiber-positioning robots and a bank of spectrographs fed by the fibers \citep{DESI2016paper,Silber2023DESI}. The DESI survey started in 2021 and has already provided millions of redshifts for different tracers with unprecedented precision up to $z\sim 1.5$, allowing to study the nature of dark energy and dark matter through Baryon Acoustic Oscillation (BAO) \citep{DESI2024bao1,DESI2024bao} and Redshift-Space Distortions (RSD) \citep{DESI2024rsd}. DESI-II, the next survey campaign of DESI will start around 2029 and aims to collect 40 million redshifts at higher densities/redshifts than DESI \citep{Schlegel2022DESI2} to address the problems of dark energy and inflation \citep{Ferraro2019inflation,Sailer2021cosmo,Chaussidon2024fnl}. LBGs at $ 2 < z < 4.5$ are of particular interest for DESI-II, expected to represent $\sim $ 2.5 million reconstructed redshifts. Optimized target selections for these high redshift tracers over the DESI footprint can be achieved by using completed and forthcoming multi-band and wide imaging surveys, including the DESI Legacy Surveys such as the Dark Energy Survey (DES, \citep{DES2005whitepaper}) and the Dark
Energy Camera Legacy Survey (DECaLS, \citep{Dey2019DESIILS}), but also the Ultraviolet Near Infrared Optical Northern Survey\footnote{\url{https://www.skysurvey.cc/}} (UNIONS, \citep{Ibata2017CFIS,Chambers2016panstarrs,Miyazaki2018HSC,Gwyn2025unions}), the Legacy Survey of Space and Time of the Vera C. Rubin Observatory\footnote{\url{https://rubinobservatory.org/about}} (LSST, \citep{LSST2009whitepaper}), and the \textit{Euclid} mission\footnote{\url{https://www.esa.int/Science_Exploration/Space_Science/Euclid}} \citep{Euclid2011whitepaper}, which will deliver imaging data in various optical and near-IR bands, of unprecedented depth and area overlapping with DESI footprint for this purpose.

This paper aims to use wide, broadband photometric surveys, such as the Canada-France Imaging Survey (CFIS, \citep{Ibata2017cfisu}), DECaLS as well as the ongoing UNIONS to select high-redshift LBGs in preparation for DESI-II. Our work is focused on UNIONS, but a similar strategy could be adopted with the future Rubin LSST that will provide $u$, $g$, $r$, $i$, $z$, and $y$-bands. In this work, we build a list of targets using a Random Forest algorithm based on UNIONS-like imaging on COSMOS field, which will be observed by DESI. We use the resulting LBG redshift distribution of spectroscopically-confirmed LBGs to conduct Fisher forecast on several cosmological parameters. Our work is focused on UNIONS, but similar strategy could be adopted with the future Rubin LSST that will provide $u$, $g$, $r$, $i$, $z$ and $y$-bands. 

The paper is organized as follows: In \cref{sec:available _dataset} we present the different datasets that are, or will be, available for LBG selection on large footprints promised by wide photometric surveys. We also present a method to degrade existing deep photometry to shallower depths to simulate the properties of future wide imaging surveys. \cref{sec:selection_method} presents the Random Forest method used in this work to select LBGs from multi-band $ugriz$ photometry, which we compare to the more standard color-color box selection used in~\cite{RuhlmannKleider2024LBGCLAUDS}. In \cref{sec:validation}, we first present how we create a sample of LBG targets from simulated photometry at UNIONS depth on the COSMOS field. First, we use a method to degrade existing deep photometry to shallower depths to simulate the properties of future wide imaging surveys.  Then, we validate the degradation method by studying the performance of a Random Forest-based target selection using CFIS+DECaLS imaging available on the XMM-LSS field. 
Then, we create a sample of UNIONS-like LBG targets on the COSMOS field and describe the sample of LBGs observed by the DESI pilot survey. 
We present in \cref{sec:forecast} several forecasts on the precision of several cosmological parameters that can be achieved with this new selection and future spectroscopic surveys. We conclude in \cref{sec:conclusion}.
\section{Optical photometry from imaging surveys}
\label{sec:available _dataset}
\begin{figure*}[t]
    \centering
    \includegraphics[width=1\textwidth]{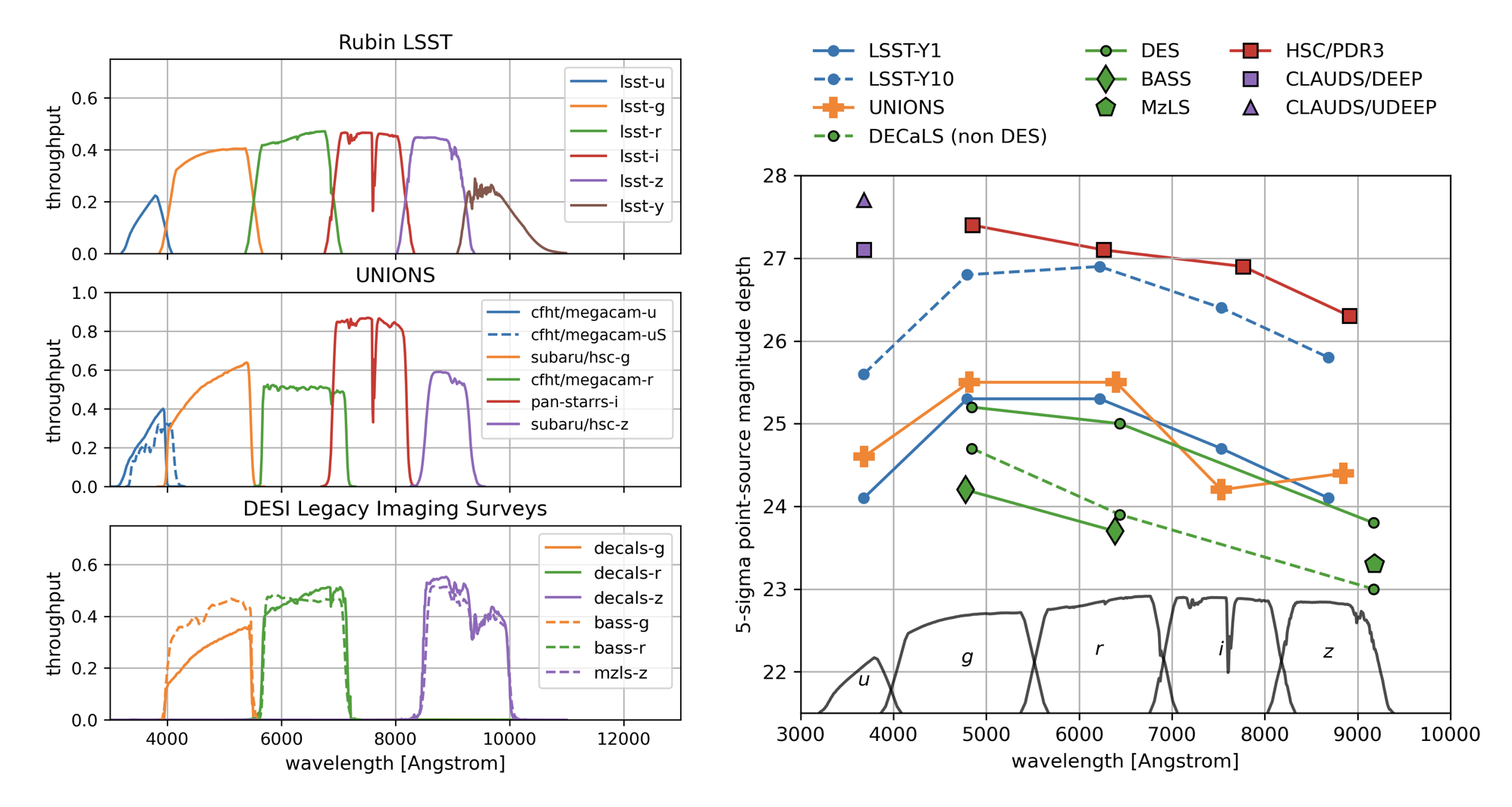}
    \caption{\textit{Left}: Total system throughput for various surveys. Top: LSST; Middle: UNIONS (CFHT/CFIS/MegaCam-$u(u^{\star})r$, HSC-$gz$, and Pan-STARRS-$i$); Bottom: DESI Legacy Imaging Surveys (DECaLS-$grz$, BASS-$gr$, and MzLS-$z$).  \textit{Right}: 5$\sigma$ point-source magnitude depth in the $ugriz$ bands as a function of the mean wavelength for each band's system throughput, across the photometric surveys (green: DESI Legacy Imaging Surveys, blue: LSST, orange: UNIONS, red: HSC/PDR3, violet: CLAUDS). LSST system throughput is included for reference (arbitrary $y$-scaling), aiding visual identification of the different bands. }
    \label{fig:response} 
\end{figure*}

This section presents current and future photometric surveys providing $ugriz$ imaging. For illustration, we show in \cref{fig:response} (left panel) the system throughput\footnote{These curves account for the contribution from the telescope (mirrors, lenses, filter), the detector quantum efficiency, and the zenith atmosphere absorption. For DECaLS (bottom left panel), the atmosphere absorption is given by the airmass of $X = 1.4$ (see \url{https://www.legacysurvey.org/dr7/description/}). For the other transmission curves, $X$ is set to $1.2$, and they are taken from the Spanish Virtual Observatory, available at \url{http://svo2.cab.inta-csic.es/theory/fps/index.php}.} for the different experiments (LSST, UNIONS and the DESI Legacy Imaging Surveys) discussed in \cref{sec:wide_imaging_surveys}. The \cref{fig:response} (right panel) displays the 5$\sigma$ magnitude depth (point source) in the $ugriz$-band, for the different surveys we use through this paper, and \cref{tab:depth_mag_photo_surveys} summarizes the tabulated values. We refer the reader to Table 2 of \citep{Zhao2024must}, which provides a more exhaustive list of per-band magnitude depths for various ongoing and future imaging surveys. Our study relies on the COSMOS and XMM-LSS fields, which are both covered by the dataset we use hereafter (CLAUDS, HSC, DECaLS, and CFIS). These fields are well-studied areas of the sky to bench test target selections for new tracers. 

\subsection{Deep imaging surveys}
In this section, we review the various deep imaging surveys that will be used throughout this paper as part of our photometry degradation method to simulate UNIONS-like imaging.
\subsubsection{HSC/PDR3}
The Hyper Suprime Camera (HSC, \citep{Aihara2019HSC})  is mounted on the 8-meter Subaru Telescope at the Mauna Kea Observatory in Hawaii. The last Public Data Release 3 (PDR3) of HSC Subaru Strategic Program (HSC-SSP) in 2021 includes deep multi-band data over 36 deg$^2$. The 5$\sigma$ point source magnitude depths listed in \cref{tab:depth_mag_photo_surveys} are taken from Table 1 in \citep{Aihara2019HSC} (Deep/UltraDeep). In this work, we focus on the HSC $griz$ data which were combined with CLAUDS (CLAUDS is detailed hereafter) on deep (XMM-LSS, COSMOS, ELAIS-N1, DEEP2-F3) and ultra-deep (E-COSMOS in COSMOS, and SXDS in XMM-LSS) fields, over 26 and 4 deg$^2$, respectively. The combination of CLAUDS and HSC-SSP data over the XMM-LSS and COSMOS fields is detailed in \citep{Desprez2023CLAUDSHSCSSP}. We will use a combination of deep and ultra-deep fields, and from \citep{Desprez2023CLAUDSHSCSSP} (see their Table 2), the magnitude depths for the $griz$-bands on XMM-LSS range from 26.5 to 24.0 (resp. from 26.9 to 24.9) for the deep (resp. ultra-deep) field. Moreover, the magnitude depths for the $griz$-bands on COSMOS range from 26.4 to 24.7 (resp. from 27.0 to 25.5) for the deep (resp. ultra-deep) field.

\subsubsection{CLAUDS}
HSC was supplemented by the U-band imaging from the Canada-France-Hawaii Telescope Large Area U-band Deep Survey (CLAUDS, \citep{Sawicki2019CLAUDS}). CLAUDS used the MegaCam imager mounted on the 3.6-meter CFHT telescope to provide very deep $U$-band imaging (depth in $U$ of 27.1 and 27.7 AB in the deep and ultra-deep fields, respectively)
over $18.60$ deg$^2$. CLAUDS was designed to follow the HSC
deep fields to similar depth, astrometric solution, and pixel scale. Moreover, CLAUDS $U$-imaging uses the original MegaCam $u^*$ filter and an upgraded $u$ filter installed in 2014, which has better throughput and covers the entire MegaCam mosaic of 40 CCDs (compared to 36 with the original filter). The different characteristics of the $u$ and $u^*$ filters are detailed in \citep{Sawicki2019CLAUDS} (see their Figure 2) and filter parameters are available on the MegaCam filter database\footnote{In the table at \url{https://www.cfht.hawaii.edu/Instruments/Filters/megaprime.html}, $u^*$ is labeled as $U'$, and $u$ as $u'$.}. The XMM-LSS field was imaged with the $u^*$ filter, while both $u$ and $u^*$ filters were used for the COSMOS field. The 5$\sigma$ point source (in 2 arcsec apertures) magnitude depths listed in \cref{tab:depth_mag_photo_surveys} are taken from \citep{Sawicki2019CLAUDS}.

\subsection{Wide imaging surveys}
\label{sec:wide_imaging_surveys}
In the following, we examine the features of various current and upcoming wide broadband imaging surveys. The footprints of DECaLS, as well as the other DESI Legacy Imaging Surveys BASS and MzLS, are displayed in \citep{Dey2019DESIILS} (see their Figure 1). The UNIONS footprint (combining the CFHT/CFIS, HSC, and Pan-STARRS) can be found on the official UNIONS public webpage\footnote{\url{https://www.skysurvey.cc/survey/}}. The LSST footprint can be found in \citep{Bianco2022lsst} (see their Figure 1). Throughout this paper, the existing DeCaLS photometry on XMM will be used to validate our degradation method and machine-learning-based target selection, while UNIONS depth will be employed to compile a list of targets for DESI spectroscopic confirmation.
\subsubsection{DECaLS}
The Dark Energy Camera Legacy Survey (DECaLS) is one of the three public projects of the DESI Legacy Imaging Surveys\footnote{\url{http://legacysurvey.org/}} \citep{Dey2019LegacySurvey} with the Beijing–Arizona Sky Survey (BASS) and the Mayall z-band Legacy Survey (MzLS), that aim to provide the targets for the DESI survey over 14,500 deg$^2$. DECaLS makes use of the Dark Energy Camera (DECam~\citep{Flaugher2015DECAM}) on the Blanco 4-meter telescope (in Cerro Tololo, Chile) that was initially built to conduct the Dark Energy Survey (DES), to provide the optical imaging in $grz$ bands over 9,000 deg$^2$. Through this paper, we use the data release 9 (DR9) of the Legacy Imaging Surveys and whose magnitude depths are shown in \cref{tab:depth_mag_photo_surveys}.
\begin{table}
\begin{center}
\begin{tabular}{l||c|c|c|c|c}
Surveys& $u$ &$g$ & $r$ &$i$&$z$\\
\hline 
\hline 
 CLAUDS (udeep)& 27.1(27.7)  &-& -&-&-\\
HSC/PDR3& - &27.4 &27.1 &26.9 &26.3\\
CFIS& 24.6&-&25.5&-&-\\
DECaLS& -  &24.7&23.9&-&23.0\\
UNIONS& 24.6&25.5&25.5&24.2&24.4\\
LSST Y1(Y10)& 24.1(25.6)&25.3(26.8)&25.3(26.9)&24.7(26.4)&24.1(25.8)\\
\end{tabular}
\end{center}
\caption{Magnitude depth for different present and future imaging surveys. Numbers in parentheses indicate either the depth of the ultra-deep fields included in the CLAUDS program (compared to the depth of deep fields), or the depth after 10 years of observations with LSST (compared to 1 year).}

\label{tab:depth_mag_photo_surveys}
\end{table}

\subsubsection{UNIONS}

UNIONS is a collaboration between the Hawaiian observatories Canada-France-Hawaii Telescope (CFHT, Mauna Kea), the Panoramic Survey Telescope and Rapid Response System (Pan-STARRS, Maui), and the Subaru telescope (Mauna Kea). UNIONS is currently providing $ugriz$ imaging for 5,000 deg$^2$ of the northern sky. First, the CFHT Canada-France Imaging Survey (CFHT/CFIS) is targeting the $u$-band and $r$-band with the Megacam imager at CFHT and will provide competitive image quality to all other current, large, ground-based facilities, up to a magnitude depth of 25 for the $r$-band over 5,000 deg$^2$ and to a $u$-band depth of 24.6 over 9,000 deg$^2$. At the same time, Pan-STARRS is obtaining the $i$-band, and the Wide Imaging with Subaru Hyper Suprime-Cam of the Euclid Sky (WISHES) will provide the $z$-band. UNIONS has already completed more than $80\%$ of the survey, and the first data release was available on the XMM-LSS field as we ended this manuscript. So the 5-6 year program of DESI-II that will start in 2029 could benefit from these datasets in the Northern sky. Expected magnitude depths to be reached by the ongoing UNIONS in the $u$, $g$, $r$, $i$, and $z$ bands are listed in \cref{tab:depth_mag_photo_surveys} \citep{Gwyn2025unions,EuclidCollaboration2022survey}.

Moreover, the XMM-LSS field was observed with a strategy strictly identical to CFIS-$u$. These observations on XMM-LSS aim at testing LBG selection at a shallower depth than CLAUDS (2-3 magnitudes deeper), to extend the LBG science program over the extensive CFIS-$u$ sky coverage of several thousands square degrees. In the latter, observations of XMM-LSS \textit{à-la} CFIS will be referred to as CFIS observations for simplicity.

\subsubsection{LSST}
The Vera C. Rubin Observatory, still under construction, is designed to conduct a 10-year wide-area, deep, multi-band optical imaging survey of the night sky visible from Chile. The Legacy Survey of Space and Time (LSST~\citep{LSST2012whitepaper}) will catalog about 20,000 deg$^2$ of the southern sky starting in 2026 in the $ugrizy$ bands. After 1 year and after 10 years of observations, the survey will reach respective magnitude depths\footnote{The 5-sigma point source depths for LSST are reported from the latest Rubin simulation v3.6, whose details can be found at \href{https://usdf-maf.slac.stanford.edu/}{https://usdf-maf.slac.stanford.edu/}} as indicated in \cref{tab:depth_mag_photo_surveys}.

\subsection{Simulating shallower depth}
\label{sec:shallowing_method}
In this section, we present a method to degrade the imaging of a deep survey to mimic a shallower imaging. This step is mandatory since we want to test the LBG selection at UNIONS-like depth, whose images are still being processed. In this section, we make use of the different imaging datasets available on the XMM-LSS field, namely CLAUDS+HSC imaging (the deep survey) that we will degrade to CFIS+DECaLS-depth (the shallower survey), and compare to the existing CFIS+DECaLS dataset on XMM-LSS. Let us note that XMM-LSS was not surveyed by the CLAUDS-$u$ filter but with the CLAUDS-$u^*$ one, so we consider CLAUDS-$u^*$ in this section but will use CLAUDS-$u$ on COSMOS. 

We first define the typical magnitude error, for a given flux and magnitude depth. Using the galaxy AB magnitude system given by $m = 22.5 -2.5\log_{10}(f)$, the per-galaxy magnitude error $\Delta m$ is linked to the galaxy flux noise-to-signal ratio $NSR = \Delta f/f$ via $\Delta m = s NSR$, where $s = 2.5/\log(10)\approx 1.0857$. Here, we consider the flux as a Gaussian random variable, which is generally assumed in the literature \citep{Euclid2021phootometry,Rykoff2015magnitudeerror} and is used in computing object magnitude errors using e.g. \texttt{Sextractor} \citep{Bertin1996SExtractor}. Moreover, the latter magnitude error expression is obtained by considering the flux high signal-to-noise regime \citep{Ivezic2019LSST}, such as errors are assumed to be Gaussian also in magnitude space. From \citep{Ivezic2019LSST}, the flux $NSR$ can be decomposed in two distinct contributions (i) $NSR_{\rm rand}$ is the random photometric error (associated with the intervening random processes in photon counts) and (ii) $NSR_{\rm sys}$ is the systematic photometric error (due to, e.g., imperfect modeling of the Point-Spread Function, but not including uncertainties in the absolute photometric zero-point). For the latter, we consider the optimistic case by neglecting the systematic uncertainties. The random photometric $NSR$ expresses as \citep{Ivezic2019LSST,Graham2020photometry} (see calculation details in \citep{crenshaw2024})
\begin{equation}
    NSR^2_{\rm rand} = (0.04 - \gamma) x + \gamma x^2 
\end{equation}
where $x = 10^{0.4(m-m_5)}$. Here, $m_5$ is the $5\sigma$ point-source depth in a given band and represents the magnitude of an object with a $NSR_{\rm rand}=1/5$. In the above equation, $\gamma$ depends on sky brightness, readout noise, etc., which can be determined from the system throughput. For the latter, we use the fiducial value of $\gamma = 0.04$ which sets the impact of, e.g., sky brightness, readout noise, to be zero. From these assumptions, we have adopted a simple model for the magnitude error similar to the procedure described in \citep{Euclid2021phootometry}; For a galaxy with index $k$, the error on the magnitude is then given by
\begin{equation}
    \Delta m_k = s NSR_{\rm rand} =  s\times 0.2\times 10^{0.4(m_k - m_\mathrm{depth})}.
    \label{eq:Deltam_depth}
\end{equation}
which depends only on the magnitude depth in the considered band $m_5$. Using this flux error model, we randomize the galaxy flux $f_{k, \mathrm{deep}}$ from the deep imaging to $f_{k, \mathrm{shallow}}$ following 
\begin{equation}
f_{k, \mathrm{shallow}} \sim \mathcal{N}(\mu=f_{k, \mathrm{deep}}, \sigma^2 = [\Delta f_k]_{ \mathrm{shallow}}^2 - [\Delta f_k]_{\mathrm{deep}}^2)
\end{equation}
where $[\Delta f_k]_{ \mathrm{shallow}}$ and $[\Delta f_k]_{ \mathrm{deep}}$ are the flux errors measured with the "shallow" and "deep" magnitude depths (by inverting \cref{eq:Deltam_depth}), respectively. 
Then, the degraded magnitude is given by $m_{k, \mathrm{shallow}} = 22.5 - 2.5\log_{10}(f_{k, \mathrm{shallow}})$, and its error is 
\begin{equation}
    [\widehat{\sigma}_{m, \mathrm{shallow}}]_k = [\widehat{\sigma}_{m, \mathrm{deep}}]_k \times 10^{0.4(m_{k, \mathrm{shallow}} - m_{k, \mathrm{deep}})}\times 10^{-0.4(m_{\rm depth, shallow} - m_{\rm depth, deep})}.
\end{equation}
The first term is the measured magnitude error (corresponding to the deep imaging), the second term shifts the error to a higher value if the magnitude $m_{\rm shallow} > m_{\rm depth}$, according to our error model, and the third term shifts the magnitude error to a higher value due to the difference in depths since $m_{\rm depth, shallow} < m_{\rm depth, deep}$. This magnitude error correction ensures that the "updated" magnitude error follows on average the error model in \cref{eq:Deltam_depth}, applied to the new magnitude depth. So this method can be used on a homogeneous deep field (input) to simulate shallower imaging (output), from which we can compute the randomized fluxes. 

All the assumptions we have made through these calculations enable a simple scaling between the input deep and out shallow photometry \citep{Euclid2021phootometry}. However, we recall that the scaling between two different depths works for point sources (i.e. the scaling between two different depths), since object size plays a role in rescaling the point-source $NSR$ (see e.g. \citep{vandenBusch2020photometrykids}, who proposed a scaling of galaxy magnitude between different depths including photometric aperture areas in the sky-dominated regime). Moreover, the sky-dominated regime we imposed by choosing the per-band parameter $\gamma=0.04$ in all photometric bands may be slightly optimistic, especially in the $u$-band, due to e.g. high atmospheric absorption in the ultra-violet. For instance \citep{Ivezic2019LSST} (see their Table 2) provided $\gamma = 0.38$ for LSST-like photometry in the $u$-band. Moreover, this methodology assumes that the input and output band filters are similar, which may not be the case when converting magnitudes between two different instruments (as it will be the case when moving from CLAUDS imaging to CFIS imaging). For all these reasons, our simulation procedure is however optimistic and does not include all the specific features from the input dataset and the desired characteristics of the output dataset.
\begin{figure*}[t]
    \centering
    \includegraphics[width=.49\textwidth]{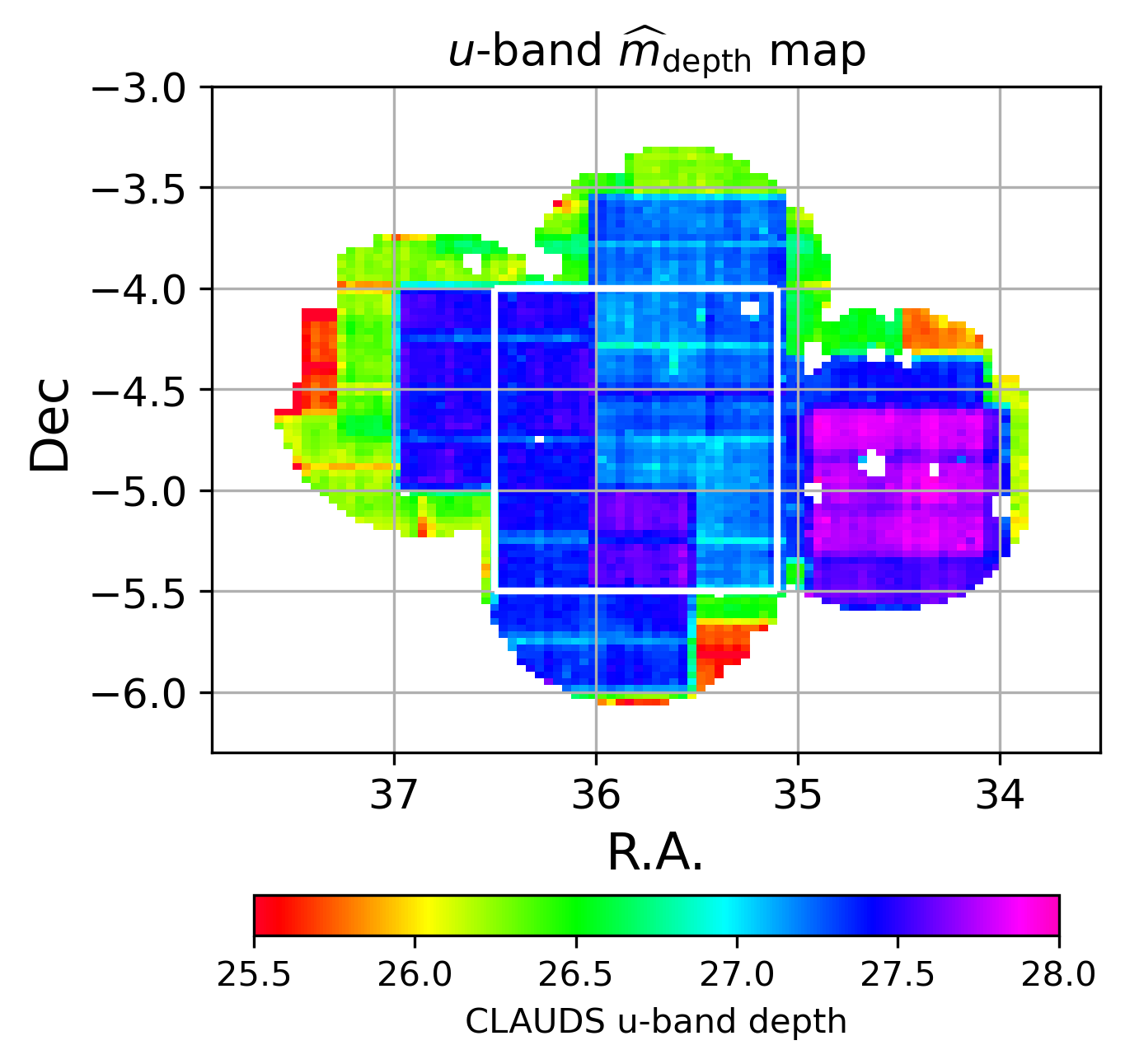}
    \includegraphics[width=.49\textwidth]{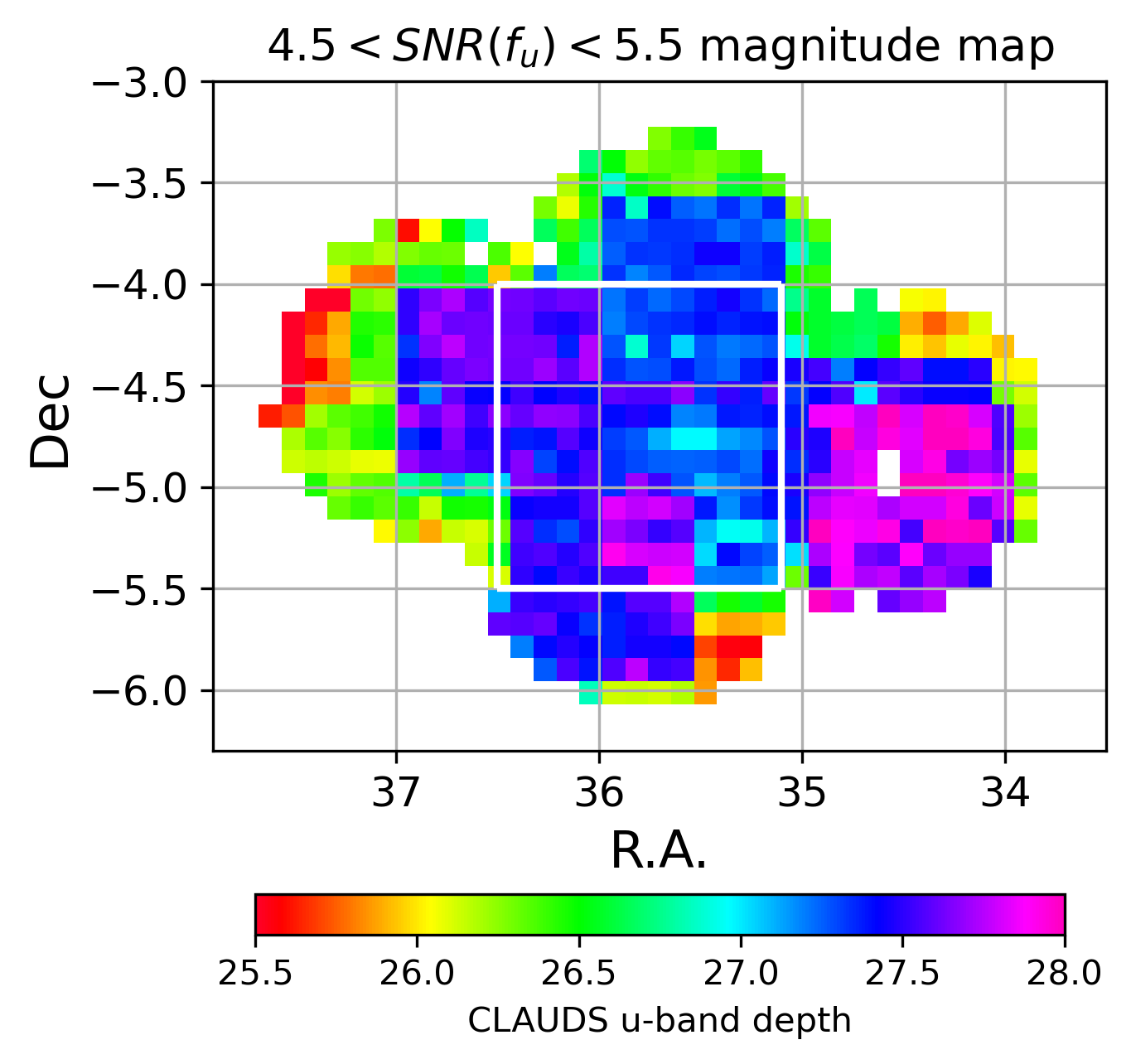}
    \caption{\textit{Left:} Binned map of the CLAUDS $u^*$-band magnitude depth estimates over the XMM-LSS field. \textit{Right:} Binned map of $u^*$-band magnitudes corresponding to a flux signal-to-noise ratio between 4.5 and 5.5.}
    \label{fig:err_u_CFIS_map}
\end{figure*}

We can apply this methodology to degrade CLAUDS+HSC to a CFIS+DECalS depth. CFIS+DECaLS magnitude depths for $u,g,r,z$ (taken from the literature) are listed in \cref{tab:depth_mag_photo_surveys}. For the three datasets, we use a cut on the magnitude error $\sigma(\mathrm{mag})<5$ in all bands and we only consider galaxies within $23 < r < 24.3$. Such magnitude cuts in the $r$-band are motivated by (i) the fact that most LBGs at a redshift $\sim 2.5-3$ are above a $r$-band magnitude of 23, then the lower limit cut removes stars and other galaxies (ii) the fact that the efficiency of the spectroscopic redshift measurement for very faint objects (with a $r$-band magnitude higher than 24.3) is low \citep{RuhlmannKleider2024LBGCLAUDS} for a typical exposure time of 2 hours (again, the details on effective time and spectroscopic redshift measurement are presented hereafter in \cref{sec:unions-like_ts_cosmos}).

The magnitude depths in \cref{tab:depth_mag_photo_surveys} are indicative only and do not take into account possible inhomogeneities in the survey depth over the XMM footprint for CLAUDS, HSC, CFIS, and DECaLS. To illustrate this, we reconstruct the two-dimensional depth map of the CLAUDS $u^*$-band imaging using two different methods. First, using our error model in \cref{eq:Deltam_depth}, we convert each measured magnitude $\widehat{m}_i$ and its error $\Delta\widehat{m}_i$ on an estimate of the survey depth $\widehat{m}_{\mathrm{depth}, i}$ in the considered band. The binned two-dimensional map of magnitude depth estimates is represented in \cref{fig:err_u_CFIS_map} (left panel). We see three different regions (the ultra-deep region in one of the HSC pointing, the deep region, and the edges with shallower depths, see Figure 3 in \citep{Sawicki2019CLAUDS}). 

The method detailed above relies on a given error model for the measured magnitude. For the second method, we can estimate the magnitude depth independently of our error model, by measuring the binned map of measured magnitudes that have a flux signal-to-noise ratio in the $u^*$-band between 4.5 and 5.5. The corresponding binned average map of these magnitudes is shown in  \cref{fig:err_u_CFIS_map} (right), displaying the same features as our previous method, however noisier since we use fewer galaxies.

\begin{figure*}[t]
    \centering
    \includegraphics[width=1\textwidth]{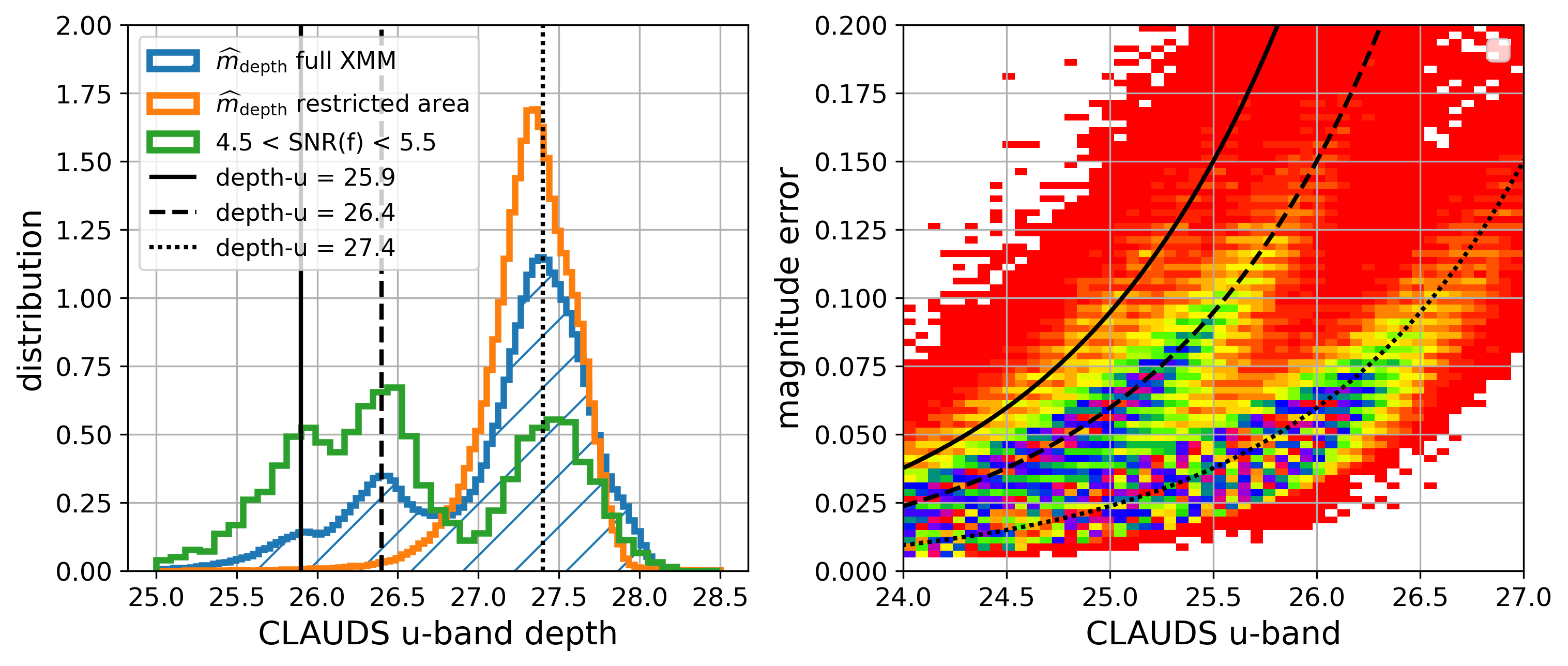}
    \caption{\textit{Left}: Distribution of magnitude $u$-depth estimated from \cref{eq:Deltam_depth}. The vertical bars indicate the average position of the three distinct peaks in the distribution. \textit{Right}: The colored histogram (arbitrary scaling) is the distribution of magnitude error in the $u$-band as a function of the $u$-band magnitude. Black lines are the magnitude error models.
    }
    \label{fig:err_model_plot_u}
\end{figure*}

In \cref{fig:err_model_plot_u} (left panel), we show in blue the histogram of the depth estimates from our first method over the full XMM footprint. We see three distinct features in the distribution, represented by the vertical lines (positioned by hand), given by $m_{\rm depth}=[25.9,  26.4, 27.4]$. Similarly, we show in green the histogram of object magnitudes with flux signal-to-noise ratio between 4.5 and 5 in the $u^*$-band, displaying the three same features. 

In \cref{fig:err_model_plot_u} (right) we show the magnitude error against magnitude for the $u$-band, and we over-plot the three error models using respectively $m_{\rm depth}=[25.9,  26.4, 27.4]$. Again, we see that the error models follow the three distinct patterns in the error-magnitude distribution, revealing that a single error model with a unique $m_{\rm depth}$ is not relevant in this case. 

For that reason, for the rest of the analysis with CLAUDS+HSC data, we restrict to the central 2 deg$^2$ region of the CLAUDS XMM-LSS imaging \cref{fig:err_u_CFIS_map}, compared to the full field of $4.13$ deg$^2$. The histogram of the $u^*$-band depth estimates in this restricted field is displayed in orange in \cref{fig:err_model_plot_u} (left panel). In this restricted area, the recovered magnitude depths for CLAUDS+HSC are presented in \cref{tab:depth_mag_photo_surveys_XMM_COSMOS} (third line),
showing a deeper average $u$-band magnitude than the full XMM footprint (second line). After these cuts, the final CLAUDS+HSC dataset comprises 41,000 objects.

Second, we similarly want to restrict to a homogeneous footprint in depth in the CFIS+ DECaLS imaging. The CFIS observations, coming from a dedicated program isolated from the CFIS footprint, have non-negligible edges with shallower coverage: we reject regions covered by one CFIS exposure only (we keep the number of exposures to be equal to 2 or 3) over the XMM footprint. After these cuts, the CFIS+DECaLS dataset comprises 71,000 objects. The recovered magnitude depths after applying this cut are given in the first line of \cref{tab:depth_mag_photo_surveys_XMM_COSMOS}. 

\begin{figure}
     \centering
 \includegraphics[width=1\textwidth]{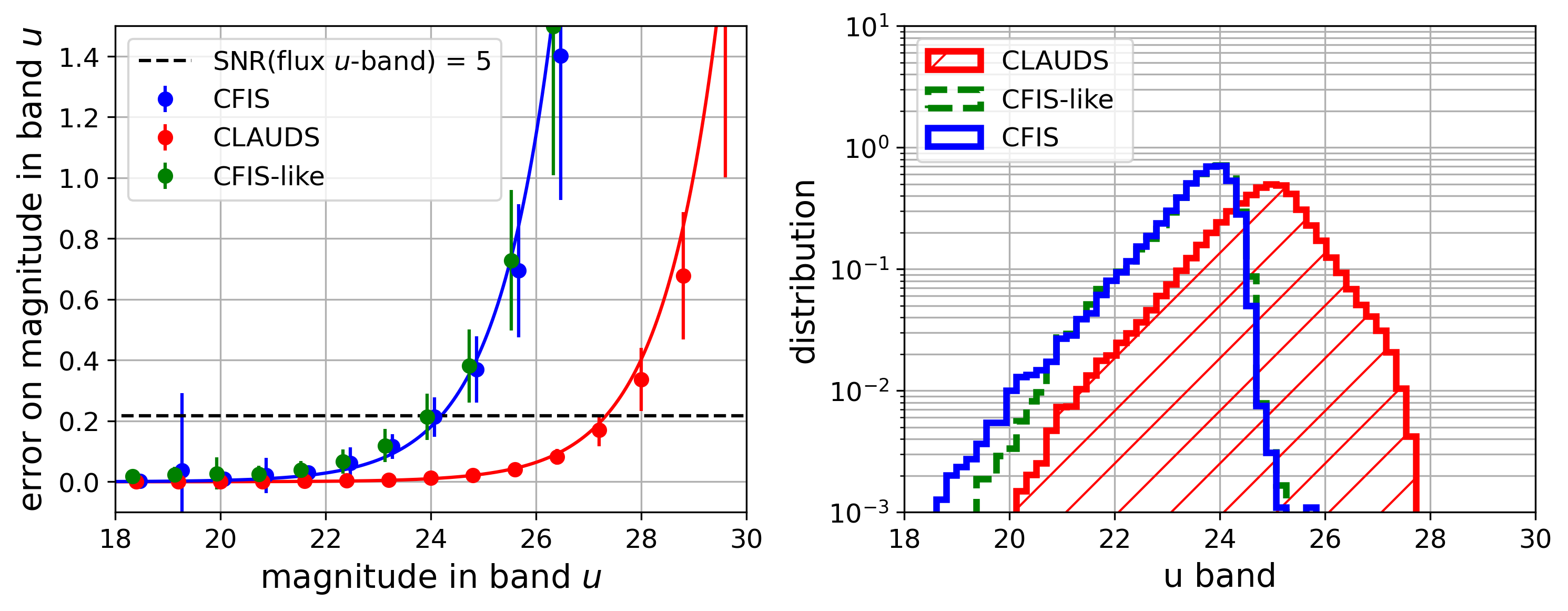}
     \caption{\textit{Left}: magnitude error as a function of $u$-band magnitude. Dots and error bars 
     present results for
     CLAUDS imaging (red), CLAUDS imaging degraded to CFIS+DECaLS depth (green), and actual CFIS+DECaLS imaging (blue). \textit{Right}: normalized magnitude distribution in the $u$-band after a cut SNR$(f_u)>5$. }
     \label{fig:validation_degraded_photom}
\end{figure}

After such a treatment and considering input (i.e., CLAUDS+HSC) and output (i.e., CFIS+DECaLS) depths in \cref{tab:depth_mag_photo_surveys_XMM_COSMOS}, we can proceed to the degradation of the CLAUDS+HSC magnitudes. We show in \cref{fig:validation_degraded_photom} (left panel) the
magnitude error as a function of the $u$-band magnitude for the three different datasets: the input CLAUDS+HSC in blue, the output CFIS+DECaLS-like (i.e. degraded CLAUDS+HSC imaging) in green, and the targeted imaging CFIS+DECaLS in blue. The error models with depths from \cref{tab:depth_mag_photo_surveys_XMM_COSMOS} match well the true CFIS+DECaLS magnitude errors.
In \cref{fig:validation_degraded_photom} (right panel), we show the $u$-band magnitude distribution of the three different datasets, after applying a $u$-band flux signal-to-noise ratio cut  $> 5$, represented by the back dashed line in \cref{fig:validation_degraded_photom} (left panel). The similarity between the CFIS+DECaLS-like and CFIS+DECaLS $u$($u^*$)-band magnitude distributions validates that the degradation method works as expected. We repeated the same procedure on the CLAUDS $grz$ bands and found the same conclusions when comparing them to true CFIS+DECaLS imaging.

\begin{table}
\begin{center}
\begin{tabular}{l||l|c|c|c|c|c} 
Survey & Photometry & $u$($u^*$) &$g$ & $r$ &$i$&$z$\\
\hline
\hline
\multirow{3}{5em}{XMM-LSS} & CFIS+DECaLS& 24.2 & 24.9 & 24.7 & - & 23.7 \\
&CLAUDS+HSC - full & (27.3) & 26.6 & 25.9 & 25.6 & 25.4 \\
&CLAUDS+HSC - partial& (27.3) & 26.6 & 25.8 & 25.5 & 25.4 \\
\hline
\multirow{2}{5em}{COSMOS}
&CLAUDS+HSC & 27.0 (26.3)&27.0& 26.7& 26.5 & 26.0\\
&UNIONS-like & 24.6 & 25.5&25.1&24.2 &24.4\\
\end{tabular}
\end{center}
\caption{Measured 5-sigma point source magnitude depth on the XMM-LSS (first line) and COSMOS (second line) fields. "full" indicates the full XMM-LSS fooprint, displayed in color in \cref{fig:err_u_CFIS_map}. The "partial" denotes for data within the white box (again in \cref{fig:err_u_CFIS_map}).}
\label{tab:depth_mag_photo_surveys_XMM_COSMOS}
\end{table}


\section{Selection methods}
\label{sec:selection_method}
This work investigates the feasibility of selecting high redshift LBGs from their imaging properties with a Random Forest algorithm. We also aim to compare Random Forest selection with the more common color-color box selection method used in the literature.
This section uses CLAUDS+HSC deep field imaging data and photometric redshifts obtained from CLAUDS and HSC-SSP imaging data \citep{Desprez2023CLAUDSHSCSSP} with the \textsc{LePHARE} template-fitting code \citep{Arnouts1999,Arnouts2002,Ilbert2006}. 

\subsection{Color-color box selection}
\label{sec:coloc-color}
\begin{figure}
    \centering
    \includegraphics[width=.48\textwidth]{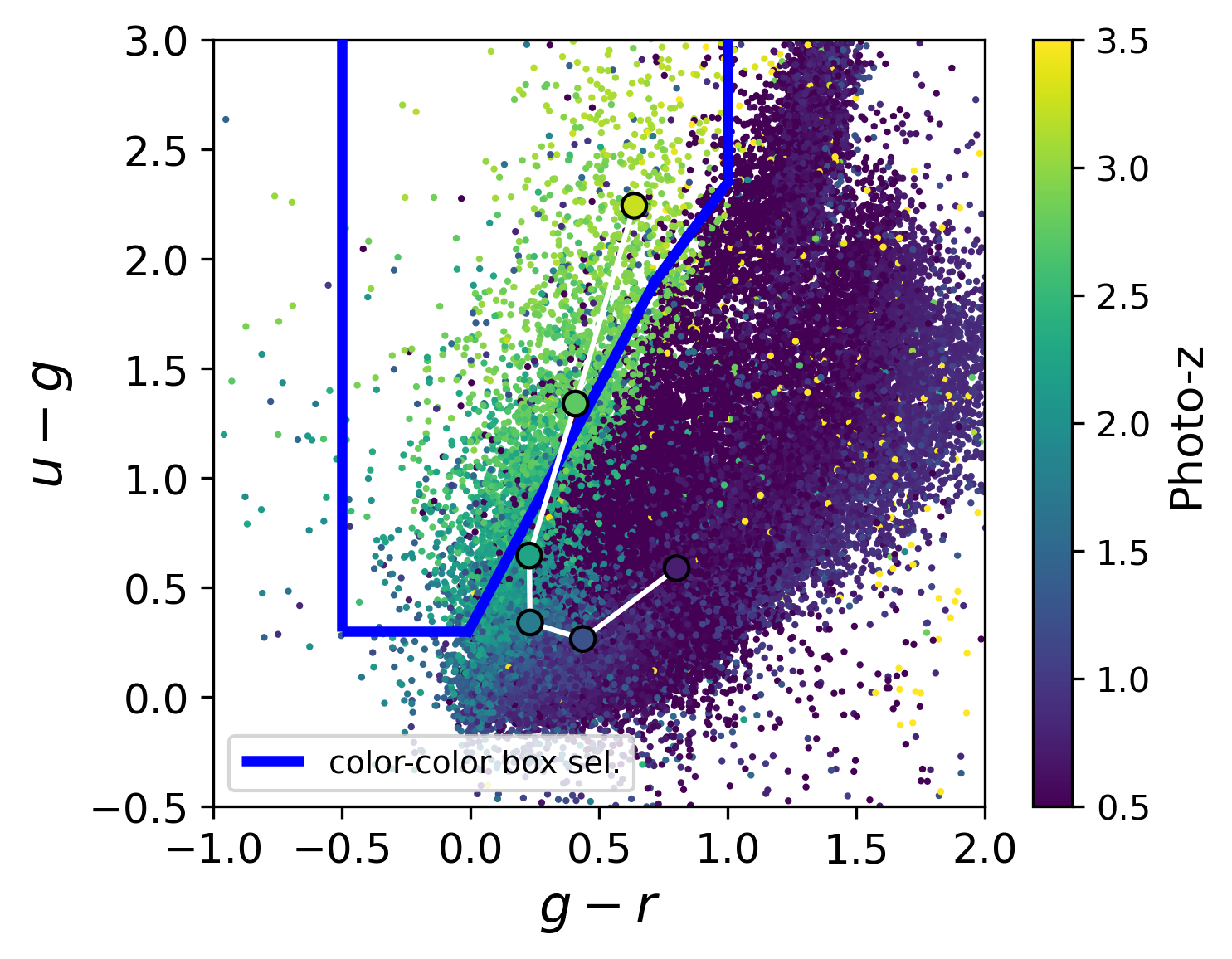}
    \includegraphics[width=.48\textwidth]{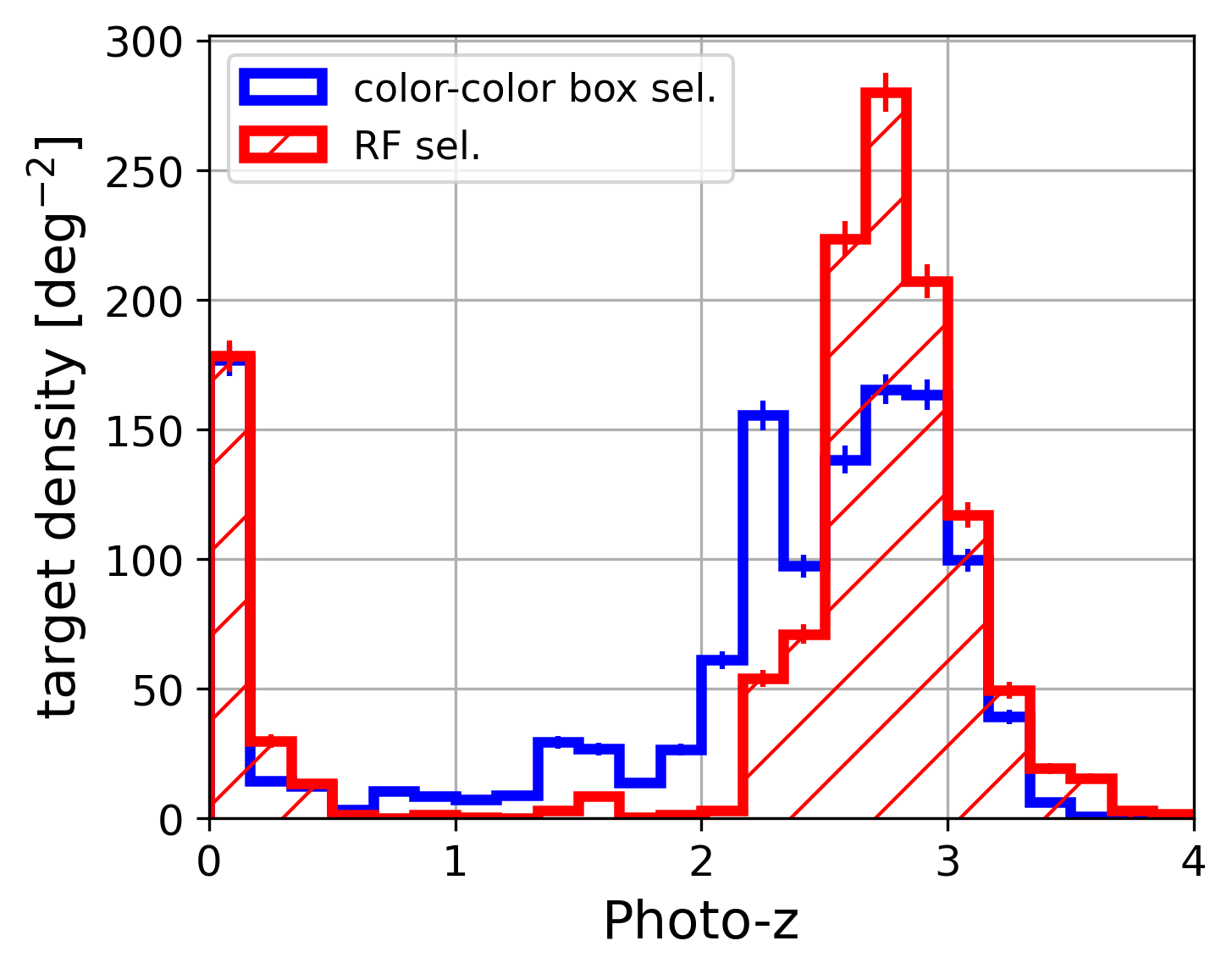}
    \caption{\textit{Left}: color-color plot illustrating the selection of LBGs used in~\cite{RuhlmannKleider2024LBGCLAUDS} (CLAUDS+HSC photometry with $22.5 < r < 23.75$). The green/yellow dots represent the parent
    sample. The blue polygon corresponds to the color-color box selection. For illustration purposes, 
    the black-edged dots are the mean positions of the parent sample in 6 different redshift bins with width $\Delta z = 0.5$, ranging from $z=0.5$ to $z=3.5$. \textit{Right}: photometric redshift distribution of LBG targets, obtained from the Random Forest classification in red, and the color-color box selection in blue.}
    \label{fig:color_color_box_nz_box_RF}
\end{figure}

The color-cut selection method for selecting high redshift LBGs uses the flux decrement blueward of the Lyman limit of LBG spectra in their rest frame, due to absorption by neutral hydrogen. High redshift LBGs are then selected for their lack of emission in the $u$-band, compared to their observed flux in other bands \citep{RuhlmannKleider2024LBGCLAUDS}. This $u$-dropout method allows LBGs to be selected in the redshift range $[2.5, 3.5]$. Similarly, the $g$-dropout (lack of flux in the $g$-band) can be used to select LBGs at redshift $3.5 < z < 4.5$. In this section, we use the color-color box selection referred to as [COSMOS: TMG $u$-dropout] in \citep{RuhlmannKleider2024LBGCLAUDS} (see their Table 1), with cut in the $r$-band given by $22.5 < r < 23.75$, and $\mathrm{err}(u) < 1$. The target selection is given by
\begin{align}
    &(i)\ u - g > 0.3\\
    &(ii)\ -0.5< g-r<1\\ 
    &(iv)\ [u - g > 2.2\times(g - r) + 0.32] \cup [u - g > 0.9 \cap u - g > 1.6\times(g - r) + 0.75]
\end{align}
where $\mathrm{err}(u)$ is the uncertainty on the $u$-band magnitude. \cref{fig:color_color_box_nz_box_RF} (left panel) shows the $ugr$ color-color plot from a fraction of the CLAUDS+HSC imaging. The blue box represents the color-color cut selection and provides a target density of $1290$ per deg$^{2}$. The selection was applied to the entire COSMOS dataset, and the photometric redshift distribution of the target density is represented in blue in \cref{fig:color_color_box_nz_box_RF} (right panel) and has a mean redshift $\langle z|z>2\rangle_{\rm box} = 2.67$ with a total target density of $z>2$ LBGs given by $n_{\rm box}(z > 2)=$ 946 deg$^{-2}$. 
\subsection{Random Forest}
\begin{figure*}[t]
\centering
\includegraphics[width=1\columnwidth,trim=4 4 4 4,clip]{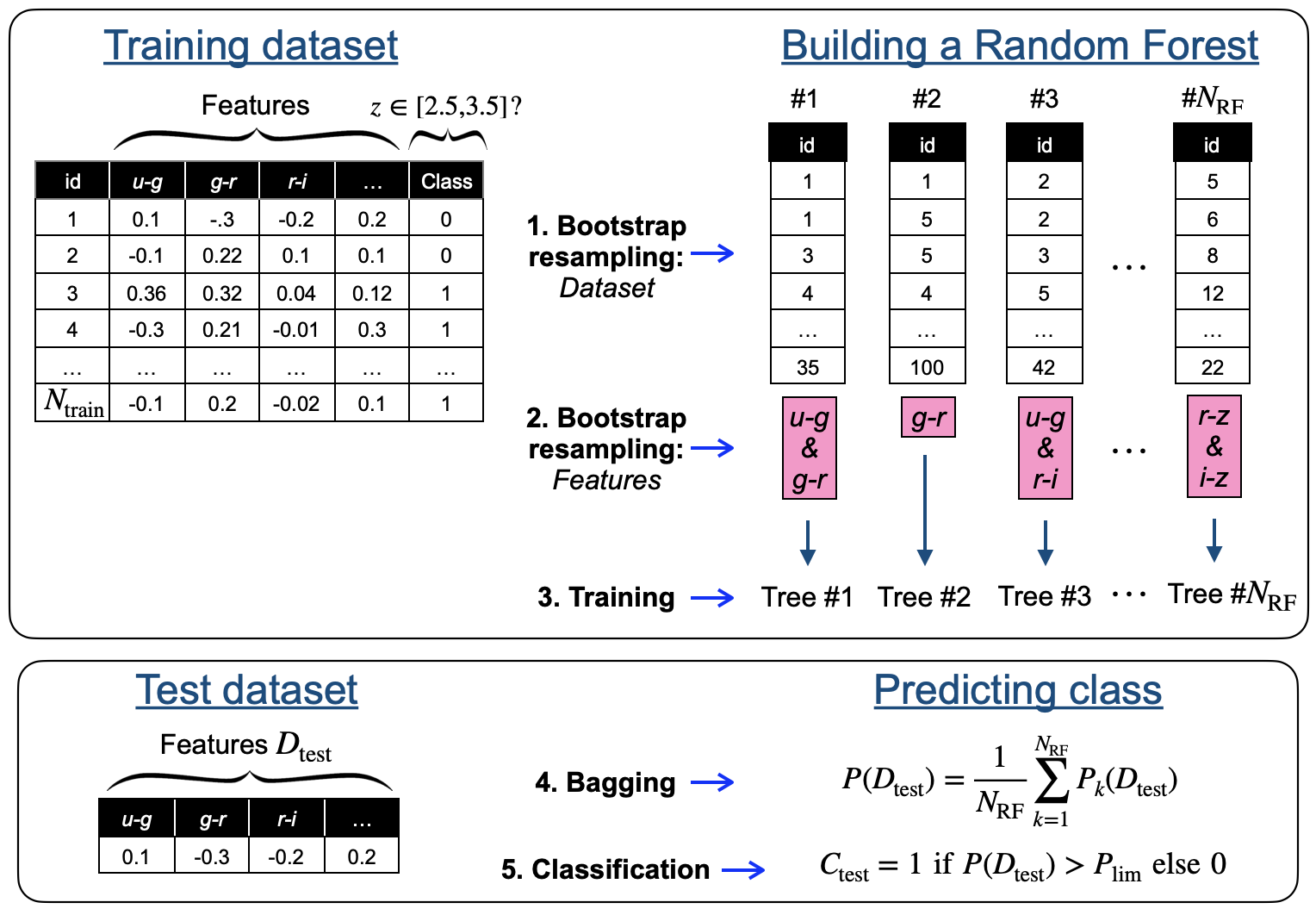}
\caption{Illustration of the training process of a Random Forest algorithm (upper panel) and object classification (lower panel) in 5 steps. Upper panel: The training involves generating $N_{\rm RF}$ bootstrap samples of the training dataset (step 1.) and $N_{\rm RF}$ bootstrap subsets of the full list of features (step 2.), which are then used to train $N_{\rm RF}$ decision trees (step 3.). Lower panel: The final classification of a test galaxy is achieved by aggregating the independent classifications from the decision trees into an average probability (\textit{bagging}, step 4.), followed by providing a prediction (e.g., $0$ or $1$) based on a quality threshold $P_{\rm lim}$ (step 5.).} 
\label{fig:RF_principle}
\end{figure*}
Machine Learning methods have proven to improve target selections for spectroscopic surveys compared to standard methods \citep{Hoyle2016TSrandomforest,Chaussidon2023TSQSODESI,Yeche2009TSQSODESI}. In this work, we use a Random Forest\footnote{We use the scikit-learn implementation available at \url{https://scikit-learn.org/stable/modules/generated/sklearn.ensemble.RandomForestClassifier.html}.} (RF) algorithm \citep{Breiman2001RandomForest} that can be used in the context of target selection for both regression tasks (i.e. trained to predict the redshift of a test galaxy from its imaging properties) and classification problems (i.e. to predict the \textit{class} of a test galaxy). In this work, we consider using RF classification to select galaxies in a given redshift interval, we will use the \cref{fig:RF_principle} as a guideline through this section.

\subsubsection{Principle}
\paragraph{Training:}
A Random Forest is a supervised, ensemble learning algorithm based on a set of $N_{\rm RF}$ decision trees. A decision tree is a flowchart-like tree structure where each internal node represents a feature, the branch represents a decision rule, and each leaf node represents the outcome. We first use a training dataset (represented by the left table in \cref{fig:RF_principle}, upper panel), for which we know the class (or classification) $\mathcal{C}_{\rm train}$ of each object. In this work, classes identify training objects in a dedicated redshift interval, namely
\begin{equation}
\mathcal{C}_{\rm train} = \left\{
    \begin{array}{ll}
        1\ &\mathrm{if}\ z \in [z_{\rm min}, z_{\rm max}] \\
        0 & \mbox{else.}
    \end{array}
\right.
\label{eq:classes_train}
\end{equation}
In this section and through the following sections until \cref{sec:pilot_obs_desi}, we refer to as "LBG" the target galaxies within the redshift range $z_{\rm min} < z < z_{\rm max}$ in the photometric sample; so it refers to the naming of the Random Forest class $\mathcal{C}=1$. We assume that the redshift only (i.e. all galaxies living in the $z_{\rm min} < z < z_{\rm max}$ range) is a good \textit{proxy} to the underlying \textit{true} LBG-type population at these redshifts, the latter will be confirmed by DESI spectroscopic observations in the next \cref{sec:pilot_obs_desi}. The spectroscopically-confirmed LBG sample will be explicitly mentioned through the text.
Here, focus exclusively on the redshift information of the targeted class in \cref{eq:classes_train}, representing a relatively straightforward approach. This choice was driven by the lack of strong distinguishing features for LBGs, such as morphological markers, that could be incorporated as additional targeted classes in our RF methodology. Future extensions of this approach could explore more advanced RF-based selection techniques that integrate color-redshift dependencies along with other features, such as morphological information. Such a combined approach holds promise for refining the selection process and broadening its applicability. In contrast, the current approach, which relies only on redshift as the target class, adopts a conservative framework. As demonstrated in \cref{sec:pilot_obs_desi}, the forecasted redshift distribution of the spectroscopically confirmed LBG sample, selected based on color-redshift criteria, aligns well with the observed distribution from dedicated DESI observations in the COSMOS field.

Next, $N_{\rm RF}$ training datasets are generated by drawing a random selection of training data (taken by replacement), called \textit{bootstrapping} (step 1. in \cref{fig:RF_principle}, upper panel). A second bootstrap resampling (step 2. in \cref{fig:RF_principle}, upper panel) is performed to associate each $k$-th sample with a random selection of training data features (e.g. only $u-g$ and $g-r$ instead of the complete list of colors).
Then, $N_{\rm RF}$ decision trees are trained independently, this process involves recursive binary splitting of data based on a criterion like Gini Impurity (ref) (step 3. in \cref{fig:RF_principle}, upper panel), such as each $k$-th tree uses the $k$-th bootstrap training dataset and the $k$-th subset of features. After training, each tree returns a probability $P_k(\mathcal{D}) = 1$ (resp. 0). 

Predicting object class: For a new test sample denoted by $\mathcal{D}_{\mathrm{test}}$ (left table in \cref{fig:RF_principle}, lower panel), its proper features are queried down the forest of trees and the prediction is obtained by aggregating the $N_{\rm RF}$ tree predictions (called \textit{bagging}, step 4. in \cref{fig:RF_principle}, lower panel) in the average probability
\begin{equation}
    P(\mathcal{D}_{\mathrm{test}}) = \frac{1}{N_{\rm RF}}\sum_{k=1}^{N_{\rm RF}}P_k(\mathcal{D}_{\mathrm{test}})
    \label{eq:RF_proba_average}
\end{equation}
The final classification $\mathcal{C}_{\rm test}$ (step 5. in \cref{fig:RF_principle}, lower panel), lower panel is given by
\begin{equation}
\mathcal{C}_{\rm test} = \left\{
    \begin{array}{ll}
        1\ &\mathrm{if}\ P(\mathcal{D}_{\mathrm{test}}) > P_{\rm lim} \\
        0 & \mbox{else}
    \end{array}
\right.
\label{eq:classes_test}
\end{equation}
where $P_{\rm lim}$ is an RF quality threshold, and is a free parameter that can be chosen to match a given target density budget. The hyper-parameters of an RF algorithm are the number of trees and nodes in each tree. 

A Random Forest can effectively model complex non-linear relationships between features and target classes, surpassing traditional color-cut methods. This capability is particularly advantageous when characteristic LBG spectral features, such as the two Lyman breaks at $912$ and $1216$ ~\r{A} (rest-frame), are observable at optical wavelengths for $z > 2.5$. RF inherently accounts for feature interactions, such as the co-linearity between $u-g$ and $g-r$ colors (\cref{fig:color_color_box_nz_box_RF}, left panel), where traditional color-color box selection techniques require strict, manually defined cuts or non-machine-learning approaches like the Fisher discriminant \citep{Raichoor2016fisher}. 
RF offers greater flexibility in defining probabilistic decision boundaries, enabling nuanced classification in probabilistic scenarios influenced by data noise or scattering from physical processes. Additionally, RF is more robust to noisy data, as it aggregates decisions from multiple trees, mitigating the impact of errors and reducing prediction variance compared to individual trees. In contrast, strict color cuts are more susceptible to errors caused by noise near decision boundaries. 
By design, RF methods also minimize over-fitting, a common issue with decision tree-based classification. Furthermore, as we will discuss later, RF provides insights into feature importance, identifying the most relevant features (e.g., colors) for classification.

The Random Forest training sample is obtained by using the magnitude cut $22.5 < r < 23.75$ and the magnitude error cut $\mathrm{err}(u)< 1$ (to be consistent with the color-color cut example). After these cuts, the total dataset is about 180,000 objects. The number of objects to be used for training is discussed in the next section. We consider the 4 colors constructed from the CLAUDS+HSC $ugriz$ magnitudes. To decide the object classes in \cref{eq:classes_train}, we choose $[z_{\rm min}, z_{\rm max}]=[2.5, 3.5]$, 
where we use photometric redshifts derived with the {\sc LePHARE} code based on the 6-band $ugriz$ data from CLAUDS+HSC \citep{Desprez2023CLAUDSHSCSSP}. However, other improved photometric redshift catalogs are available for the COSMOS field, including refined CLAUDS+HSC photometric redshifts that incorporate near-infrared data (for a total of 11 bands) \citep{Desprez2023CLAUDSHSCSSP} as well as the COSMOS2020 dataset \citep{Weaver2022cosmos2020}. In \cref{app:diff_z}, we discuss the impact of using these alternative photometric datasets on the performance metrics of the RF algorithm when using CLAUDS+HSC magnitudes, and we investigate the shape of the 11-bands and COSMOS2020 photometric redshifts of targets compared to the 6-bands one when targets are selected from a 6-bands-trained RF model. We find that the photometric redshift distribution is not significantly impacted in the $z>2$ range. For this paper, which depends on spectroscopic efficiency calibrated using CLAUDS+HSC redshifts, we will continue using the CLAUDS+HSC photometric redshift dataset for Random Forest training, but this would deserve further investigation and refinement, especially when considering shallower imaging to be used as a feature of the RF, since the robustness of our results are valid when using deep imaging data. 

In this redshift range, the spectral features of LBGs appear in the optical bands. Moreover, the spectroscopic redshift measurement is more efficient at $z > 2.5$, than in $2.0 < z < 2.5$ \citep{RuhlmannKleider2024LBGCLAUDS} (we detail this method in \cref{sec:unions-like_ts_cosmos}). From this approach, our first aim is to identify objects within the redshift range of $2.5 < z < 3.5$ from their imaging properties, without explicit mention of their type (LBGs, Quasars, etc.). Then, the minimization of the \textit{contamination} (we define this performance metric hereafter) by objects other than LBGs is not a primary goal of our Random Forest approach. In a Random Forest perspective, inspired by the $u$-dropout technique, we rather expect to find a high correlation between colors and redshift for LBG specifically in this redshift range, to be learned by the algorithm. One could improve the classification by training with other target characteristics, such as morphology, that can help mitigate the contamination from point-like objects, for instance. 

\begin{figure*}
    \centering
    \includegraphics[width=.49\textwidth]{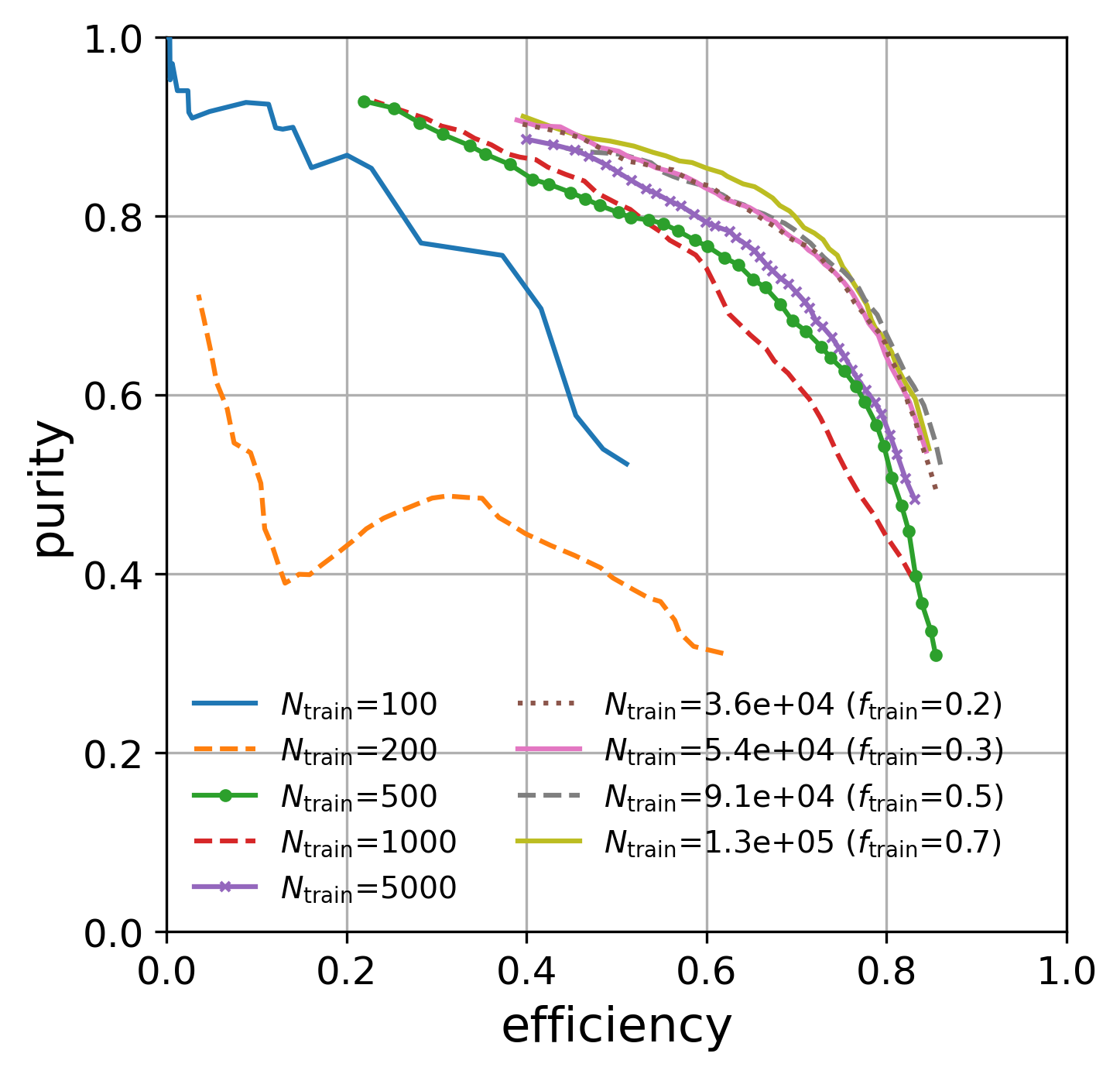}
    \includegraphics[width=.49\textwidth]{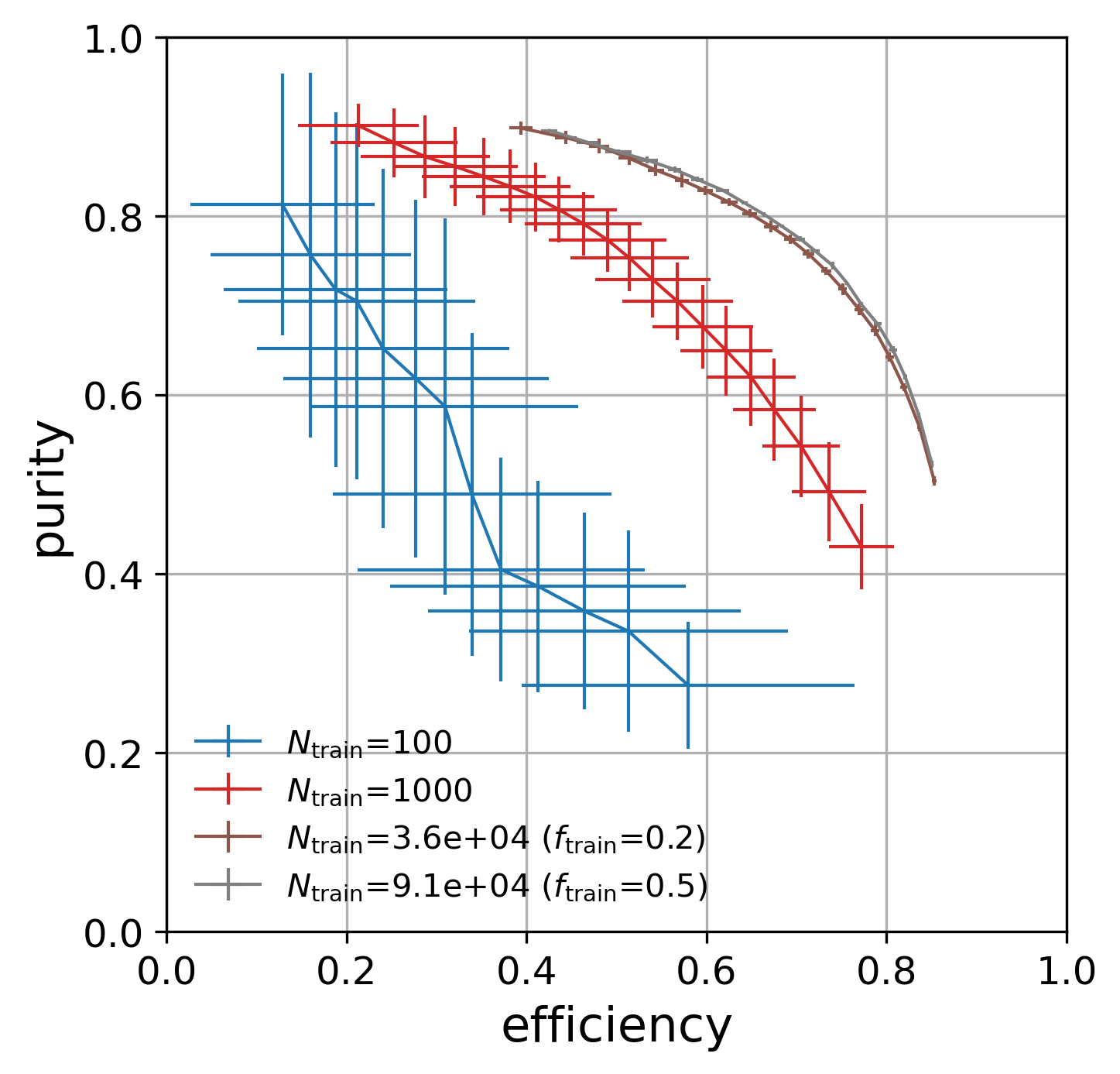}
    \caption{\textit{Left}: Random Forest purity against efficiency (the RF algorithm is trained with $ugrz$-bands). The different colors correspond to different training sample sizes $N_{\rm train}$ elements. \textit{Right}: \textit{mean} purity against \textit{mean} efficiency, with corresponding dispersions. The colors matched the ones used on the left plot. 
    }
    \label{fig:validation_RF}
\end{figure*}

\subsubsection{Performance metrics}
The algorithm uses the training data to learn the relationship between the features and the target classes. The test data are used to evaluate the performance of the model. Using a test sample from which we know the true classification, we define two parameters to evaluate the performance of the Random Forest classification. For a given threshold 
$P_{\rm lim}$, the number of selected targets of the test sample is given by $N_{\rm sel}$. The number of selected LBG is denoted by $N^{\rm LBG}_{\rm sel}$  and $N^{\rm LBG}_{\rm true}$ is the number of galaxies in the test sample that are in $[z_{\rm min}, z_{\rm max}]$. We define the \textit{purity} $p$ and the \textit{efficiency} $\epsilon$ respectively given by
\begin{equation}
    p = N^{\rm LBG}_{\rm sel}/N_{\rm sel}\hspace{0.5cm}\mathrm{and}\hspace{0.5cm}\epsilon = N^{\rm LBG}_{\rm sel}/N^{\rm LBG}_{\rm true},
    \label{eq:purity_efficiency}
\end{equation}
where $N^{\rm LBG}_{\rm sel} = N_{\rm sel}(2.5 < z < 3.5)$. The purity denotes the fraction of selected galaxies that are effectively in the desired redshift interval. The efficiency denotes our ability to select the parent distribution of galaxies in the considered redshift interval. In this work, we rather use \textit{efficiency} as a terminology instead of \textit{completeness}, the latter referring to the fraction of true selected LBGs relative to the underlying \textit{true} LBG population. However, this true population is inherently inaccessible in a photometric sample due to pre-selection biases such as detection limits and depth, which prevent defining an absolute "true" LBG population. This contrasts with studies relying on the response of mock LBG samples to dropout target selection (see e.g. \citep{malkan2017,ono2018,Harikane2023lbg}), where a "true" LBG population is assumed. On the other hand, efficiency (also called \textbf{recall} in machine-learning of classification problems) is a more relative measure, quantifying the fraction of relevant retrieved entries (e.g., selected objects with $z \in [2.5, 3.5]$) compared to all relevant entries in the photometric sample. Given the inherent limitations in defining completeness for our photometric sample, we use "efficiency" throughout this work.

First, using purity/efficiency as performance metrics for the RF training, we investigate, at CLAUDS depth, for 
which training dataset size, $N_{\rm train} = \{200, 500, ...\}$, the RF performance becomes stable. In the following, we also refer to the fraction of the total dataset used for training $f_{\rm train}=N_{\rm train}/N_{\rm tot}$. We use the remaining data for testing for the different $N_{\rm train}$. For different $N_{\rm train}$, we show in \cref{fig:validation_RF} (left panel) the purity against efficiency where $P_{\rm lim}$ runs from 0.1 (lower right part of the lines) to 0.8 (upper left part of the lines).  We see that using a few hundred objects for training results in poorer performance (i.e., lower purity and/or efficiency) compared to using a few tens of thousands of objects. From $f_{\rm train} = 30\%$ (a few tens of thousands of objects) to $f_{\rm train} = 70\%$, the RF performance is stable.

An inherent limitation of using a Random Forest is the potential inconsistency between the training and test datasets, i.e., the conditional density of the class given the input features, $P(\mathcal{D})$ in \cref{eq:RF_proba_average}, learned from the training set may not accurately represent that of the test set. This discrepancy arises due to tracer selection function effects, which introduce spatial dependencies in the relationship between broadband photometry and class membership. In our Random Forest approach, the impact of the selection function becomes significant when the training and test fields are located in different regions of the sky (then impacted differently by dust extinction, depth coverage, and other observational factors) potentially leading to variations in performance when testing on COSMOS versus XMM and introducing spurious density fluctuations in tracer distributions. These effects can result in significant biases in clustering analyses \citep{Mueller2021fnl,Chaussidon2023TSQSODESI,Chaussidon2024fnl}. Let us note that this limitation is also present in traditional color-box techniques (e.g., $u$-dropout), where a single selection criterion is applied across the full footprint despite inhomogeneities in the selection function. While this issue is particularly critical for targeting LBGs—whose selection depends on a deep and homogeneous $u$-band—it is mitigated in our case, as the training and test datasets are randomized across the full COSMOS area.

Nevertheless, robustness tests still can be conducted to assess the precision of our RF regression. Here below, we want to test the stability of the RF prediction after the resampling of the training sample. First, we define $K = N_{\rm tot}/N_{\rm train}$ independent splits, each one containing $N_{\rm train}$ objects. For instance, \cref{fig:validation_RF} (left panel) shows the RF performance metric
measured of a single split, for different fractions of the training sample $f_{\rm train}$ (colors). We have repeated the training of the RF algorithm using each independent $K$-th split as the training dataset, and we tested the algorithm with the remaining $K$-1 splits. For each independent $K$-th training, we have computed the performance metrics. This method can be compared with the "$K$-fold" cross-validation technique in machine learning, where each fold (or split) is used once as a test sample while the $K$-1 remaining folds form the training set (i.e., doing the opposite: $K$-1 folds for training and 1 fold for testing). The \textit{mean} purity against the \textit{mean} efficiency is displayed \cref{fig:validation_RF} (right panel) for different $N_{\rm train}$ values. For $N_{\rm train} = 100$ or $N_{\rm train} = 1,000$, there are $\sim 2,500$ and $\sim 250$ independent splits, respectively. For both cases, we measured the mean performance and the dispersion with only 30 splits rather than using all splits (we have tested using 50 to 100 splits, and the results are similar). We see that for these two cases, the performance is poorer and the dispersion over the $K$ independent splits is large. When considering larger fractions of the dataset used for training, such as from $f_{\rm train} = 0.2$ (from which we can construct 5 independent splits) to $f_{\rm train} = 0.5$ (2 splits), we have checked that each $K$-th RF 
performance is dispersed by less than $1\%$ around the mean performance (note that for $f_{\rm train} = {0.2, 0.5}$, they are respectively 5 and 2 independent splits) and that there is less than 1$\%$ difference between using $f_{\rm train} = 0.2$ or $f_{\rm train} = 0.5$ (namely, using from $5\times10^4$ to $10^5$ objects for training). 

\begin{figure*}
    \centering
    \includegraphics[width=.49\textwidth]{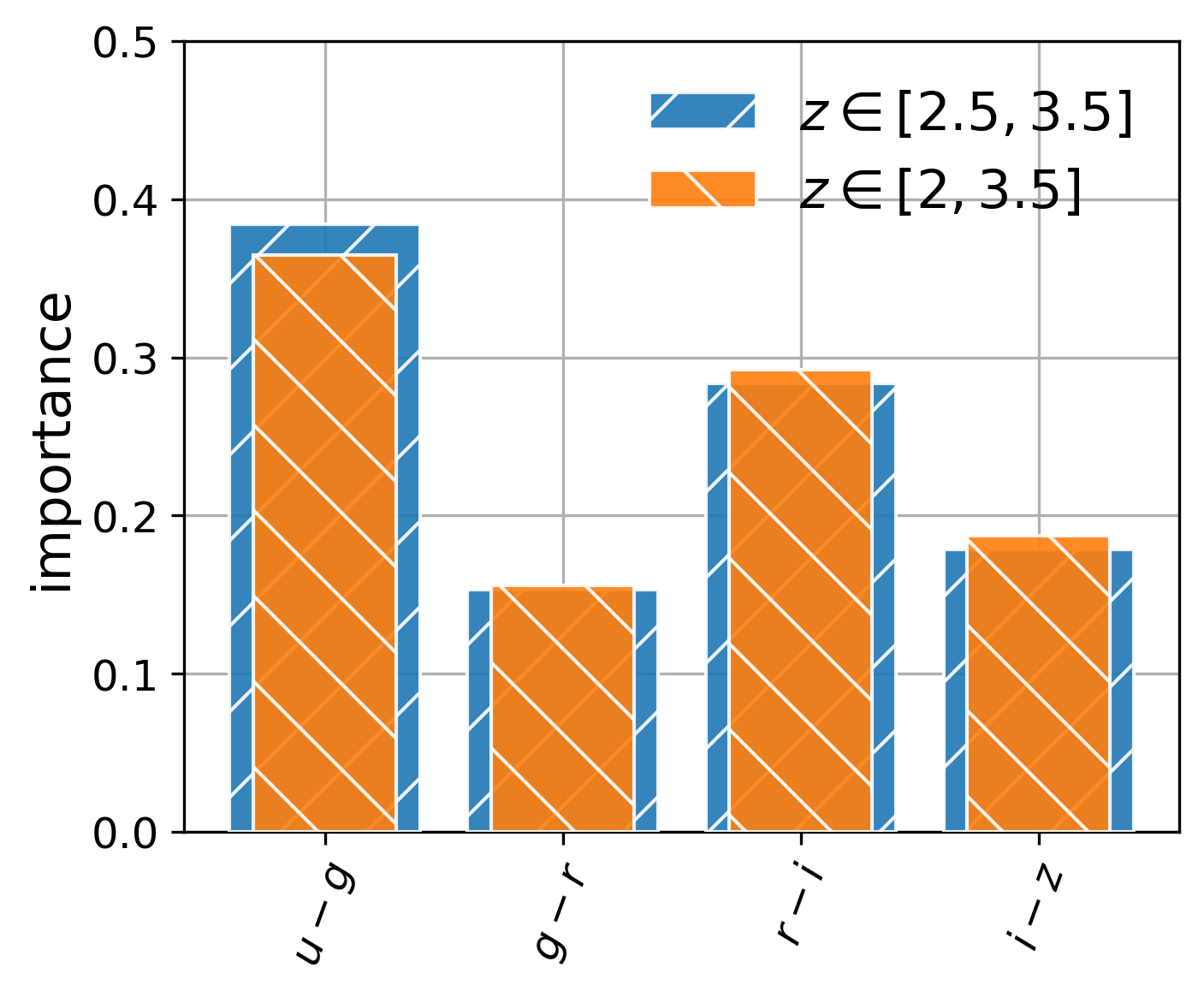}
    \includegraphics[width=.49\textwidth]{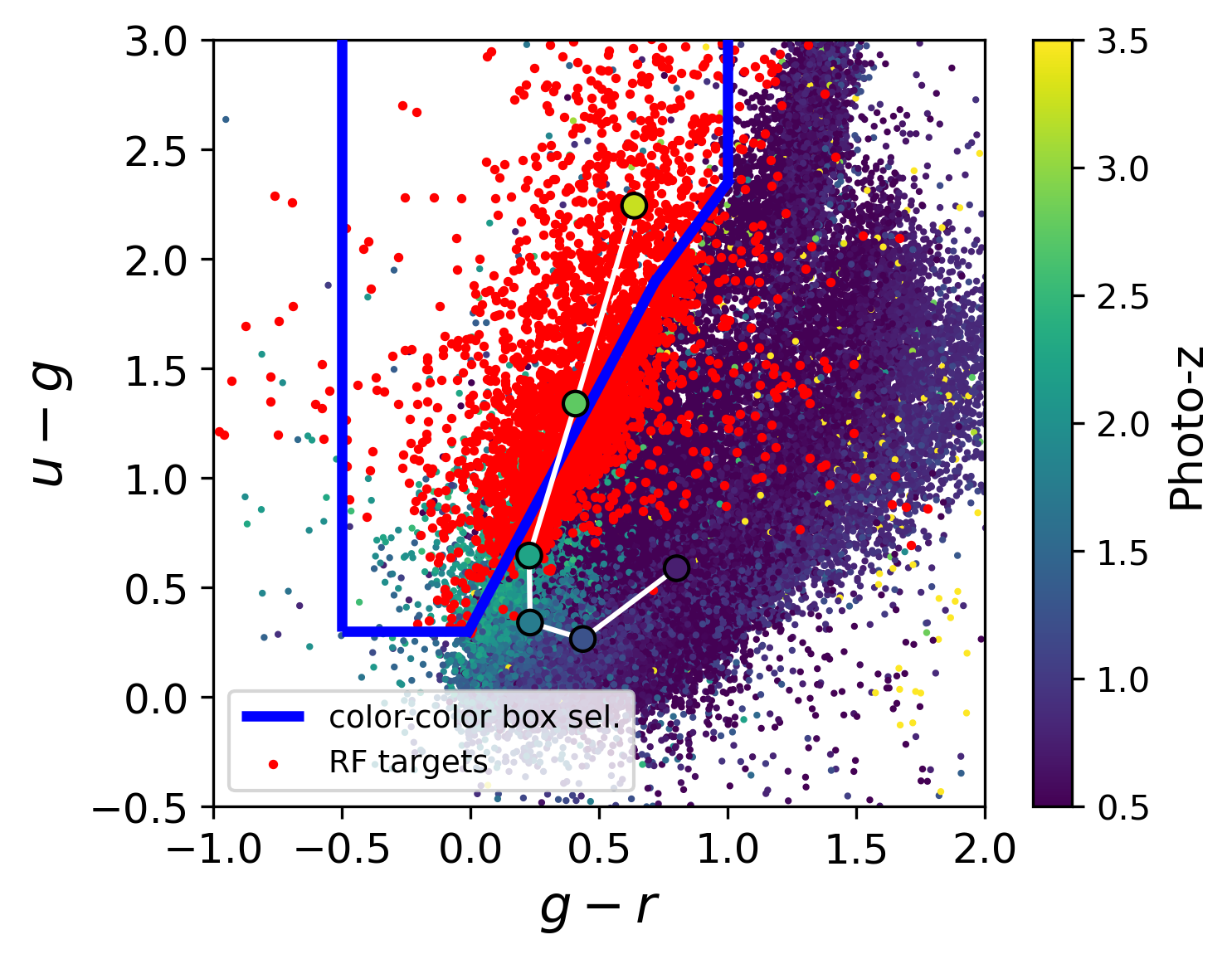}
    
    \caption{\textit{Left}: Importance of the 4 different features for the Random Forest algorithm. \textit{Right}: As in \cref{fig:color_color_box_nz_box_RF} (right), the green/yellow dots represent the parent test sample. The red dots correspond to LBG targets selected by our RF  algorithm, to match the target density provided by the color-color box selection (in blue).  }
    \label{fig:comparison_RF_color}
\end{figure*}

Alternatively, we can also look for the features used for training that most affect LBG classification. The importance features metric for the RF implemented in \texttt{scikit-learn} is based on the Gini importance, which counts the times a feature is used to split a node after the Random Forest is trained, weighted by the number of samples it splits. A feature with a higher \textit{importance} value is more discriminating in the classification than a feature with a lower importance value. The feature importances are shown in \cref{fig:comparison_RF_color} (left panel) for the four features used for training, namely the colors $u-g$, $g-r$, $r-i$, and $i-z$. Our baseline is to classify galaxies within the range $z\in [2.5, 3.5]$, for which the corresponding importance is shown in blue. We see that the $u-g$ and $r-i$ color play a key role in LBG selection since the first 
measures the $u$-dropout for $z\in [2.5, 3.5]$ galaxies, and the second encodes more complex correlations between color and redshift that the RF algorithm has learned from data.  

However, we note that interpreting the feature importance graph in detail can be challenging due to the natural co-linearity between galaxy colors, stemming from their physical interdependence. This colinearity causes importance scores to reflect combined contributions rather than isolated effects. Consequently, the high importance of one feature may partly represent the influence of its correlated counterparts, instead of revealing its high predictive power in a Random Forest procedure. Future work could mitigate this issue by employing methods such as conditional importance or de-correlation techniques.

We can test the RF regression when modifying slightly the LBG classification by including lower redshifts, namely using the interval $z\in [2.0, 3.5]$. We see that with this setup, displayed in orange in \cref{fig:validation_RF} (left panel), feature importances are not enhanced/decreased at higher/lower redshift. In the following, we will 
continue with the baseline $[2.5, 3.5]$. 

\subsubsection{Comparison with color-color box selection}

We use $P_{\rm lim} = 0.24$ in \cref{eq:classes_test} to select LBGs, to match the color-color box selection target density\footnote{The target density for the RF approach is calculated as $N_{\rm sel}^{\rm RF}/(f_{\rm test} S)$, where $S$ is the surface area in deg$ ^2$, to "remove" the test sample from the target density estimation, whereas it is given as $N_{\rm sel}^{\rm box}/S$ for the color-color box selection.} of 1290 deg$^{-2}$. This gives a purity $p = 0.68$ and an efficiency $\epsilon = 0.78$. The LBG targets selected by our Random Forest algorithm are represented in red in \cref{fig:comparison_RF_color} (right panel). Moreover, $\sim 60\%$ of RF-selected targets are also selected by the color-color cut method, and reversely. \cref{fig:color_color_box_nz_box_RF} (right panel) presents the corresponding photometric redshift distribution in red, compared to that of the color-color box selection in blue. We find that the RF selection gives a mean redshift $\langle z| z > 2 \rangle_{\rm RF} = 2.79$, and a target density $n_{\rm RF}(z > 2)= 1040$ deg$^{2}$ so providing an improvement of 100 deg$^{-2}$ compared to the color-color box selection and a higher mean redshift by 0.1. 


The LBG selection we have presented above is based on CLAUDS+HSC imaging on the COSMOS deep field, offering deep $u$-band imaging with a depth larger than 27. By the end of 2027, CFIS and the Rubin LSST will provide a 5-sigma point source depth in the $u$-band of 24.1 over a few thousand square degrees. By 2035, with 10 years of LSST, we can anticipate a depth in the $u$ band of around 25.6.
\section{UNIONS-like LBG selection with a Random Forest algorithm}
\label{sec:validation}
In this section, we begin by validating our degradation method and Random Forest-based target selection approach. For this, we applying our Random Forest algorithm on a simulated CFIS+DeCaLS imaging from the deeper CLAUDS+HSC on the XMM field, that we compare to the results we obtain on the true CFIS+DeCaLS imaging. Next, we propose a UNIONS-like target selection by degrading CLAUDS+HSC imaging on the COSMOS field. These selected targets were observed by DESI, enabling us to derive a LBG spectroscopically confirmed redshift distribution.
\subsection{Validation with CLAUDS imaging degraded to CFIS depth on XMM}
\label{sec:cfis_decals_RF}
\begin{figure}[t]
    \centering
    \includegraphics[width=.48\textwidth]{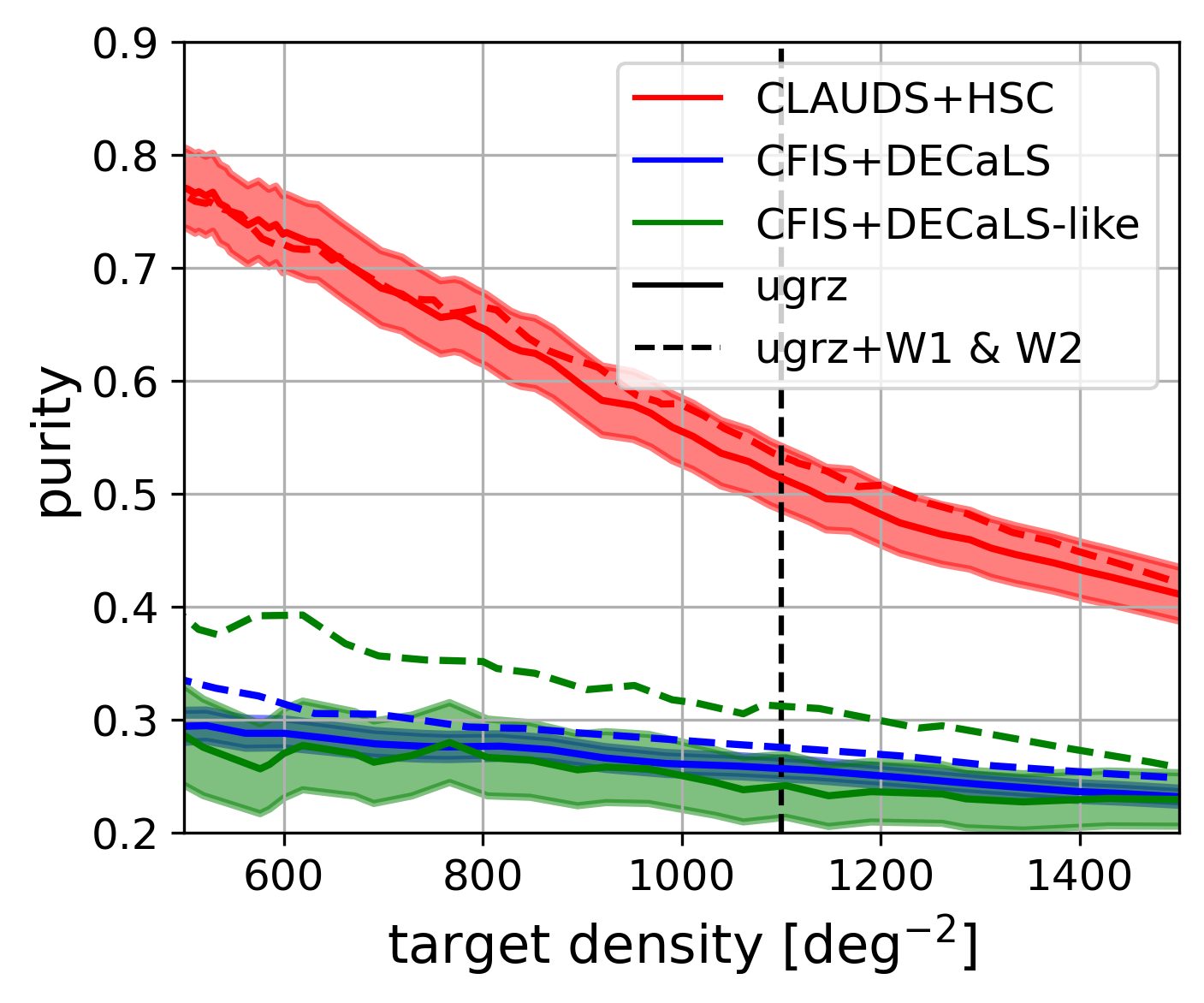}
    \includegraphics[width=.48\textwidth]{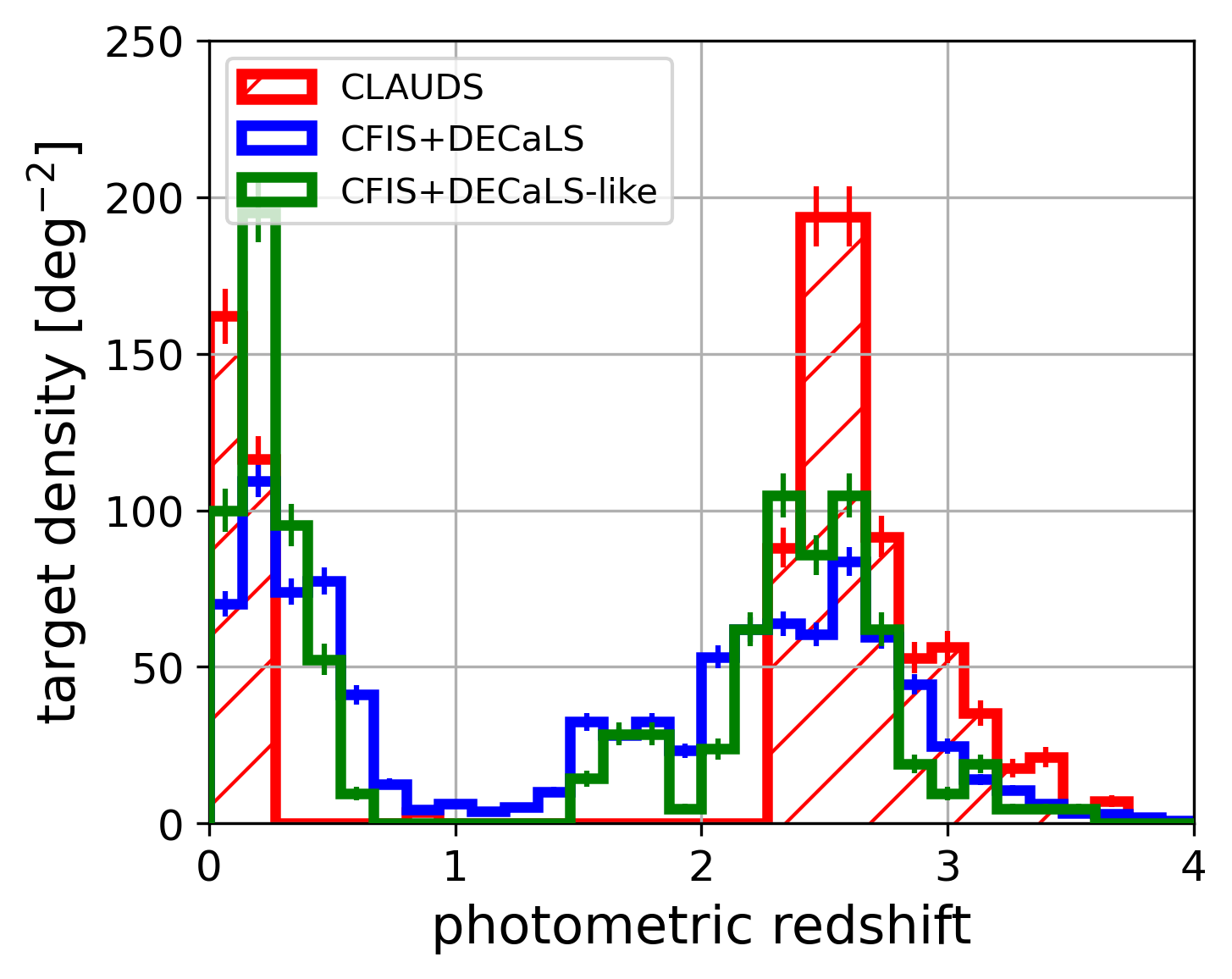}
    \caption{\textit{Left}: Purity (with 1$\sigma$ band) as a function of target density for the CLAUDS+HSC imaging (red), CFIS+DECaLS imaging (blue), and CLAUDS+HSC imaging degraded to CFIS+DECaLS depth. \textit{Right}: Corresponding photometric redshift distribution of targets for a density of $n_{\rm LBG} = 1,100$ deg$^{-2}$.}
    \label{fig:comparison_color_degraded_cfis_decals}
\end{figure}
We have presented in \cref{sec:available _dataset} a method to degrade magnitudes to shallower depth. We also have illustrated this method by degrading CLAUDS($u$)+HSC($grz$) magnitudes towards CFIS($u$)+DECaLS($grz$) depths on the XMM-LSS. In this section, we compare the corresponding Random Forest performances for selecting LBGs when using a CLAUDS+HSC-to-CFIS+DECaLS degraded dataset on XMM-LSS. 

We train a Random Forest classifier on three different configurations, first based on CFIS+ DECaLS imaging, second on CLAUDS+HSC imaging, and third on the CFIS+DECaLS-like imaging obtained by degrading CLAUDS+HSC data. For the CFIS-DECaLS training dataset, since we want to compare with the CFIS+DECaLS-like case, we use a cut on the $u$-band number of exposures introduced earlier to estimate the $u$-band depth. This ensures that the magnitude depth of the CFIS+DECaLS data is the same as the CFIS+DECaLS-like data for a fair comparison. 

For consistency, the three RF algorithms have been trained on the three configurations\footnote{Again, the three configurations are (i) CLAUDS+HSC (ii) CFIS+DECaLS (iii) CFIS+DECaLS-like, where the latter is obtained by degrading CLAUDS+HSC photometry to a shallower depth. Each RF algorithm is trained on $ugrz$-bands.} using $\sim 35,000$ galaxies after applying
\begin{itemize}
\item $23<r<24.3$ 
\item err$(m_x)<5$, $x=\{ugrz\}$.
\end{itemize}
For CFIS+DECaLS, we consider the full XMM footprint since the cut on the number of exposures removes too many galaxies. For CLAUDS+HSC and CFIS+DECaLS-like, we use the data within the field represented by the white box in \cref{fig:err_u_CFIS_map}. As a result, fewer galaxies (6,000) are used for testing in the CLAUDS+HSC and CFIS+DECaLS-like cases, compared to CFIS+DECaLS (35,000).

We trained the RF algorithms using two different sets of features, first considering the 3 colors derived from the optical bands $ugrz$, and second, adding the near-IR bands $W1$ and $W2$ provided by WISE \citep{Wright2010Wise}. We use only $ugrz$ since the $i$-band was in fact observed by DES but not included in the data release 9 of the Legacy Surveys.

\cref{fig:comparison_color_degraded_cfis_decals} (left panel) shows the Random Forest purity as a function of the obtained LBG target density\footnote{Since the three configurations were not trained on the same sky area, target densities are given by $n_i = N_{\rm sel}/(f_{\mathrm{test}, i}S_i)$, where $f_{\mathrm{test}, i}$ is the fraction of the dataset used for testing (that is different from a configuration to another) and $S_i$ is the sky area (either 4.13 or 2 deg$^2$).}, first by training only with the optical $ugrz$ colors (full lines) and second by adding the two infrared magnitudes $W1$ and $W2$ (dashed lines). The optimistic scenario using the CLAUDS+HSC photometry is displayed in red, whereas the CFIS+DECaLS scenario is in blue and the training from the CFIS+DECaLS-like dataset is in green. For clarity, we only show the error bar on the purity for the $ugrz$ case, but they are similar in the $ugrz+W1W2$ case. We see that the larger number of galaxies used for testing in the CFIS+DECaLS case gives much smaller error bars (in blue).  

From \cref{fig:comparison_color_degraded_cfis_decals} (left panel), for the same target density of $n_{\rm LBG} = 1,100$ deg$^{-2}$ (dashed vertical line in \cref{fig:comparison_color_degraded_cfis_decals}, left panel), CLAUDS+HSC gives a purity of 0.5 when CFIS+DECaLS provides around 0.3. The RF performance when trained on the CLAUDS-degraded dataset fairly reproduces the CFIS+DECaLS behavior as a function of the target density for the $ugrz$ case but provides slightly better performance 
when using $W1$ and $W2$. The difference is not meaningful if errors are taken into account. 

Once fixing the target density to $n_{\rm LBG} = 1,100$ deg$^{-2}$ for all selections, \cref{fig:comparison_color_degraded_cfis_decals} (right panel) shows the photometric redshift distribution of LBG targets for our three configurations. The CLAUDS+HSC optimistic case is represented in red, the CFIS+DECaLS-like is represented in green mimics fairly well the targeted one (in blue). 

Keys numbers such as the mean redshift and density on different redshift ranges for these three selections are given in \cref{tab:key_numbers_XMM} (first three columns). First, restricting to  $2.5 < z < 3.5$ (i.e. the redshift range used for training the RF), the mean LBG redshift and LBG density\footnote{Let us note that the number of LBGs selected in the redshift range considered for the Random Forest algorithm (i.e. $2.5 < z < 3.5$) are obtained by $p(1,100 $ deg$^{-2})\times 1,100$ deg$^{-2}$, where $p$ is the purity.} are labeled as $\langle z |2.5 < z < 3.5\rangle_{\rm RF}$ and $n_{\rm RF}(2.5 < z < 3.5) = N_{\rm sel}(2.5 < z < 3.5)/f S$, respectively. In practice, we would possibly use the LBG sample starting from $z\sim z_{\rm min}$, we indicate the LBG density $n(z_{\rm min} <z) = N_{\rm sel}(z_{\rm min} <z)/f S$ for different value of $z_{\rm min}$. 
\begin{table}
    \centering
    \resizebox{0.99\textwidth}{!}{%
    \begin{tabular}{c||c|c|c|c}
& CLAUDS+HSC & CFIS+DECaLS-like & CFIS+DECaLS & UNIONS-like\\
\hline
\hline
$\langle z |2.5 < z < 3.5\rangle_{\rm RF}$ & $2.74\pm 0.25$ &$2.39 \pm 0.41$ &$2.2.77 \pm 0.22$& 2.85 (2.88) \\
$\langle z|z > 1 \rangle$ & $2.66\pm 0.31$ &$2.41 \pm 0.39$ &$2.35 \pm 0.50$& - \\
$\langle z|z > 2 \rangle$ & $2.66\pm 0.29$&$2.52 \pm 0.29$& $2.55 \pm 0.35$&2.76 (2.83)\\
\hline
$n_{\rm RF}(2.5<z<3.5)$ & 503 deg$^{-2}$ &247 deg$^{-2}$& 277 deg$^{-2}$& 683(429) deg$^{-2}$ \\
$n(z > 1)$ & 792 deg$^{-2}$ &619 deg$^{-2}$& 669 deg$^{-2}$& - \\
$n(z > 2)$ & 789 deg$^{-2}$&538 deg$^{-2}$&522 deg$^{-2}$&873(493) deg$^{-2}$ \\
$n(z > 2.5)$ & 479 deg$^{-2}$&233 deg$^{-2}$&270 deg$^{-2}$&-\\
    \end{tabular}}
    \caption{First block: Mean redshift for (i) RF-selected LBGs within the redshift range $2.5<z<3.5$ (ii) $z > 1$ RF-selected LBGs (iii) $z > 2$ RF-selected LBGs. We consider a target density budget of $1,100$ deg$^{-2}$. Second block: corresponding LBG densities. For the last column (UNIONS-like), the numbers in parenthesis are obtained after applying the correction of the total spectroscopic redshift efficiency (2 hours of exposure). }
    \label{tab:key_numbers_XMM}
\end{table}
Moreover, we have tested the impact of cuts applied to data on the RF performance. We have tested different magnitude error cuts (from $\mathrm{err}(m)<5$ in all bands to higher values, and for all datasets), different numbers of exposures for the CFIS $u$-band imaging (changing the output magnitude depth) or the maximum $r$-band magnitude for all datasets. All of these changes affect the training samples for each dataset, and thus RF performances. These tests gave at most a 10$\%$ difference with the baseline that we presented above. We also mention that the error model we use considers that the data are in the sky-limited noise regime, and does not account for specific systematics in the calibration for the input and output surveys, in addition to the possible difference in shape and response between the input (CLAUDS+HSC) and output (CFIS+DECaLS) photometric pass-bands.

In this section we have presented the Random Forest classification of LBGs using degraded imaging data from CLAUDS-HSC imaging on XMM-LSS, to mimic a shallower depth survey (CFIS+DECaLS). We demonstrated that the performance of the RF we get using the simulated dataset compares fairly to the ones obtained using the true dataset, thus validating our approach of mimicking shallower photometry from deep photometry. 
\subsection{LBG selection from UNIONS-like imaging on COSMOS}\label{sec:unions-like_ts_cosmos}
In this section, we degrade CLAUDS +HSC imaging on COSMOS to UNIONS depths (see \cref{tab:depth_mag_photo_surveys}), and we train a Random Forest classifier to extract a list of LBG targets. We also study the impact of spectroscopic redshift efficiency on the recovered photometric redshift distribution. 

\begin{figure}[t]
\centering
\includegraphics[width=0.49\textwidth]{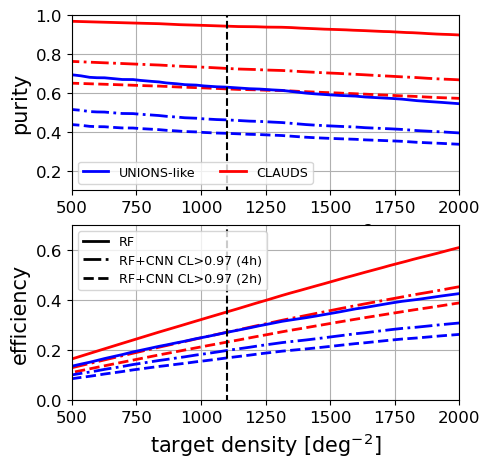}
\includegraphics[width=0.49\textwidth]{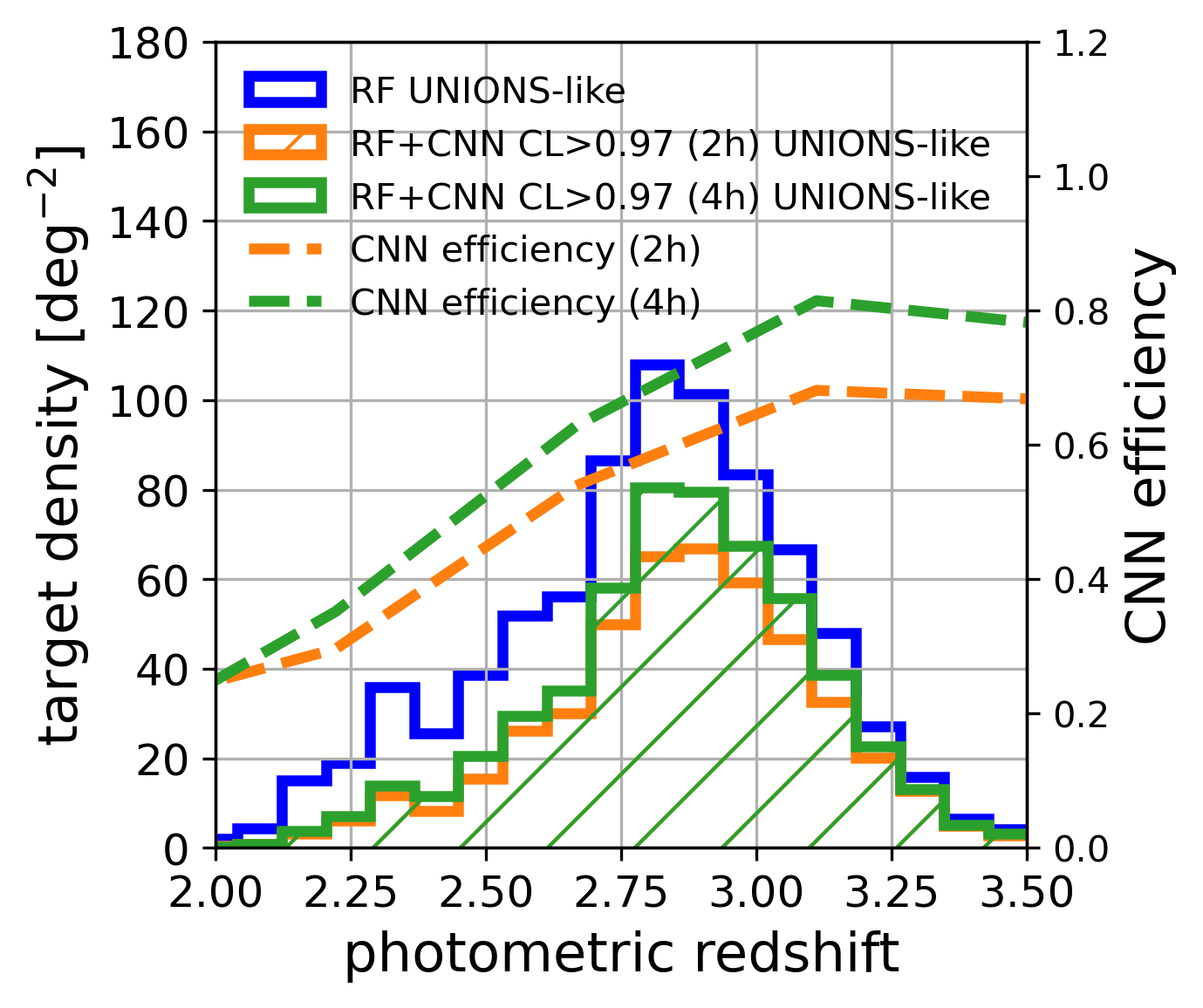}
\caption{\textit{Left}: purity (upper panel) and efficiency (lower panel) as a function of target density for CLAUDS and UNIONS-like LBG Random Forest selections. The pure RF selection is displayed in full lines, whereas the CNN redshift efficiency is plotted in dotted-dashed (resp. dashed) lines for 4h (resp. 2h) of effective time. \textit{Right}: target density distribution for a UNIONS-like selection for a density budget of $1,100$ deg$^{-2}$ in blue, corrected for CNN redshift efficiency in green (4h) and orange (2h). The right y-label indicates the CNN redshift efficiency whose behavior as a function of redshift is given by the dashed lines. 
}
\label{fig:COSMOS_purity_nz_TS}
\end{figure}

UNIONS depths from literature are summarized in \cref{tab:depth_mag_photo_surveys}. To degrade the CLAUDS+ HSC imaging on the COSMOS field, we use input depth in the first and second lines in \cref{tab:depth_mag_photo_surveys} and UNIONS-like output depth still in \cref{tab:depth_mag_photo_surveys}. Since the CLAUDS+HSC depths are not strictly equal to the ones in the COSMOS field (COSMOS CLAUDS+HSC depths are listed in the 4-th line in \cref{tab:depth_mag_photo_surveys_XMM_COSMOS}), the effective depths that we use to obtain the UNIONS-like dataset are slightly different of a few percent and are listed in \cref{tab:depth_mag_photo_surveys_XMM_COSMOS}.
We use the method of \cref{sec:available _dataset} to degrade CLAUDS photometry to UNIONS depths and then train the RF with the complete set of $ugriz$ degraded data. The full dataset is obtained after (as for the validation with XMM in \cref{sec:cfis_decals_RF})
\begin{itemize}
\item $23<r<24.3$ 
\item err$(m_x)<5$, $x=\{ugriz\}$.
\end{itemize}
in all bands gives 280,000 objects, where 55,000 are used for training and the remaining (80$\%$) is used for testing. Curves of purity as a function of target density for UNIONS-like and CLAUDS-based classifications (full lines) are shown in \cref{fig:COSMOS_purity_nz_TS} (left panel). As expected, the UNIONS-like selection provides poorer results than CLAUDS+HSC, going from $p=0.9$ to $p=0.6$ for $n_{\rm LBG} = 1,100$ deg$^{-2}$ (vertical dashed line). Let us note that CLAUDS+HSC purity is significantly enhanced compared to the one measured in \cref{sec:cfis_decals_RF} (see \cref{fig:comparison_color_degraded_cfis_decals}, left panel) using CLAUDS+HSC photometry on XMM-LSS. Indeed, we have found that CLAUDS+HSC $ugriz$-band are deeper than those measured on XMM-LSS (see \cref{tab:depth_mag_photo_surveys_XMM_COSMOS}). Moreover, compared to the RF classification on XMM-LSS, we add the $i$-band to the existing $ugrz$ bands on COSMOS. It does not impact our degradation method, since the updated UNIONS-like magnitudes are computed with input CLAUDS+HSC photometry on COSMOS to reach the desired depth in \cref{tab:depth_mag_photo_surveys_XMM_COSMOS}. Fixing a target budget of $n_{\rm LBG} = 1,100$ deg$^{-2}$, the photometric distribution of UNIONS-like targets are shown in blue in \cref{fig:COSMOS_purity_nz_TS} (right panel).

So far, we have evaluated the efficiency of our target selection method—defined as the recovered fraction of the parent distribution in 
$2.5 < z < 3.5$—based on broad-band photometry rather than intrinsic galaxy properties. At this stage, our method does not explicitly target LBG-type galaxies, and a high selection efficiency in broad-band photometry (as measured by our Random Forest) does not necessarily translate to high completeness for LBG-type objects, as other galaxy types may exhibit similar colors. In that sense, photometry alone cannot be sufficient to determine the type of the galaxy. In this section, we assess the \textit{effective} or \textit{total} performance of our Random Forest method in selecting LBG-type galaxies, which is ultimately determined during the phase of spectroscopic redshift measurement in DESI. The performance of the DESI spectroscopic measurement method, detailed hereafter, needs to be accounted for to estimate the final density of spectroscopically confirmed LBGs. 

This work is conducted in the context of the next phase of the Dark Energy Spectroscopic Instrument (DESI, already introduced). For the spectroscopic redshift determination from LBG spectra used in DESI, we consider the work presented in detail in \cite{RuhlmannKleider2024LBGCLAUDS} who investigated the feasibility of a color-cut LBG target selection with CLAUDS and HSC imaging, supplemented by several dedicated spectroscopic observation campaigns, two on the COSMOS field in 2021 and 2023, and one on the XMM-LSS field in 2022. First, a convolutional neural network (CNN) derived from QuasarNET\footnote{QuasarNET is the algorithm used for QSOs in DESI as part of the spectroscopic classification pipeline.} \citep{Busca2018quasarnet} is applied to each DESI spectrum to perform successively a classification (LBG type or not) and a redshift regression task. The training of the CNN is based on the identification of 14 absorption lines and two emission lines. For each emission/absorption line, the CNN returns a confidence level (CL). The 16 CLs are ranked in decreasing order, and a spectrum is declared as classified as an LBG by CNN if the fifth CL exceeds a given threshold. The CNN output redshifts correspond to the maximum CL. A more precise LBG redshift is then obtained using the RedRock (RR) software\footnote{\url{https://github.com/desihub/redrock}} \citep{Guy2023RROCK} that uses the CNN output redshift as a prior and refines its measurement. In this analysis, LBG-specific templates are used for RR, created from 840 visually inspected LBG spectra from the DESI pilot survey of the XMM-LSS (see the full details of the CNN and RR architectures in \citep{RuhlmannKleider2024LBGCLAUDS}). For the color-color box selection from CLAUDS photometry used in~\cite{RuhlmannKleider2024LBGCLAUDS} and for a CNN CL threshold of 0.97, the  
redshift determination efficiency of the CNN+RR procedure (the number of CNN classified LBGs divided by the number of proposed targets) was found to vary with redshift, from 25$\%$ at redshift 2.1 up to a plateau of 80$\%$ between redshift 3.0 and 3.4, for a fiber exposure effective time\footnote{Exposure time refers to the amount of time that a single DESI fiber collects light from a celestial target during an observation. An effective time of 4 hours means that spectra from different exposures were co-added for a total effective time of 4 hours. More technically, effective exposure time denotes the amount of time necessary to reach a certain uncertainty in ’nominal’ observational conditions for DESI, defined to be a 1.1” seeing, a sky background of 21.07 AB magnitude per square arc second in $r$-band, photometric conditions, observations at zenith, through zero Galactic dust reddening \citep{Schlafly2023effectivetime}.} of 4 hours (see their Figure 18, right panel in \citep{RuhlmannKleider2024LBGCLAUDS}). These results from~\cite{RuhlmannKleider2024LBGCLAUDS} are shown as dashed lines in \cref{fig:COSMOS_purity_nz_TS} (right panel), in orange for an effective time of 4 hours and in green for 2 hours, as a function of the photometric redshift. The drop in efficiency at low redshift originates from different effects, such as the difficult redshift determination for non-emitting LBGs, combined with a possible evolution effect of the LBG population which would disadvantage emitters at redshifts lower than 3. At fixed redshift, the spectroscopic efficiency depends slightly on the effective time. So, the overall performance in determining precise spectroscopic LBG redshifts depends on the observing time, as well as the type of LBGs selected across redshift $z>2$.
\begin{figure} [t]
 \centering
 \includegraphics[width=\textwidth]{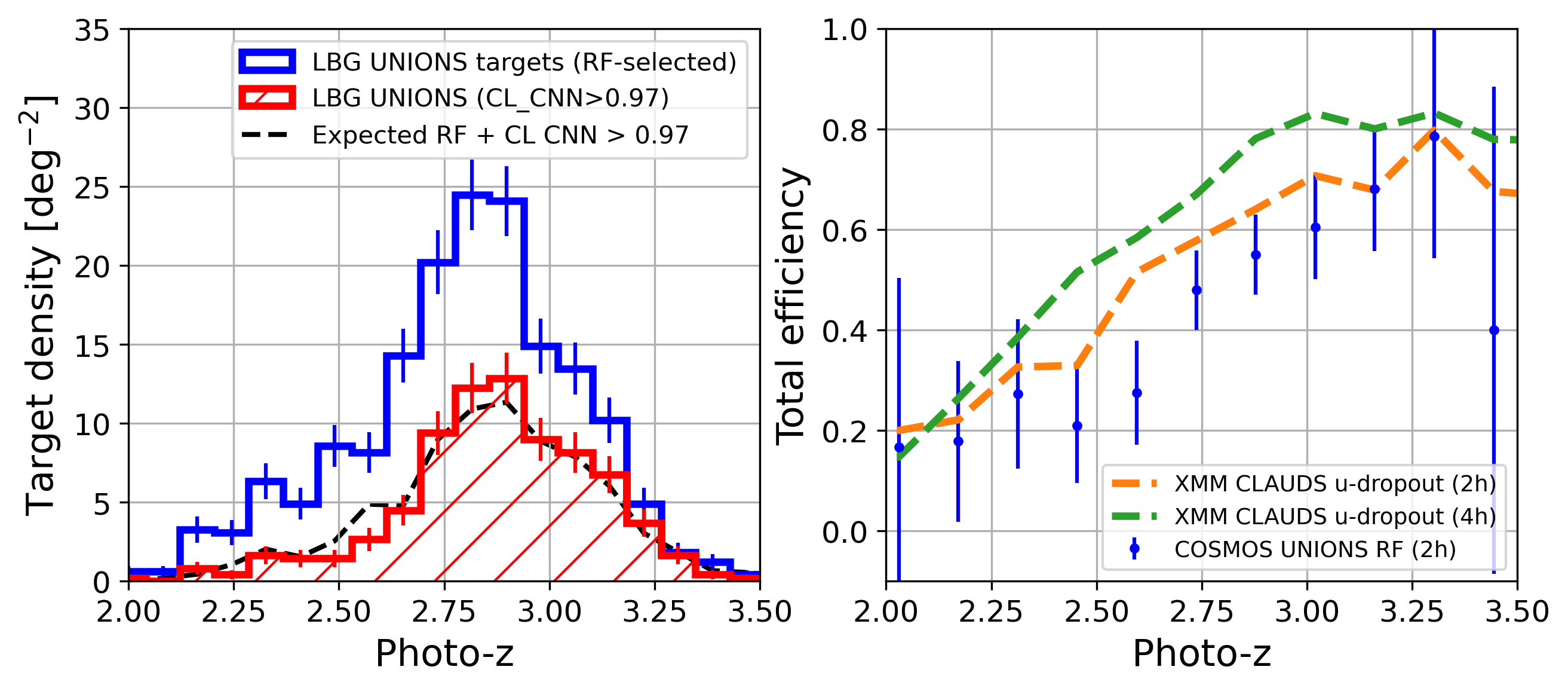}
 \caption{\textit{Left}: photometric redshift distribution of DESI LBG targets (1000 targets from the UNIONS-like RF selection): in blue for the total target sample, in red for those targets with spectroscopic confirmation by the CNN using a CL$>0.97$ selection. 
 The black line is the expected distribution from the full UNIONS-like sample corrected from the expected CNN+RR redshift efficiency and rescaled to the CNN CL$>0.97$ sample size. \textit{Right}: total efficiency as a function of the photometric redshift. Green and orange correspond to the total efficiency obtained on XMM (with CLAUDS+HSC imaging) for two different effective exposure times. Blue points correspond to the efficiency measured on COSMOS after DESI observations of the UNIONS-like RF LBG selection. }
 \label{fig:DESIcampaign_nz_total_eff}
 \end{figure}

The spectroscopic redshift determination efficiency degrades the overall performance of the RF LBG selection, as shown in \cref{fig:COSMOS_purity_nz_TS} (left panel), where we define the \textit{effective} purity
\begin{equation}
    p_{\rm eff} = \frac{1}{N_{\rm sel}}\sum_{k=1}^{N_{\rm sel}^{\rm LBG}}w(z_k),
\end{equation}
that is now the fraction of \textit{true} LBGs selected from the RF classification, and whose spectroscopic redshifts are successfully reconstructed from our CNN+RR approach with a CNN CL $>0.97$ (the above equation corresponds to \cref{eq:purity_efficiency}, where $N_{\rm sel}^{\rm LBG}$ has been replaced by the sum of $w(z_k)$). Here above, $w(z_k)$ is the spectroscopic redshift efficiency evaluated at the object's redshift $z_k$. The effective efficiency expresses similarly as the effective purity, by replacing the denominator $N_{\rm sel}$ by $N_{\rm sel}^{\rm LBG}$. The effective purity (upper panel) and completeness (lower panel) are displayed in \cref{fig:COSMOS_purity_nz_TS} (left panel) as dotted-dashed lines for 4 hours of effective time and as dashed lines for 2 hours of effective time. First, considering RF-only selection with a target density budget of $1,100$ deg$^{-2}$ decreases the purity by a factor $\sim 1.6$ and efficiency by a factor $\sim 1.4$ when going from CLAUDS+HSC to UNIONS. Through $N_{\rm sel}^{\rm LBG} = N_{\rm sel}\times p =1,100\times p$, effective purity gives the number of LBGs with confirmed type and spectroscopic redshifts in the redshift range $[2.5, 3.5]$. Considering UNIONS as a reference, the purity is degraded by a factor $\sim 1.2$ (resp. 1.5) when considering 4 hours (2 hours) of effective fiber exposure time. The same numbers stand for the impact of DESI spectroscopic confirmation on the efficiency, by construction.

After convolution with the redshift determination efficiency curves\footnote{Doing that, we assume that the galaxy population probed by our RF-based approach compares fairly to the one selected in \citep{RuhlmannKleider2024LBGCLAUDS}.}, we obtain the photometric redshift distributions in orange in \cref{fig:COSMOS_purity_nz_TS} (resp. green) for an effective time of 2 hours (resp. 4 hours). We see that the spectroscopic efficiency lowers the LBG population at lower redshift, and the effect is as important as the effective time is small. While it is challenging to assess the precise impact of DESI spectroscopic confirmation on overall efficiency—specifically, the connection between observed LBGs and the underlying LBG population—we observe that the mean LBG redshift increases due to the removal of a relatively large fraction of low-redshift LBGs within the range [2.0, 2.75]. Although this reduces the total number of LBGs, it fortunately does not lower the sample's mean redshift. This outcome is advantageous for cosmological applications, as it targets higher LBG bias values (see \cref{sec:forecast}).

Since the LBG spectroscopic redshift reconstruction is limited below $z=2$ due to spectroscopic redshift estimation, we investigate the key numbers for the selected sample at $z>2$, which are presented in \cref{tab:key_numbers_XMM} (last column). The numbers without parenthesis are obtained after RF selection only, whereas the numbers in parenthesis are obtained after applying the spectroscopic redshift efficiency (2 hours of effective time). From an initial budget of $1,100$ deg$^{-2}$ we can have a LBG density of $493$ deg$^{-2}$ with confirmed redshifts at $z >2$. For comparison, the authors in \citep{RuhlmannKleider2024LBGCLAUDS} proposed a spectroscopically confirmed LBG sample density of 620 deg$^{-2}$ in the redshift range $2.3 < z < 3.5$ for $r < 24.2$, using $u$-dropout techniques applied to CLAUDS+HSC data. While the requirements for DESI-II are not yet finalized, a reasonable threshold suggests a target budget of 1000 deg$^{-2}$, with a spectroscopically confirmed LBG sample density of 300 deg$^{-2}$ for $z > 2$.

\subsection{Pilot observations of COSMOS with DESI}
\label{sec:pilot_obs_desi}

DESI has completed a pilot survey on COSMOS in 2024 to test the UNIONS-like LBG selection. We proposed a density budget of 1,100 deg$^{-2}$, and a total of 1,000 targets were retained for the spectroscopic follow-up by DESI. Galaxy spectra were measured with an effective time of 2 hours. In this section, we present the results of this pilot survey, i.e. the spectroscopic redshift distribution of the targeted LBGs.

The photometric redshift distribution of the 1,000 DESI targets is represented in blue in \cref{fig:DESIcampaign_nz_total_eff} (left panel), within the redshift range of interest $z\in [2, 3.5]$. After spectroscopic redshift determination with the CNN+RR procedure described in \cref{sec:unions-like_ts_cosmos}, the photometric redshift distribution of the final sample, corresponding to a CNN confidence level CL$>0.97$ is represented in red. This provides a mean redshift $\langle z_{\rm phot}|z_{\rm phot} > 2\rangle = 2.87\pm 0.23$. The fraction of secure LBGs over the $z_{\rm phot} > 2$ sample of DESI targets is 0.44, a bit lower than the expected ratio from last \cref{sec:unions-like_ts_cosmos} which is  493/873=0.56, with spectroscopic redshift efficiency taken from \citep{RuhlmannKleider2024LBGCLAUDS}. This difference can be seen in \cref{fig:DESIcampaign_nz_total_eff} (right panel), since our measured spectroscopic efficiency per redshift bin is slightly lower.

To obtain the black dashed line, 
we use the photometric redshift distribution of the full sample of RF-selected LBG targets provided to DESI (i.e. with $n_{\rm LBG}=1,100$ deg$^{-2}$) and apply the CNN+RR spectroscopic redshift efficiency for 2 hours of effective time described in \cref{sec:unions-like_ts_cosmos}. This distribution has a mean redshift $\langle z_{\rm phot}|z_{\rm phot} > 2\rangle = 2.86\pm 0.24$. Rescaling this distribution to the size of the CNN CL$>0.97$ sample (red histogram) produces the black dashed line. The agreement between the two is very good. 

The total CNN efficiency for this RF-based LBG selection is represented in blue in \cref{fig:DESIcampaign_nz_total_eff} (right panel) as a function of the photometric redshift, showing roughly good agreement with the CNN+RR efficiency described in \cref{sec:unions-like_ts_cosmos} for 2 hours of effective time (orange), 
though slightly lower.

\begin{figure} [t]
 \centering
\includegraphics[width=\textwidth]{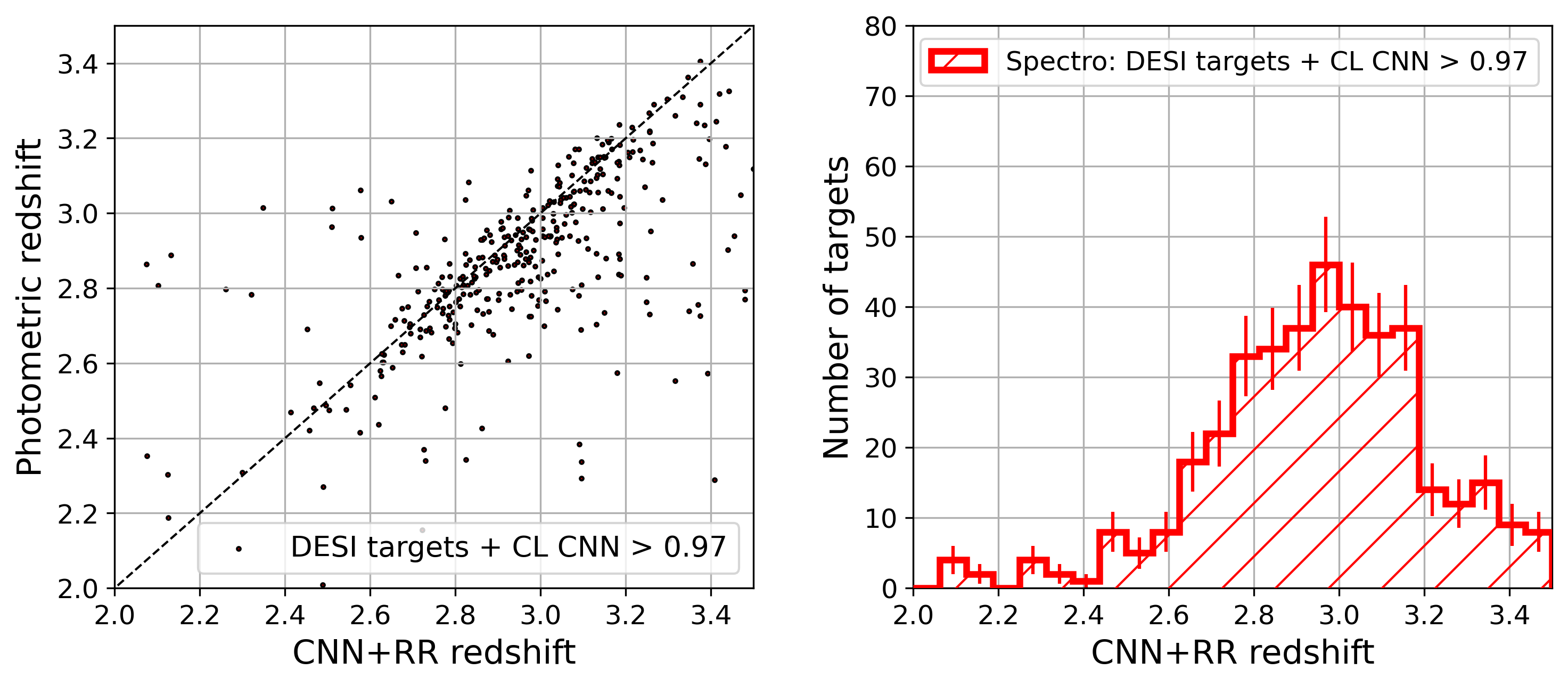}
 \caption{\textit{Left}: photometric redshift as a function of the CNN+RR spectroscopic redshift for a CNN CL threshold of CL$>0.97$. 
 \textit{Right}: CNN+RR spectroscopic redshift distribution for the observed RF LBG targets, using a CNN CL threshold of CL$>0.97$. 
 }
 \label{fig:DESI_ZZ}
 \end{figure}

\cref{fig:DESI_ZZ} (left panel) shows the photometric redshifts as a function of the CNN+RR spectroscopic ones, for the CNN CL$>0.97$ sample. The spectroscopic redshift distribution for the CL$>0.97$ sample is represented in \cref{fig:DESI_ZZ} (right panel). The mean redshift is $\langle z_{\rm spec}|z_{\rm spec} > 2\rangle = 3.0\pm 0.3$. 

\section{Forecasts}
\label{sec:forecast}
In this section, we present forecasts on the Alcock-Paczynski (A.P.) parameters from Baryon Acoustic Oscillations (BAO), the fraction of dark energy, and local Primordial non-Gaussianities (NG) from the LBG redshift distribution derived in \cref{sec:unions-like_ts_cosmos} and in \cref{sec:pilot_obs_desi}, respectively. We restrict the forecast to the redshift-space galaxy power spectrum and do not explore the possibility of using LBG Lyman-$\alpha$ forests. However, valuable information might be extracted from this observable, thanks to the large increase in lines of sight \cite{McQuinn_White_Ly_forest}. Such forests have been detected in \cite{LBG_Ly_forest_1st_obs} and by the CLAMATO survey \citep{CLAMATO_I,CLAMATO_II}. Forecasting Lyman-$\alpha$ forest constraints is more complicated than those for the galaxy power spectrum, and we leave this possibility for future work. The observed nonlinear-redshift-space power spectrum (including shot noise) is denoted $P_{gg}(k,\mu,z)$. We indexed all pairs $(k,\mu)$ by $i$, divide a survey into redshift bins $b_n$, and evaluate the Fisher matrix for each bin as \citep{FishLSS_NSailer,2023MNRAS.521.3648D}: 
\begin{align}
    &F_{\alpha \beta}(b_n)=\sum_{ij}\frac{\partial P_{gg,i}(z_n)}{\partial \theta_\alpha} C_{ij}^{-1}(z_n)\frac{\partial P_{gg,i}(z_n)}{\partial \theta_\beta},\\
    &\text{ where } C_{ij}(z_n)=\delta_{ij}^{K}\frac{4\pi^2}{k_i^2\, V_n \Delta\mu_i \,\Delta k_i}(P_{gg,j}(z_n))^2.
\end{align}
We introduced $V_n$ and $z_n$ the comoving volume and the central redshift of the bin $b_n$, and $\theta $s which represent cosmological or nuisance parameters. We use the standard approximation that correlations between non-overlapping redshift bins are negligible, and the Fisher matrix for a full survey is 
\begin{equation}
    F_{\alpha\beta}=\sum_nF_{\alpha\beta}(b_n).
\end{equation}

For our forecasts, we adopted a bias value $b_{\rm LBG}=3.3$, as measured in  \citep{RuhlmannKleider2024LBGCLAUDS}, 
without any redshift evolution because of the lack of current constraints. We fix the DESI-II sky coverage to 5,000 deg$^2$ and test both 800 and 1,100 deg$^{-2}$ target densities. We adopted a redshift binning $\Delta z=0.1$. 
For the spectroscopic distribution with an 800 deg$^2$ budget, we re-scaled the $1,100$ deg$^2$ one by the appropriate factor, without accounting for a possible shift, since the purity is roughly similar. The integral of each densities over the range $[2, 3.5]$ is respectively $0.44 \times 800 = 353$ deg$^{^2}$ and $0.44 \times 1,100 = 486$ deg$^{^2}$. We make use of the forecasting tool  \texttt{FishLSS}\footnote{\url{https://github.com/NoahSailer/FishLSS}} whose implementation is described in \citep{FishLSS_NSailer}.
We compare these DESI-II forecasts with the full 14,000 deg$^2$ DESI program ones. We use the tracer biases and distribution reported in \cite{DESI2023_science-validation} (and we re-do the forecasts with \texttt{FishLSS} for consistency), as well as the Ly-$\alpha$ forest BAO forecasts (which were done with a code from \cite{DESI_BAO_forecast2014}). We also compute the cosmic variance (CV) limit associated with both survey volumes. 

Acoustic oscillations in the baryon-photon plasma before recombination imprint a characteristic scale in the large-scale structure of matter. Although the 3D matter distribution is not directly observable, galaxies, quasars, and the Lyman-$\alpha$ forest faithfully trace the BAO feature \citep{DESI2024_BAO_gal,DESI_2024_bao_ly}. Combined with constraints from the cosmic microwave background (CMB) or Big Bang nucleosynthesis (BBN), the apparent size of the BAO standard ruler perpendicular and parallel to the line of sight enables to probe the angular diameter distance $D_A(z)$ and the Hubble parameter $H(z)$.
 Using a fiducial power spectrum template as a ruler, these measurements are frequently expressed in terms of Alcock-Paczynski-like (AP) dilation parameters
 \begin{equation}
     \alpha_\parallel=\frac{H^{\rm fid}r_{d}^{\rm fid}}{H(z)r_{d}},  \hspace{1cm}
     \alpha_\perp=\frac{D_A(z)r_d^{\rm fid}}{D_A^{\rm fid}r_d}.
 \end{equation}
By comparing BAO measurements at different redshifts, we can track the relative evolution of these parameters over cosmic time, ultimately providing powerful constraints on the properties of dark energy \citep{2024_BAO_cosmo}.

\begin{figure}
    \centering
    \includegraphics[width=1\textwidth]{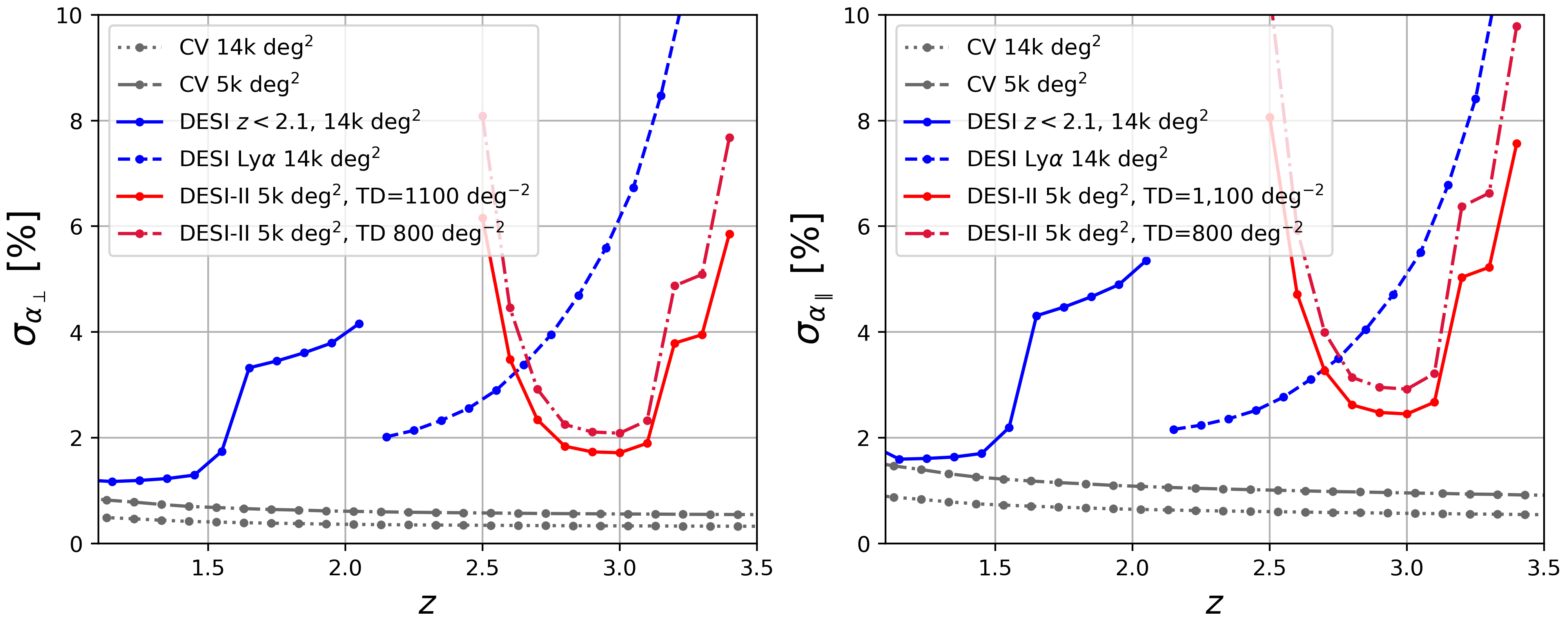}
    \caption{Forecasts of the Alcock-Paczynski (A.P.) parameters for the  DESI survey (in blue, with LRGs+ELGs+QSO for $z<2.1$ and Ly-$\alpha$ auto- and cross-correlations for $z>2.1$), the two DESI-II proposed survey of this article (red), compared with the cosmic variance limit for 14,000 and 5,000 deg$^2$ surveys (grey). We adopted a redshift binning $\Delta z=0.1$ for every configuration. TD stands for target density.}
    \label{fig:BAO_forecast}
\end{figure}

We first report in Fig. \ref{fig:BAO_forecast} the uncertainties on the Alcock-Paczynski parameters, obtained from the measurements of the BAO feature. The relative errors on $\alpha_\perp$ and $\alpha_\parallel$ can be interpreted as relative error on $D_{\rm A}/r_{\rm s}$ and $r_{\rm s}\,H$. The methods used involved marginalizing over a linear bias and 15 “broadband” polynomials, reproducing the experimental procedure (for details we refer to \cite{FishLSS_NSailer}). 
Our results show that the DESI-II survey we are proposing will reach few-percent precision in both distance measures in the redshift range $2.45 < z < 3.45$, for a binning $\Delta z=0.1$. The lowest relative uncertainties are 1.7$\%$ ($\alpha_\perp$) and 2.4$\%$ ($\alpha_\parallel$) for bins around z$\sim 3$ for the target density 1,100 deg$^{-2}$. For every bin, the highest density sample benefits a significant gain of around $20 \%$ over the lower density one. In both cases, these samples provide highly complementary measurements to the DESI ones, peaking in the redshift range where the Ly-$\alpha$ forest constraint becomes loose ($z>2.6$). A single bin forecast gives a combined uncertainty for $2.6<z<3.2$ of 0.73$\%$ (0.89) for $\alpha_{\perp}$ and 1.0$\%$ (1.2) for $\alpha_\parallel$, for 1,100 deg$^{-2}$ target density (800 deg$^{-2}$). These measurements are more than a factor of two higher than the cosmic variance limit. Consequently, they can be improved, at the cost of a new facility with higher multiplexing power. 

\begin{figure}
    \centering
    \includegraphics[width=1\textwidth]{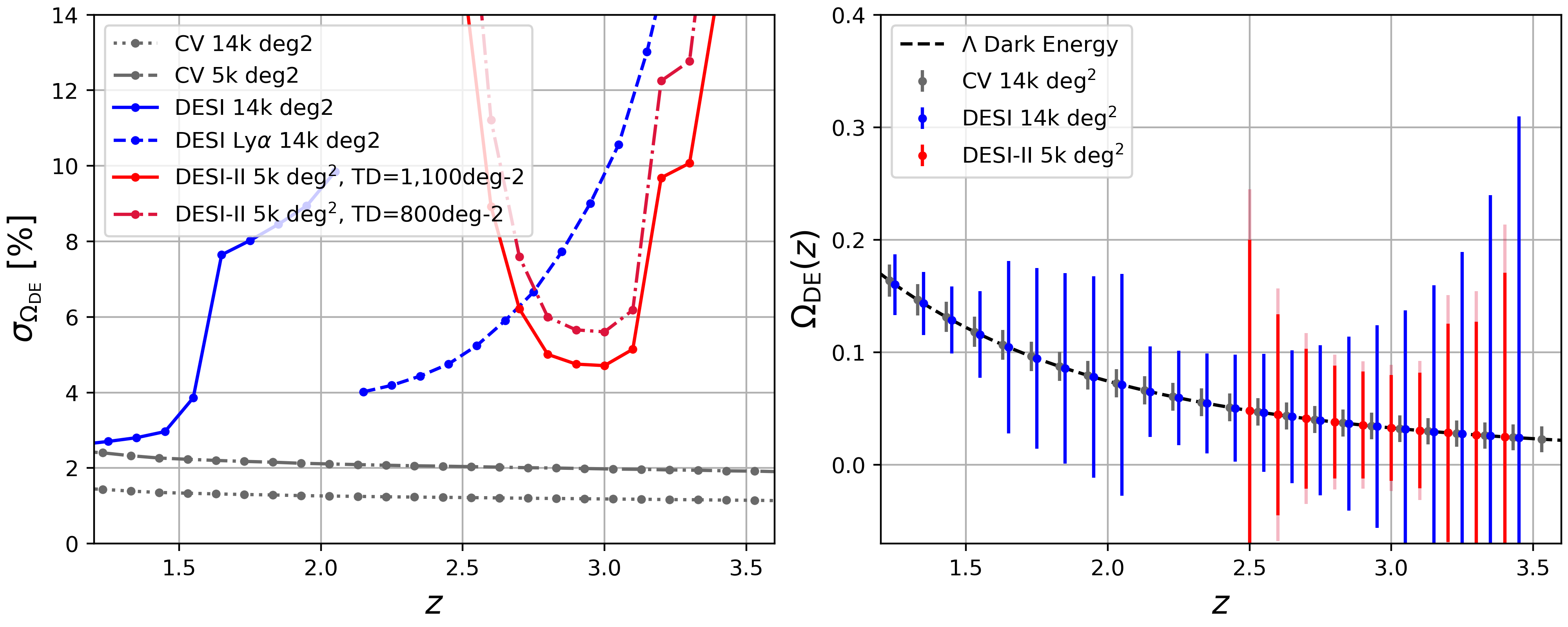}
    \caption{\textit{Left}: Forecast on the fraction of dark energy $\Omega_{\rm DE}$ for the same configurations as Fig. \ref{fig:BAO_forecast}, from the radial BAO parameters $\sigma_{\alpha_\parallel}$, with redshift bins $\Delta z=0.1$. \textit{Right:} Evolution of $\Omega_{\rm DE}$ for constant $\Lambda$, and the associated errors for C.V. (grey, for a 14,000 deg$^{2}$ survey), DESI (blue, galaxy and Ly-$\alpha$), and DESI-II (red for a target density of 1100 deg$^{-2}$ and light red for 800 deg$^{-2}$). TD stands for target density.}
    \label{fig:DE_forecast}
\end{figure}

The recent measurement of BAO from the DESI collaboration has shown a preference for dynamical dark energy at more than 2$\sigma$ (when combined with CMB or Type1a Supernovae) over the standard constant $\Lambda$ model \citep{DESI2024_BAO_gal,DESI_2024_bao_ly,2024_BAO_cosmo}. Thus it is crucial to investigate the potential of a DESI-II survey to constrain dark energy models, particularly for a redshift range loosely constrained by DESI. In Fig. \ref{fig:DE_forecast}, we forecast the uncertainty on the dark energy fraction for the different surveys, with a simplified procedure.  We use the first Friedman equation, which for $z<4$ is $\Omega_{\rm m}(z)+\Omega_{\rm DE}(z)=1$. We neglect the covariance between $h(z)$ and $\omega_{\rm m}$, we assume the uncertainty from CMB+DESI on $\omega_{\rm m}$ to be $\sigma_{\omega_{\rm m}}=6.5\times 10^{-4}$ (cf. Table 3 of \cite{FishLSS_NSailer}), and we take the relative error on $h(z)$ from the radial BAO measurements: $\sigma_{\ln{h}}=\sigma_{\ln\alpha_{\parallel}}$ \citep{DDE_forecast_Linder,FishLSS_NSailer}. Our forecasts predict at best 5-6 $\%$ constraint for individual bins, and a 2$\%$ one for an effective measurement $2.6<z<3.2$. As illustrated by the right panel, it will not be sufficient to measure a $\Lambda$ contribution which is predicted to be a few-percent component at $z>2$. Still, we gain sensitivity compared to the DESI Ly-$\alpha$ forest to measure a high redshift ($z\sim 3$) departure of a dark energy component from the $\Lambda$CDM framework. In particular, these types of early dark energy models have the potential to relieve the $H_0$ tension between early and late time measurements. 

Extending the precise observation of large-scale structure 3D modes to the high-redshift universe, DESI-II will be competitive (and complementary through lensing) with CMB observations \citep{Planck2020} as a probe of primordial perturbations and, therefore, a important survey to test inflationary models beyond the standard single-field, slow-roll scenario. Addressing the full potential of a DESI-II survey to investigate primordial features is beyond the scope of our work, and we humbly limit our forecast to the study the scale-dependent correction to linear galaxy bias by Primordial Non-Gaussianities, $\Delta b (k)\propto f_{\rm NL}b_\phi/\alpha(k)$, that is particularly significant at large scales \citep{NG_Dalal2008, Matarrese_Verde_2008}, consequently associated with systematic challenges not explored here. Such correction, parametrized by $f_{\rm NL}$, is induced by “local”-type primordial non-Gaussianity, characterized by a large signal in squeezed bispectrum configurations, a specific prediction of multi-field inflation. The current best constraint, $f_{\rm NL}^{\rm loc} = -0.9 \pm 5.1$, derived from \textit{Planck} CMB data \citep{Planck2020}, is consistent with Gaussian initial conditions and single-field inflation ($f_{\rm NL}^{\rm loc}=0$). Recent constraints on $f_{\rm NL}$ based on the scale-dependent bias are presented in \cite{fnl_DESI_CMB} for DESI QSO and CMB lensing, and \cite{fnl_chaussidon} for the DESI Y1 QSO and LRG samples.

For forecasting details, we refer the reader to \citep{2023MNRAS.521.3648D}. We fix the minimal mode to $k_{\rm min}=2\pi V_{\rm bin}^{-\frac{1}{3}}$, and evaluate our Fisher matrix for a single redshift bin $2.5<z<3.5$, neglecting the evolution of the galaxy distribution. We do so to avoid under-evaluating long-range mode by dividing the redshift range into bins. Furthermore, high density has in practice relatively small impact on NG, since the $nP>1$ is rapidly reached, and the volume is usually the limiting factor. Marginalizing over the galaxy bias, we predict $\sigma_{f_{\rm NL}}=6.9$ and $7.6$ for 1,100 and 800 deg$^{-2}$ densities. 
In the dense limit (evaluated for $n\sim 10^{-2}$ $h^3$ Mpc$^{-3}$, $b\sim 3.3$), the predicted uncertainty on NG is $\sigma_{f_{\rm NL}} \approx 4.7$ for our DESI-II volume. Thus our predicted measurement is close to the lower limit, and the asymptotic possible gain would require a large increase of the density. Still, our prediction is at a similar level as \textit{Planck} result $f_{\rm NL}=-0.9\pm 5.1$ \cite{Planck2020}, and DESI Y1 result $f_{\rm NL}=-3.6 \pm 9.1$. Our forecast suggests that achieving the precision required to place stringent constraints on inflationary models ($\sigma_{f_{\rm NL}} \approx 1$) using spectroscopic galaxy surveys necessitates the development of a next-generation (Stage V) spectroscopic survey.

\begin{figure}
    \centering
    \includegraphics[width=1\linewidth]{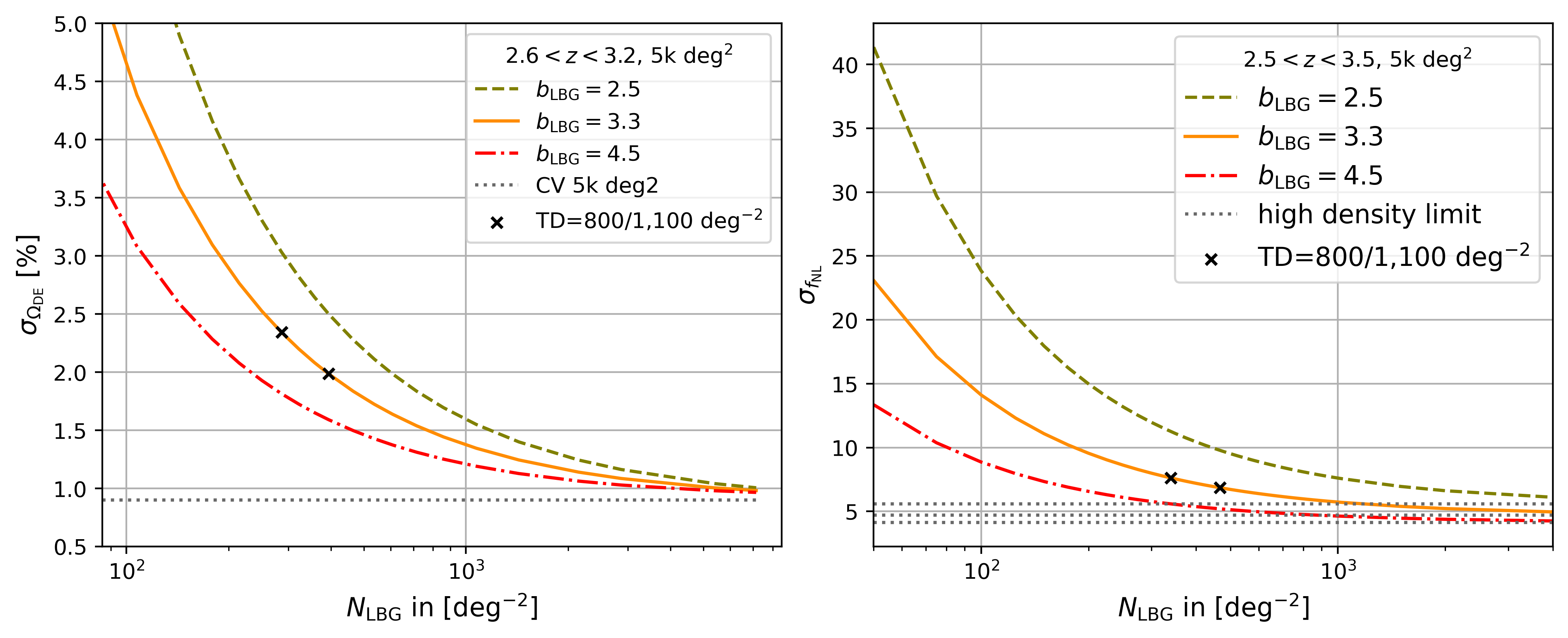}
    \caption{Impact of the sample size, through the number of LBG within the redshift range, and galaxy bias $b_{\rm LBG}$, for two different science cases: the fraction of dark energy (\textit{left panel}), and local non-Gaussianity amplitude (\textit{right panel}). The two proposed survey forecasts are represented by black crosses. We assume a DESI-II sky coverage of 5,000 deg$^2$. TD stands for target density.}
    \label{fig:DE_fnl_Ntarg_b}
\end{figure}

In the Fig. \ref{fig:DE_fnl_Ntarg_b},  we further explore the impact of the increase of the LBG sample, forecasting the constraints on the fraction of dark energy, and NG, respectively for redshift bins $2.6<z<3.2$ and $2.5<z<3.5$, varying the observed LBG numbers within each bin, for three galaxy biases: $b_{\rm LBG}=2.5;\,3.3;\,4.5$. The points corresponding to the (full) target density of 800 and 1,100 deg$^{-2}$ are highlighted in black. For NG, the asymptotic regime is rapidly reached, (for $N_{\rm LBG}$ twice smaller than our 'smaller' fiducial survey), and there is no significant gain in doubling the target density of our proposed survey. One can see the improvement with higher biases: not only is the asymptotic convergence faster, but the limit is also lower with higher biases. This difference is due to our model $b_{\phi}=2(b_{\rm LBG}-p)$, which increases the derivative $\partial \Delta b/\partial f_{\rm NL}$ with galaxy bias, and so some coefficients of the Fisher matrix. For dark energy, and more generally BAO measurement, the asymptotic regime starts for larger densities than NG. For instance, we gain a $50\%$ (resp. $30\%$) with a target density of 2,000 deg$^{-2}$ versus 800 (resp. 1,100). For comparison, the NG gain is limited to $25\%$ (resp $10\%$). As for NG, higher bias is associated with better measurements, but with the same CV limit. This figure illustrates the possibility of adjusting the target density, depending on the science cases, to optimize observational time. 

\begin{figure}
    \centering
    \includegraphics[width=.48\textwidth]{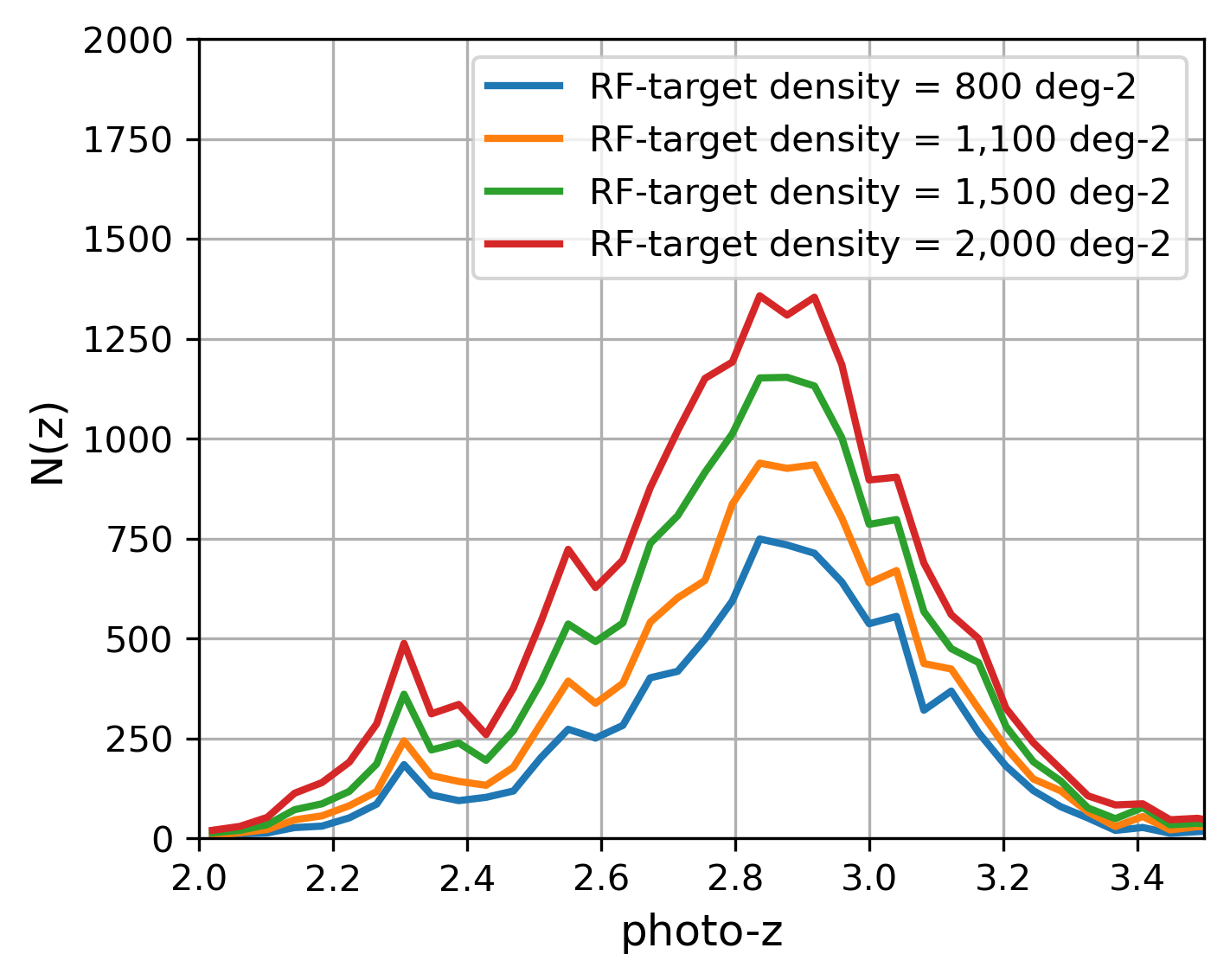}
    \includegraphics[width=.46\textwidth]{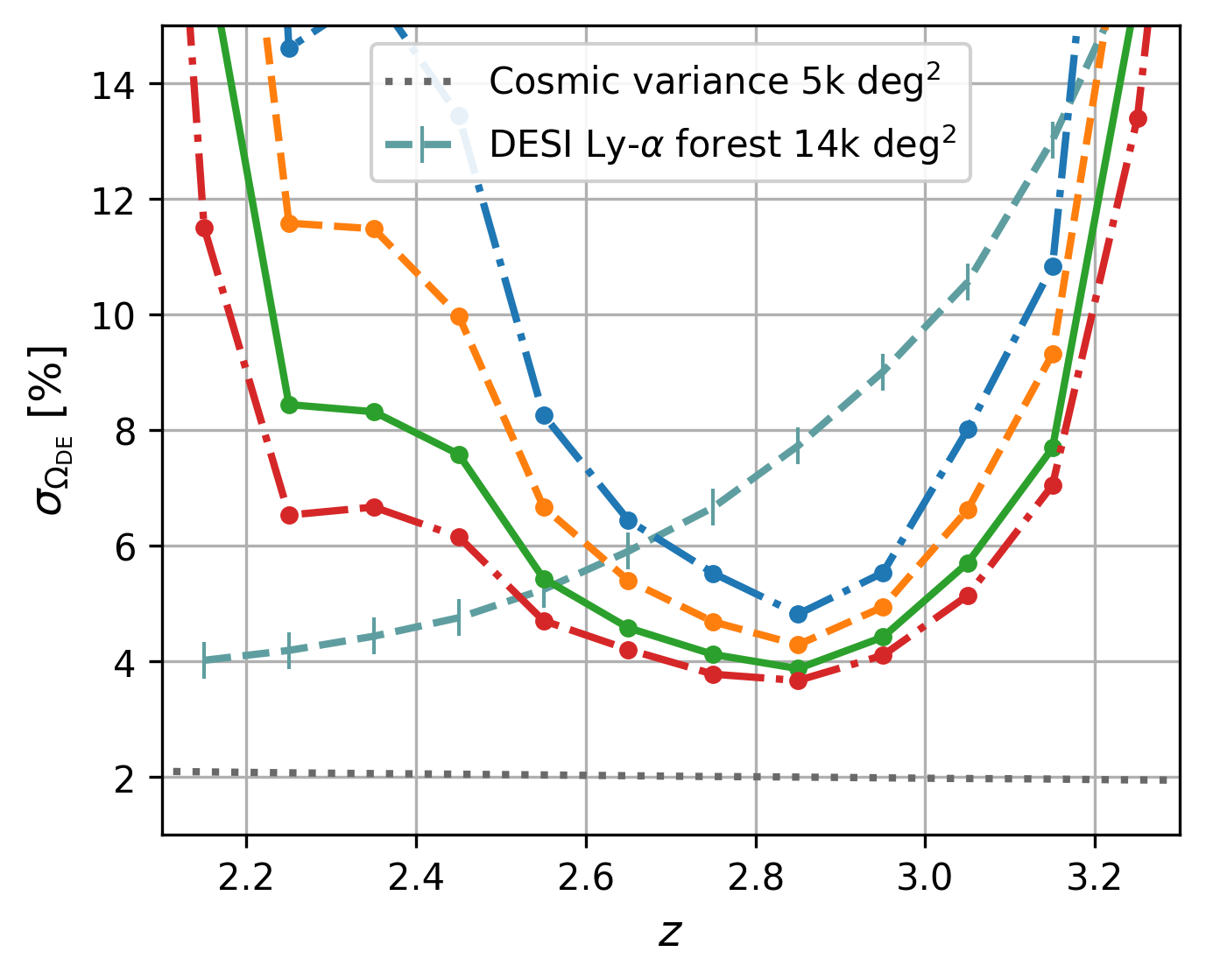}
    \caption{\textit{Left}: Photometric redshift distribution for three different Random Forest quality cuts, after accounting for spectroscopic redshift efficiency. \textit{Right}: Associated forecasts on the dark energy fraction for a 5,000 deg$^2$ survey, on top of DESI Ly-$\alpha$ forest ones.}
    \label{fig:DE_forecast_photoz}
\end{figure}

In the above, we have computed the forecast considering two target densities $1,100$ deg$^{-2}$ and $800$ deg$^{-2}$, where we assumed that the recovered spectroscopic distribution obtained in \cref{sec:pilot_obs_desi} with a target density of $1,100$ deg$^{-2}$ can be re-scaled to match the $n(z)$ corresponding to a target density of $800$ deg$^{-2}$, with higher purity. Since DESI observations were made for $1,100$ deg$^{-2}$, we cannot change easily the level of purity of targets, so we explore more in detail the effect of purity on the shape of the $n(z)$ and then on forecasts, by considering photometric redshifts, as performed in \cref{sec:unions-like_ts_cosmos}. \cref{fig:DE_forecast_photoz} (left panel) shows the four different photometric distributions for four different target densities, namely 800, 1,100,  1,500, and 2,000 deg$^{-2}$, that we obtain by applying different quality cut $P_{\rm lim}$ to the test dataset in \cref{sec:unions-like_ts_cosmos}. Moreover, we convolve the distribution with the spectroscopic efficiency presented in \cref{sec:unions-like_ts_cosmos} (with 2 hours of exposure), and we consider the redshift range $[2, 3.5]$, providing a density of LBG of 375, 495, 645 and 795 deg$^{-2}$, respectively. As expected, the recovered $n(z)$ are very similar (up to a factor), since the purity is stable in that target density range, demonstrating that target density can be increased up to $2,000$ deg$^{-2}$ and providing roughly the same fraction of confirmed LBGs for the LBG science program. \cref{fig:DE_forecast_photoz} (left panel) shows the forecasts of the uncertainty on the dark energy fraction with these three different $n(z)$ in \cref{fig:DE_forecast_photoz} (right panel), showing that the constraining power is maximum at a redshift of $\sim 2.85$. The mean redshift is not perfectly consistent with the one recovered from spectroscopy (see \cref{fig:DESI_ZZ}, left panel). Another difference is the lower uncertainties for $2.2<z<2.6$ compared to Fig \ref{fig:DE_forecast}, as expected from the larger tail of the distributions. 

\section{Conclusions}
\label{sec:conclusion}
Lyman-Break Galaxies are promising tracers for exploring the properties of dark energy in the high-redshift Universe. In light of DESI-II, the next phase of DESI set to begin in 2029, the role of broadband-wide photometric surveys such as UNIONS or the Vera C. Rubin LSST is essential to design the optimal selection of these tracers.

In this paper, we have explored the feasibility of selecting high-redshift Lyman Break Galaxies from these wide-broadband imaging surveys, with a focus on the ongoing UNIONS. For that, we have introduced a method to mimic a shallower imaging dataset by degrading the observed galaxy magnitudes of a deeper survey, allowing us to test the impact of shallower depth on LBG target selection. For that, we have considered using a Random Forest approach, to select high-redshift Lyman Break Galaxies in the {\sc LePHARE} photometric redshift range $[2.5, 3.5]$, using colors based on $ugriz$ magnitudes. Random Forest classification can effectively model complex non-linear relationships between observed galaxy colors and redshift, often surpassing traditional color-cut methods. 
We first tested this methodology on the XMM-LSS field by degrading existing deep CLAUDS+HSC imaging to the shallower CFIS+DECaLS depth. We showed that the Random Forest-based LBG target using the CFIS+DECaLS-like simulated dataset compares fairly to the existing one.

We then used this method on the COSMOS field, in preparation for a DESI pilot survey in May 2024, where we degraded CLAUDS+HSC imaging to match the magnitude depths of UNIONS. We forecasted the performance of this UNIONS-like selection and found that a target density budget of $n_{\rm LBG} =1,100$ deg$^{-2}$ provides $n_{\rm LBG}=873$ deg$^{-2}$ with $z>2$ for $23 < r <24.3$. This density turns to $n_{\rm LBG}(z > 2)=493$ deg$^{-2}$ with a mean photometric redshift of $\langle z|z>2\rangle = 2.83$ after accounting for the DESI spectroscopic redshift confirmation, based on a convolutional neural network and template fitting, giving an overall performance of the selection of $\sim 45\%$.

This selection was tested in a dedicated observation campaign of COSMOS with DESI. The 1,000 UNIONS-like observed targets were initially proposed for a target budget of $1,100$ deg$^{-2}$, from which have derived the corresponding spectroscopic redshift distribution based on the 420 spectroscopically-confirmed LBGs, providing a mean spectroscopic redshift of $\langle z|z>2\rangle = 3.0$, and the expected combined efficiency (Random Forest+spectroscopic redshift confirmation).

Although not addressed in this work, we emphasize the importance of correcting imaging systematics arising from the Random Forest selection function of tracers; Spatial variations in observational conditions, such as dust extinction and depth coverage in the relevant photometric bands, can introduce biases when training and test fields are sourced from different sky regions, potentially affecting the homogeneity of tracer large-scale distributions and thus, clustering analyses. Future research should examine the impact of these RF-sourced imaging systematics on LBG clustering analyses and explore corrective methods (see, for example, \cite{Chaussidon2023TSQSODESI,Chaussidon2024fnl}, who investigated the effect of imaging systematics on Quasars and Luminous Red Galaxies at large-scale clustering scales and their influence on $f_{\rm NL}$ measurements).
    
For cosmology, the interest of LBG samples is in measuring the galaxy clustering amplitudes and redshift space distortions at high redshift. Based on the recovered spectroscopic redshift distribution, we have explored the constraining power of using the LBG redshift-space galaxy power spectrum on the Alcock-Paczynski parameters, the fraction of dark energy, and local Primordial non-Gaussianities, displaying a maximum at redshift 2.8-3, within a redshift range $z > 2$. These redshifts are currently probed by the Ly$\alpha$ forest. Our forecasts predict at best 5-6 $\%$ constraint for individual bins, and a 2$\%$ one for an effective measurement $2.6<z<3.2$.

We also have explored the impact of different target budgets on the forecasts for the fraction of dark energy, by considering lower purity values in the LBG sample. We found that the slow decrease in purity, resulting in an LBG target density going from $800$ to $2,000$ deg$^{-2}$ does not strongly affect the shape of the photometric redshift distribution to lower redshifts. The constraining power is then increased due to higher LBG density, with a constant maximum at $z\sim 2.8-3.0$ (photometric redshift) over target budgets.

This study offers some prospects for the DESI-II phase, focusing on selecting LBGs from wide broadband imaging surveys using a Random Forest approach, which holds the potential to deliver competitive cosmological constraints at redshifts currently explored by the Ly$\alpha$ forest. While we believe this work provides reasonable estimates for the approximate magnitude of LBG number densities and redshift coverage, further advancements are possible within the framework of this analysis. We outline a non-exhaustive list of potential improvements, that would increase the precision and robustness of the conclusions of this paper. Our degradation method detailed in \cref{sec:shallowing_method} relies on some relatively simple hypotheses, to rescale with low-cost deeper imaging to a shallower one. In particular, we have neglected the effect of the relative contribution of readout noise in the magnitude error model, which may be non-negligible in the $u$-band. As $u$-band is of particular importance in LBG target selection, improved modeling of $u$-band imaging noise contributions would be required to better assess the impact of $u$-band imaging strategy.
For the training of the Random Forest algorithm, we choose a conservative approach by using only the color-redshift dependency. This selection could be enhanced by adding other interesting LBG-related features, such as their morphology. 

\section*{Data availability}
All the material needed to reproduce the figures of this publication is available at this site: 
\url{https://doi.org/10.5281/zenodo.13709754}.

\acknowledgments

The authors thank the anonymous reviewer for their insightful comments and suggestions. This paper has undergone internal review by the Dark Energy Spectroscopic Instrument collaboration. The internal reviewers were Allyson Brodzeller and Rongpu Zhou.

This material is based upon work supported by the U.S. Department of Energy (DOE), Office of Science, Office of High-Energy Physics, under Contract No. DE–AC02–05CH11231, and by the National Energy Research Scientific Computing Center, a DOE Office of Science User Facility under the same contract. Additional support for DESI was provided by the U.S. National Science Foundation (NSF), Division of Astronomical Sciences under Contract No. AST-0950945 to the NSF’s National Optical-Infrared Astronomy Research Laboratory; the Science and Technology Facilities Council of the United Kingdom; the Gordon and Betty Moore Foundation; the Heising-Simons Foundation; the French Alternative Energies and Atomic Energy Commission (CEA); the National Council of Humanities, Science and Technology of Mexico (CONAHCYT); the Ministry of Science, Innovation and Universities of Spain (MICIU/AEI/10.13039/501100011033), and by the DESI Member Institutions: \url{https://www.desi.lbl.gov/collaborating-institutions}. 
C. P. and C. Y. acknowledge support from grant ANR-22-CE31-0009 for the HZ-3D-MAP project and from grant ANR-22-CE92-0037 for the DESI-Lya project. W. d’A. acknowledges support from the  MICINN projects PID2019-111317GB-C32, PID2022-141079NB-C32 as well as predoctoral program AGAUR-FI ajuts (2024 FI-1 00692) Joan Oró.

The DESI Legacy Imaging Surveys consist of three individual and complementary projects: the Dark Energy Camera Legacy Survey (DECaLS), the Beijing-Arizona Sky Survey (BASS), and the Mayall z-band Legacy Survey (MzLS). DECaLS, BASS, and MzLS together include data obtained, respectively, at the Blanco telescope, Cerro Tololo Inter-American Observatory, NSF’s NOIRLab; the Bok telescope, Steward Observatory, University of Arizona; and the Mayall telescope, Kitt Peak National Observatory, NOIRLab. NOIRLab is operated by the Association of Universities for Research in Astronomy (AURA) under a cooperative agreement with the National Science Foundation. Pipeline processing and analyses of the data were supported by NOIRLab and the Lawrence Berkeley National Laboratory. Legacy Surveys also uses data products from the Near-Earth Object Wide-field Infrared Survey Explorer (NEOWISE), a project of the Jet Propulsion Laboratory/California Institute of Technology, funded by the National Aeronautics and Space Administration. Legacy Surveys was supported by: the Director, Office of Science, Office of High Energy Physics of the U.S. Department of Energy; the National Energy Research Scientific Computing Center, a DOE Office of Science User Facility; the U.S. National Science Foundation, Division of Astronomical Sciences; the National Astronomical Observatories of China, the Chinese Academy of Sciences and the Chinese National Natural Science Foundation. LBNL is managed by the Regents of the University of California under contract to the U.S. Department of Energy. The complete acknowledgments can be found at https://www.legacysurvey.org/.

Any opinions, findings, and conclusions or recommendations expressed in this material are those of the author(s) and do not necessarily reflect the views of the U. S. National Science Foundation, the U. S. Department of Energy, or any of the listed funding agencies.

This work also uses data obtained and processed as part of the CFHT Large Area U-band Deep Survey (CLAUDS), which is a collaboration between astronomers from Canada, France, and China. CLAUDS is based on observations obtained with MegaPrime/MegaCam, a joint project of CFHT and CEA/Irfu, at the CFHT which is operated by the National Research Council (NRC) of Canada, the Institut National des Science de l’Univers of the Centre National de la Recherche Scientifique (CNRS) of France, and the University of Hawaii. CLAUDS uses data obtained in part through the Telescope Access Program (TAP), which has been funded by the National Astronomical Observatories, Chinese Academy of Sciences, and the Special Fund for Astronomy from the Ministry of Finance of China. CLAUDS uses data products from TERAPIX and the Canadian Astronomy Data Centre (CADC) and was carried out using resources from Compute Canada and Canadian Advanced Network For Astrophysical Research (CANFAR).

This paper is also based on data collected at the Subaru Telescope by the Hyper Suprime-Cam (HSC) collaboration and retrieved from the HSC data archive system, which is operated by the Subaru Telescope and Astronomy Data Center (ADC) at NAOJ. Their data analysis was in part carried out with the cooperation of Center for Computational Astrophysics
(CfCA), NAOJ. The HSC collaboration includes the astronomical communities of Japan and Taiwan, and Princeton University. The HSC instrumentation and software were developed by the National Astronomical Observatory of Japan (NAOJ), the Kavli Institute for the Physics and Mathematics of the Universe (Kavli IPMU), the University of Tokyo, the High Energy Accelerator Research Organization (KEK), the Academia Sinica Institute for Astronomy and Astrophysics in Taiwan (ASIAA), and Princeton University. Funding was contributed by the FIRST program from the Japanese Cabinet Office, the Ministry of Education, Culture, Sports, Science and Technology (MEXT), the Japan Society for the Promotion of Science (JSPS), Japan Science and Technology Agency (JST), the Toray Science Foundation, NAOJ, Kavli IPMU, KEK, ASIAA, and Princeton University.

The HSC collaboration members are honored and grateful for the opportunity of observing the Universe from Maunakea, which has the cultural, historical and natural significance in Hawaii.

\bibliographystyle{JHEP}
\bibliography{main.bib}{}

\let\jnlstyle=\rm\def\jref#1{{\jnlstyle#1}}\def\aj{\jref{AJ}} \def\araa{\jref{ARA\&A}} \def\apj{\jref{ApJ}\ } \def\apjl{\jref{ApJ}\ } \def\apjs{\jref{ApJS}} \def\ao{\jref{Appl.~Opt.}} \def\apss{\jref{Ap\&SS}} \def\aap{\jref{A\&A}} \def\aapr{\jref{A\&A~Rev.}} \def\aaps{\jref{A\&AS}} \def\azh{\jref{AZh}} \def\baas{\jref{BAAS}} \def\jrasc{\jref{JRASC}} \def\memras{\jref{MmRAS}} \def\mnras{\jref{MNRAS}\ } \def\pra{\jref{Phys.~Rev.~A}\ } \def\prb{\jref{Phys.~Rev.~B}\ } \def\prc{\jref{Phys.~Rev.~C}\ } \def\prd{\jref{Phys.~Rev.~D}\ } \def\pre{\jref{Phys.~Rev.~E}} \def\prl{\jref{Phys.~Rev.~Lett.}} \def\pasp{\jref{PASP}} \def\pasj{\jref{PASJ}} \def\qjras{\jref{QJRAS}} \def\skytel{\jref{S\&T}} \def\solphys{\jref{Sol.~Phys.}} \def\sovast{\jref{Soviet~Ast.}} \def\ssr{\jref{Space~Sci.~Rev.}} \def\zap{\jref{ZAp}} \def\nat{\jref{Nature}\ } \def\iaucirc{\jref{IAU~Circ.}} \def\aplett{\jref{Astrophys.~Lett.}} \def\apspr{\jref{Astrophys.~Space~Phys.~Res.}} \def\bain{\jref{Bull.~Astron.~Inst.~Netherlands}}
  \def\fcp{\jref{Fund.~Cosmic~Phys.}} \def\gca{\jref{Geochim.~Cosmochim.~Acta}} \def\grl{\jref{Geophys.~Res.~Lett.}} \def\jcp{\jref{J.~Chem.~Phys.}} \def\jgr{\jref{J.~Geophys.~Res.}} \def\jqsrt{\jref{J.~Quant.~Spec.~Radiat.~Transf.}} \def\memsai{\jref{Mem.~Soc.~Astron.~Italiana}} \def\nphysa{\jref{Nucl.~Phys.~A}} \def\physrep{\jref{Phys.~Rep.}} \def\physscr{\jref{Phys.~Scr}} \def\planss{\jref{Planet.~Space~Sci.}} \def\prd{\jref{Phys.~Rev.~D}} \def\procspie{\jref{Proc.~SPIE}} \let\astap=\aap \let\apjlett=\apjl \let\apjsupp=\apjs \let\applopt=\ao \def\jcap{\jref{JCAP}} \def\pasa{\jref{PASA}}

\providecommand{\href}[2]{#2}\begingroup\raggedright\begin{thebibliography}{10}

\bibitem{Steidel1996LBG}
C.C.~{Steidel}, M.~{Giavalisco}, M.~{Pettini}, M.~{Dickinson} and K.L.~{Adelberger}, \emph{{Spectroscopic Confirmation of a Population of Normal Star-forming Galaxies at Redshifts Z > 3}}, \href{https://doi.org/10.1086/310029}{\emph{\apjl} {\bfseries 462} (1996) L17} [\href{https://arxiv.org/abs/astro-ph/9602024}{{\ttfamily astro-ph/9602024}}].

\bibitem{RuhlmannKleider2024LBGCLAUDS}
V.~{Ruhlmann-Kleider}, C.~{Y{\`e}che}, C.~{Magneville}, H.~{Coquinot}, E.~{Armengaud}, N.~{Palanque-Delabrouille} et~al., \emph{{High redshift LBGs from deep broadband imaging for future spectroscopic surveys}}, \href{https://doi.org/10.48550/arXiv.2404.03569}{\emph{arXiv e-prints} (2024) arXiv:2404.03569} [\href{https://arxiv.org/abs/2404.03569}{{\ttfamily 2404.03569}}].

\bibitem{malkan2017}
M.A.~Malkan, D.P.~Cohen, M.~Maruyama, N.~Kashikawa, C.~Ly, S.~Ishikawa et~al., \emph{Lyman-break {{Galaxies}} at z {$\sim$} 3 in the {{Subaru Deep Field}}: {{Luminosity Function}}, {{Clustering}}, and [{{O III}}] {{Emission}}}, \href{https://doi.org/10.3847/1538-4357/aa9331}{\emph{The Astrophysical Journal} {\bfseries 850} (2017) 5}.

\bibitem{ono2018}
Y.~Ono, M.~Ouchi, Y.~Harikane, J.~Toshikawa, M.~Rauch, S.~Yuma et~al., \emph{Great {{Optically Luminous Dropout Research Using Subaru HSC}} ({{GOLDRUSH}}). {{I}}. {{UV}} luminosity functions at z {$\sim$} 4-7 derived with the half-million dropouts on the 100 deg2 sky}, \href{https://doi.org/10.1093/pasj/psx103}{\emph{Publications of the Astronomical Society of Japan} {\bfseries 70} (2018) S10}.

\bibitem{harikane2022}
Y.~Harikane, Y.~Ono, M.~Ouchi, C.~Liu, M.~Sawicki, T.~Shibuya et~al., \emph{{{GOLDRUSH}}. {{IV}}. {{Luminosity Functions}} and {{Clustering Revealed}} with 4,000,000 {{Galaxies}} at z 2-7: {{Galaxy-AGN Transition}}, {{Star Formation Efficiency}}, and {{Implication}} for {{Evolution}} at z {$>$} 10}, \href{https://doi.org/10.3847/1538-4365/ac3dfc}{\emph{The Astrophysical Journal Supplement Series} {\bfseries 259} (2022) 20}.

\bibitem{Steidel1999LBG}
C.C.~{Steidel}, K.L.~{Adelberger}, M.~{Giavalisco}, M.~{Dickinson} and M.~{Pettini}, \emph{{Lyman-Break Galaxies at z>\raisebox{-0.5ex}\textasciitilde4 and the Evolution of the Ultraviolet Luminosity Density at High Redshift}}, \href{https://doi.org/10.1086/307363}{\emph{\apj} {\bfseries 519} (1999) 1} [\href{https://arxiv.org/abs/astro-ph/9811399}{{\ttfamily astro-ph/9811399}}].

\bibitem{Giavalisco2004LBG}
M.~{Giavalisco}, M.~{Dickinson}, H.C.~{Ferguson}, S.~{Ravindranath}, C.~{Kretchmer}, L.A.~{Moustakas} et~al., \emph{{The Rest-Frame Ultraviolet Luminosity Density of Star-forming Galaxies at Redshifts z>3.5}}, \href{https://doi.org/10.1086/381244}{\emph{\apjl} {\bfseries 600} (2004) L103} [\href{https://arxiv.org/abs/astro-ph/0309065}{{\ttfamily astro-ph/0309065}}].

\bibitem{Reddy2008LBG}
N.A.~{Reddy}, C.C.~{Steidel}, M.~{Pettini}, K.L.~{Adelberger}, A.E.~{Shapley}, D.K.~{Erb} et~al., \emph{{Multiwavelength Constraints on the Cosmic Star Formation History from Spectroscopy: The Rest-Frame Ultraviolet, H{\ensuremath{\alpha}}, and Infrared Luminosity Functions at Redshifts 1.9 lesssim z lesssim 3.4}}, \href{https://doi.org/10.1086/521105}{\emph{\apjs} {\bfseries 175} (2008) 48} [\href{https://arxiv.org/abs/0706.4091}{{\ttfamily 0706.4091}}].

\bibitem{Hildebrandt2009lbg}
H.~{Hildebrandt}, J.~{Pielorz}, T.~{Erben}, L.~{van Waerbeke}, P.~{Simon} and P.~{Capak}, \emph{{CARS: The CFHTLS-Archive-Research Survey II. Weighing dark matter halos of Lyman-break galaxies at z=3-5}}, \href{https://doi.org/10.48550/arXiv.0903.3951}{\emph{arXiv e-prints} (2009) arXiv:0903.3951} [\href{https://arxiv.org/abs/0903.3951}{{\ttfamily 0903.3951}}].

\bibitem{Finkelstein2022lbg}
S.L.~{Finkelstein}, M.B.~{Bagley}, P.~{Arrabal Haro}, M.~{Dickinson}, H.C.~{Ferguson}, J.S.~{Kartaltepe} et~al., \emph{{A Long Time Ago in a Galaxy Far, Far Away: A Candidate z {\ensuremath{\sim}} 12 Galaxy in Early JWST CEERS Imaging}}, \href{https://doi.org/10.3847/2041-8213/ac966e}{\emph{\apjl} {\bfseries 940} (2022) L55} [\href{https://arxiv.org/abs/2207.12474}{{\ttfamily 2207.12474}}].

\bibitem{Harikane2023lbg}
Y.~{Harikane}, M.~{Ouchi}, M.~{Oguri}, Y.~{Ono}, K.~{Nakajima}, Y.~{Isobe} et~al., \emph{{A Comprehensive Study of Galaxies at z 9-16 Found in the Early JWST Data: Ultraviolet Luminosity Functions and Cosmic Star Formation History at the Pre-reionization Epoch}}, \href{https://doi.org/10.3847/1538-4365/acaaa9}{\emph{\apjs} {\bfseries 265} (2023) 5} [\href{https://arxiv.org/abs/2208.01612}{{\ttfamily 2208.01612}}].

\bibitem{Miyatake2022lbgcmblensing}
H.~{Miyatake}, Y.~{Harikane}, M.~{Ouchi}, Y.~{Ono}, N.~{Yamamoto}, A.J.~{Nishizawa} et~al., \emph{{First Identification of a CMB Lensing Signal Produced by 1.5 Million Galaxies at z {\ensuremath{\sim}}4 : Constraints on Matter Density Fluctuations at High Redshift}}, \href{https://doi.org/10.1103/PhysRevLett.129.061301}{\emph{\prl} {\bfseries 129} (2022) 061301} [\href{https://arxiv.org/abs/2103.15862}{{\ttfamily 2103.15862}}].

\bibitem{Schmittfull2018fnl}
M.~{Schmittfull} and U.~{Seljak}, \emph{{Parameter constraints from cross-correlation of CMB lensing with galaxy clustering}}, \href{https://doi.org/10.1103/PhysRevD.97.123540}{\emph{\prd} {\bfseries 97} (2018) 123540} [\href{https://arxiv.org/abs/1710.09465}{{\ttfamily 1710.09465}}].

\bibitem{Chaussidon2024fnl}
E.~{Chaussidon}, C.~{Y{\`e}che}, A.~{de Mattia}, C.~{Payerne}, P.~{McDonald}, A.J.~{Ross} et~al., \emph{{Constraining primordial non-Gaussianity with DESI 2024 LRG and QSO samples}}, \href{https://doi.org/10.48550/arXiv.2411.17623}{\emph{arXiv e-prints} (2024) arXiv:2411.17623} [\href{https://arxiv.org/abs/2411.17623}{{\ttfamily 2411.17623}}].

\bibitem{Yu2018neutrinomass}
B.~{Yu}, R.Z.~{Knight}, B.D.~{Sherwin}, S.~{Ferraro}, L.~{Knox} and M.~{Schmittfull}, \emph{{Towards Neutrino Mass from Cosmology without Optical Depth Information}}, \href{https://doi.org/10.48550/arXiv.1809.02120}{\emph{arXiv e-prints} (2018) arXiv:1809.02120} [\href{https://arxiv.org/abs/1809.02120}{{\ttfamily 1809.02120}}].

\bibitem{Tudorica2017lbgmagnificationclusters}
A.~{Tudorica}, H.~{Hildebrandt}, M.~{Tewes}, H.~{Hoekstra}, C.B.~{Morrison}, A.~{Muzzin} et~al., \emph{{Weak lensing magnification of SpARCS galaxy clusters}}, \href{https://doi.org/10.1051/0004-6361/201731267}{\emph{\aap} {\bfseries 608} (2017) A141} [\href{https://arxiv.org/abs/1710.06431}{{\ttfamily 1710.06431}}].

\bibitem{Cross2024IGGL}
D.N.~{Cross} and C.~{S{\'a}nchez}, \emph{{Inverse galaxy-galaxy lensing: Magnification, intrinsic alignments, and cosmology}}, \href{https://doi.org/10.1103/PhysRevD.110.123534}{\emph{\prd} {\bfseries 110} (2024) 123534} [\href{https://arxiv.org/abs/2410.00714}{{\ttfamily 2410.00714}}].

\bibitem{DESI2022instrument}
{DESI Collaboration}, B.~{Abareshi}, J.~{Aguilar}, S.~{Ahlen}, S.~{Alam}, D.M.~{Alexander} et~al., \emph{{Overview of the Instrumentation for the Dark Energy Spectroscopic Instrument}}, \href{https://doi.org/10.3847/1538-3881/ac882b}{\emph{\aj} {\bfseries 164} (2022) 207} [\href{https://arxiv.org/abs/2205.10939}{{\ttfamily 2205.10939}}].

\bibitem{DESI2016paper}
{DESI Collaboration}, A.~{Aghamousa}, J.~{Aguilar}, S.~{Ahlen}, S.~{Alam}, L.E.~{Allen} et~al., \emph{{The DESI Experiment Part II: Instrument Design}}, \href{https://doi.org/10.48550/arXiv.1611.00037}{\emph{arXiv e-prints} (2016) arXiv:1611.00037} [\href{https://arxiv.org/abs/1611.00037}{{\ttfamily 1611.00037}}].

\bibitem{Silber2023DESI}
J.H.~{Silber}, P.~{Fagrelius}, K.~{Fanning}, M.~{Schubnell}, J.N.~{Aguilar}, S.~{Ahlen} et~al., \emph{{The Robotic Multiobject Focal Plane System of the Dark Energy Spectroscopic Instrument (DESI)}}, \href{https://doi.org/10.3847/1538-3881/ac9ab1}{\emph{\aj} {\bfseries 165} (2023) 9} [\href{https://arxiv.org/abs/2205.09014}{{\ttfamily 2205.09014}}].

\bibitem{DESI2024bao1}
{DESI Collaboration}, A.G.~{Adame}, J.~{Aguilar}, S.~{Ahlen}, S.~{Alam}, D.M.~{Alexander} et~al., \emph{{DESI 2024 III: Baryon Acoustic Oscillations from Galaxies and Quasars}}, \href{https://doi.org/10.48550/arXiv.2404.03000}{\emph{arXiv e-prints} (2024) arXiv:2404.03000} [\href{https://arxiv.org/abs/2404.03000}{{\ttfamily 2404.03000}}].

\bibitem{DESI2024bao}
{DESI Collaboration}, A.G.~{Adame}, J.~{Aguilar}, S.~{Ahlen}, S.~{Alam}, D.M.~{Alexander} et~al., \emph{{DESI 2024 VI: Cosmological Constraints from the Measurements of Baryon Acoustic Oscillations}}, \href{https://doi.org/10.48550/arXiv.2404.03002}{\emph{arXiv e-prints} (2024) arXiv:2404.03002} [\href{https://arxiv.org/abs/2404.03002}{{\ttfamily 2404.03002}}].

\bibitem{DESI2024rsd}
{DESI Collaboration}, A.G.~{Adame}, J.~{Aguilar}, S.~{Ahlen}, S.~{Alam}, D.M.~{Alexander} et~al., \emph{{DESI 2024 V: Full-Shape Galaxy Clustering from Galaxies and Quasars}}, \href{https://doi.org/10.48550/arXiv.2411.12021}{\emph{arXiv e-prints} (2024) arXiv:2411.12021} [\href{https://arxiv.org/abs/2411.12021}{{\ttfamily 2411.12021}}].

\bibitem{Schlegel2022DESI2}
D.J.~{Schlegel}, S.~{Ferraro}, G.~{Aldering}, C.~{Baltay}, S.~{BenZvi}, R.~{Besuner} et~al., \emph{{A Spectroscopic Road Map for Cosmic Frontier: DESI, DESI-II, Stage-5}}, \href{https://doi.org/10.48550/arXiv.2209.03585}{\emph{arXiv e-prints} (2022) arXiv:2209.03585} [\href{https://arxiv.org/abs/2209.03585}{{\ttfamily 2209.03585}}].

\bibitem{Ferraro2019inflation}
S.~{Ferraro} and M.J.~{Wilson}, \emph{{Inflation and Dark Energy from spectroscopy at z > 2}}, \href{https://doi.org/10.48550/arXiv.1903.09208}{\emph{\baas} {\bfseries 51} (2019) 72} [\href{https://arxiv.org/abs/1903.09208}{{\ttfamily 1903.09208}}].

\bibitem{Sailer2021cosmo}
N.~{Sailer}, E.~{Castorina}, S.~{Ferraro} and M.~{White}, \emph{{Cosmology at high redshift - a probe of fundamental physics}}, \href{https://doi.org/10.1088/1475-7516/2021/12/049}{\emph{\jcap} {\bfseries 2021} (2021) 049} [\href{https://arxiv.org/abs/2106.09713}{{\ttfamily 2106.09713}}].

\bibitem{DES2005whitepaper}
T.D.E.S.~Collaboration, \emph{The dark energy survey},  2005.

\bibitem{Dey2019DESIILS}
A.~{Dey}, D.J.~{Schlegel}, D.~{Lang}, R.~{Blum}, K.~{Burleigh}, X.~{Fan} et~al., \emph{{Overview of the DESI Legacy Imaging Surveys}}, \href{https://doi.org/10.3847/1538-3881/ab089d}{\emph{\aj} {\bfseries 157} (2019) 168} [\href{https://arxiv.org/abs/1804.08657}{{\ttfamily 1804.08657}}].

\bibitem{Ibata2017CFIS}
R.A.~{Ibata}, A.~{McConnachie}, J.-C.~{Cuillandre}, N.~{Fantin}, M.~{Haywood}, N.F.~{Martin} et~al., \emph{{The Canada-France Imaging Survey: First Results from the u-Band Component}}, \href{https://doi.org/10.3847/1538-4357/aa855c}{\emph{\apj} {\bfseries 848} (2017) 128} [\href{https://arxiv.org/abs/1708.06356}{{\ttfamily 1708.06356}}].

\bibitem{Chambers2016panstarrs}
K.C.~{Chambers}, E.A.~{Magnier}, N.~{Metcalfe}, H.A.~{Flewelling}, M.E.~{Huber}, C.Z.~{Waters} et~al., \emph{{The Pan-STARRS1 Surveys}}, \href{https://doi.org/10.48550/arXiv.1612.05560}{\emph{arXiv e-prints} (2016) arXiv:1612.05560} [\href{https://arxiv.org/abs/1612.05560}{{\ttfamily 1612.05560}}].

\bibitem{Miyazaki2018HSC}
S.~{Miyazaki}, Y.~{Komiyama}, S.~{Kawanomoto}, Y.~{Doi}, H.~{Furusawa}, T.~{Hamana} et~al., \emph{{Hyper Suprime-Cam: System design and verification of image quality}}, \href{https://doi.org/10.1093/pasj/psx063}{\emph{\pasj} {\bfseries 70} (2018) S1}.

\bibitem{Gwyn2025unions}
S.~{Gwyn}, A.W.~{McConnachie}, J.-C.~{Cuillandre}, K.C.~{Chambers}, E.A.~{Magnier}, M.J.~{Hudson} et~al., \emph{{UNIONS: The Ultraviolet Near-Infrared Optical Northern Survey}}, {\emph{arXiv e-prints} (2025) arXiv:2503.13783} [\href{https://arxiv.org/abs/2503.13783}{{\ttfamily 2503.13783}}].

\bibitem{LSST2009whitepaper}
{LSST Science Collaboration}, P.A.~{Abell}, J.~{Allison}, S.F.~{Anderson}, J.R.~{Andrew}, J.R.P.~{Angel} et~al., \emph{{LSST Science Book, Version 2.0}}, {\emph{arXiv e-prints} (2009) arXiv:0912.0201} [\href{https://arxiv.org/abs/0912.0201}{{\ttfamily 0912.0201}}].

\bibitem{Euclid2011whitepaper}
R.~{Laureijs}, J.~{Amiaux}, S.~{Arduini}, J.L.~{Augu{\`e}res}, J.~{Brinchmann}, R.~{Cole} et~al., \emph{{Euclid Definition Study Report}}, {\emph{arXiv e-prints} (2011) arXiv:1110.3193} [\href{https://arxiv.org/abs/1110.3193}{{\ttfamily 1110.3193}}].

\bibitem{Ibata2017cfisu}
R.A.~{Ibata}, A.~{McConnachie}, J.-C.~{Cuillandre}, N.~{Fantin}, M.~{Haywood}, N.F.~{Martin} et~al., \emph{{The Canada-France Imaging Survey: First Results from the u-Band Component}}, \href{https://doi.org/10.3847/1538-4357/aa855c}{\emph{\apj} {\bfseries 848} (2017) 128} [\href{https://arxiv.org/abs/1708.06356}{{\ttfamily 1708.06356}}].

\bibitem{Zhao2024must}
C.~{Zhao}, S.~{Huang}, M.~{He}, P.~{Montero-Camacho}, Y.~{Liu}, P.~{Renard} et~al., \emph{{MUltiplexed Survey Telescope: Perspectives for Large-Scale Structure Cosmology in the Era of Stage-V Spectroscopic Survey}}, \href{https://doi.org/10.48550/arXiv.2411.07970}{\emph{arXiv e-prints} (2024) arXiv:2411.07970} [\href{https://arxiv.org/abs/2411.07970}{{\ttfamily 2411.07970}}].

\bibitem{Aihara2019HSC}
H.~{Aihara}, Y.~{AlSayyad}, M.~{Ando}, R.~{Armstrong}, J.~{Bosch}, E.~{Egami} et~al., \emph{{Third data release of the Hyper Suprime-Cam Subaru Strategic Program}}, \href{https://doi.org/10.1093/pasj/psab122}{\emph{\pasj} {\bfseries 74} (2022) 247} [\href{https://arxiv.org/abs/2108.13045}{{\ttfamily 2108.13045}}].

\bibitem{Desprez2023CLAUDSHSCSSP}
G.~{Desprez}, V.~{Picouet}, T.~{Moutard}, S.~{Arnouts}, M.~{Sawicki}, J.~{Coupon} et~al., \emph{{Combining the CLAUDS and HSC-SSP surveys. U + grizy(+YJHK$_{s}$) photometry and photometric redshifts for 18M galaxies in the 20 deg$^{2}$ of the HSC-SSP Deep and ultraDeep fields}}, \href{https://doi.org/10.1051/0004-6361/202243363}{\emph{\aap} {\bfseries 670} (2023) A82} [\href{https://arxiv.org/abs/2301.13750}{{\ttfamily 2301.13750}}].

\bibitem{Sawicki2019CLAUDS}
M.~{Sawicki}, S.~{Arnouts}, J.~{Huang}, J.~{Coupon}, A.~{Golob}, S.~{Gwyn} et~al., \emph{{The CFHT large area U-band deep survey (CLAUDS)}}, \href{https://doi.org/10.1093/mnras/stz2522}{\emph{\mnras} {\bfseries 489} (2019) 5202} [\href{https://arxiv.org/abs/1909.05898}{{\ttfamily 1909.05898}}].

\bibitem{Bianco2022lsst}
F.B.~{Bianco}, {\v{Z}}.~{Ivezi{\'c}}, R.L.~{Jones}, M.L.~{Graham}, P.~{Marshall}, A.~{Saha} et~al., \emph{{Optimization of the Observing Cadence for the Rubin Observatory Legacy Survey of Space and Time: A Pioneering Process of Community-focused Experimental Design}}, \href{https://doi.org/10.3847/1538-4365/ac3e72}{\emph{\apjs} {\bfseries 258} (2022) 1} [\href{https://arxiv.org/abs/2108.01683}{{\ttfamily 2108.01683}}].

\bibitem{Dey2019LegacySurvey}
A.~{Dey}, D.J.~{Schlegel}, D.~{Lang}, R.~{Blum}, K.~{Burleigh}, X.~{Fan} et~al., \emph{{Overview of the DESI Legacy Imaging Surveys}}, \href{https://doi.org/10.3847/1538-3881/ab089d}{\emph{\aj} {\bfseries 157} (2019) 168} [\href{https://arxiv.org/abs/1804.08657}{{\ttfamily 1804.08657}}].

\bibitem{Flaugher2015DECAM}
B.~{Flaugher}, H.T.~{Diehl}, K.~{Honscheid}, T.M.C.~{Abbott}, O.~{Alvarez}, R.~{Angstadt} et~al., \emph{{The Dark Energy Camera}}, \href{https://doi.org/10.1088/0004-6256/150/5/150}{\emph{\aj} {\bfseries 150} (2015) 150} [\href{https://arxiv.org/abs/1504.02900}{{\ttfamily 1504.02900}}].

\bibitem{EuclidCollaboration2022survey}
{Euclid Collaboration}, R.~{Scaramella}, J.~{Amiaux}, Y.~{Mellier}, C.~{Burigana}, C.S.~{Carvalho} et~al., \emph{{Euclid preparation. I. The Euclid Wide Survey}}, \href{https://doi.org/10.1051/0004-6361/202141938}{\emph{\aap} {\bfseries 662} (2022) A112} [\href{https://arxiv.org/abs/2108.01201}{{\ttfamily 2108.01201}}].

\bibitem{LSST2012whitepaper}
{LSST Dark Energy Science Collaboration}, \emph{{Large Synoptic Survey Telescope: Dark Energy Science Collaboration}}, \href{https://doi.org/10.48550/arXiv.1211.0310}{\emph{arXiv e-prints} (2012) arXiv:1211.0310} [\href{https://arxiv.org/abs/1211.0310}{{\ttfamily 1211.0310}}].

\bibitem{Euclid2021phootometry}
{Euclid Collaboration}, A.~{Pocino}, I.~{Tutusaus}, F.J.~{Castander}, P.~{Fosalba}, M.~{Crocce} et~al., \emph{{Euclid preparation. XII. Optimizing the photometric sample of the Euclid survey for galaxy clustering and galaxy-galaxy lensing analyses}}, \href{https://doi.org/10.1051/0004-6361/202141061}{\emph{\aap} {\bfseries 655} (2021) A44} [\href{https://arxiv.org/abs/2104.05698}{{\ttfamily 2104.05698}}].

\bibitem{Rykoff2015magnitudeerror}
E.S.~{Rykoff}, E.~{Rozo} and R.~{Keisler}, \emph{{Assessing Galaxy Limiting Magnitudes in Large Optical Surveys}}, \href{https://doi.org/10.48550/arXiv.1509.00870}{\emph{arXiv e-prints} (2015) arXiv:1509.00870} [\href{https://arxiv.org/abs/1509.00870}{{\ttfamily 1509.00870}}].

\bibitem{Bertin1996SExtractor}
E.~{Bertin} and S.~{Arnouts}, \emph{{SExtractor: Software for source extraction.}}, \href{https://doi.org/10.1051/aas:1996164}{\emph{\aaps} {\bfseries 117} (1996) 393}.

\bibitem{Ivezic2019LSST}
{\v{Z}}.~{Ivezi{\'c}}, S.M.~{Kahn}, J.A.~{Tyson}, B.~{Abel}, E.~{Acosta}, R.~{Allsman} et~al., \emph{{LSST: From Science Drivers to Reference Design and Anticipated Data Products}}, \href{https://doi.org/10.3847/1538-4357/ab042c}{\emph{\apj} {\bfseries 873} (2019) 111} [\href{https://arxiv.org/abs/0805.2366}{{\ttfamily 0805.2366}}].

\bibitem{Graham2020photometry}
M.L.~{Graham}, A.J.~{Connolly}, W.~{Wang}, S.J.~{Schmidt}, C.B.~{Morrison}, {\v{Z}}.~{Ivezi{\'c}} et~al., \emph{{Photometric Redshifts with the LSST. II. The Impact of Near-infrared and Near-ultraviolet Photometry}}, \href{https://doi.org/10.3847/1538-3881/ab8a43}{\emph{\aj} {\bfseries 159} (2020) 258} [\href{https://arxiv.org/abs/2004.07885}{{\ttfamily 2004.07885}}].

\bibitem{crenshaw2024}
J.F.~Crenshaw, J.B.~Kalmbach, A.~Gagliano, Z.~Yan, A.J.~Connolly, A.I.~Malz et~al., \emph{Probabilistic forward modeling of galaxy catalogs with normalizing flows},  2024.

\bibitem{vandenBusch2020photometrykids}
J.L.~{van den Busch}, H.~{Hildebrandt}, A.H.~{Wright}, C.B.~{Morrison}, C.~{Blake}, B.~{Joachimi} et~al., \emph{{Testing KiDS cross-correlation redshifts with simulations}}, \href{https://doi.org/10.1051/0004-6361/202038835}{\emph{\aap} {\bfseries 642} (2020) A200} [\href{https://arxiv.org/abs/2007.01846}{{\ttfamily 2007.01846}}].

\bibitem{Arnouts1999}
S.~{Arnouts}, S.~{Cristiani}, L.~{Moscardini}, S.~{Matarrese}, F.~{Lucchin}, A.~{Fontana} et~al., \emph{{Measuring and modelling the redshift evolution of clustering: the Hubble Deep Field North}}, \href{https://doi.org/10.1046/j.1365-8711.1999.02978.x}{\emph{\mnras} {\bfseries 310} (1999) 540} [\href{https://arxiv.org/abs/astro-ph/9902290}{{\ttfamily astro-ph/9902290}}].

\bibitem{Arnouts2002}
S.~{Arnouts}, L.~{Moscardini}, E.~{Vanzella}, S.~{Colombi}, S.~{Cristiani}, A.~{Fontana} et~al., \emph{{Measuring the redshift evolution of clustering: the Hubble Deep Field South}}, \href{https://doi.org/10.1046/j.1365-8711.2002.04988.x}{\emph{\mnras} {\bfseries 329} (2002) 355} [\href{https://arxiv.org/abs/astro-ph/0109453}{{\ttfamily astro-ph/0109453}}].

\bibitem{Ilbert2006}
O.~{Ilbert}, S.~{Arnouts}, H.J.~{McCracken}, M.~{Bolzonella}, E.~{Bertin}, O.~{Le F{\`e}vre} et~al., \emph{{Accurate photometric redshifts for the CFHT legacy survey calibrated using the VIMOS VLT deep survey}}, \href{https://doi.org/10.1051/0004-6361:20065138}{\emph{\aap} {\bfseries 457} (2006) 841} [\href{https://arxiv.org/abs/astro-ph/0603217}{{\ttfamily astro-ph/0603217}}].

\bibitem{Hoyle2016TSrandomforest}
B.~{Hoyle}, K.~{Paech}, M.M.~{Rau}, S.~{Seitz} and J.~{Weller}, \emph{{Tuning target selection algorithms to improve galaxy redshift estimates}}, \href{https://doi.org/10.1093/mnras/stw563}{\emph{\mnras} {\bfseries 458} (2016) 4498} [\href{https://arxiv.org/abs/1508.06280}{{\ttfamily 1508.06280}}].

\bibitem{Chaussidon2023TSQSODESI}
E.~{Chaussidon}, C.~{Y{\`e}che}, N.~{Palanque-Delabrouille}, D.M.~{Alexander}, J.~{Yang}, S.~{Ahlen} et~al., \emph{{Target Selection and Validation of DESI Quasars}}, \href{https://doi.org/10.3847/1538-4357/acb3c2}{\emph{\apj} {\bfseries 944} (2023) 107} [\href{https://arxiv.org/abs/2208.08511}{{\ttfamily 2208.08511}}].

\bibitem{Yeche2009TSQSODESI}
C.~{Yeche}, P.~{Petitjean}, J.~{Rich}, E.~{Aubourg}, N.~{Busca}, J.C.~{Hamilton} et~al., \emph{{QSO Selection and Photometric Redshifts with Neural Networks}}, \href{https://doi.org/10.48550/arXiv.0910.3770}{\emph{arXiv e-prints} (2009) arXiv:0910.3770} [\href{https://arxiv.org/abs/0910.3770}{{\ttfamily 0910.3770}}].

\bibitem{Breiman2001RandomForest}
L.~Breiman, \emph{Random forests}, \href{https://doi.org/10.1023/A:1010950718922}{\emph{Machine Learning} {\bfseries 45} (2001) 5}.

\bibitem{Raichoor2016fisher}
A.~{Raichoor}, J.~{Comparat}, T.~{Delubac}, J.P.~{Kneib}, C.~{Y{\`e}che}, H.~{Zou} et~al., \emph{{The SDSS-IV extended Baryon Oscillation Spectroscopic Survey: selecting emission line galaxies using the Fisher discriminant}}, \href{https://doi.org/10.1051/0004-6361/201526486}{\emph{\aap} {\bfseries 585} (2016) A50} [\href{https://arxiv.org/abs/1505.01797}{{\ttfamily 1505.01797}}].

\bibitem{Weaver2022cosmos2020}
J.R.~{Weaver}, O.B.~{Kauffmann}, O.~{Ilbert}, H.J.~{McCracken}, A.~{Moneti}, S.~{Toft} et~al., \emph{{COSMOS2020: A Panchromatic View of the Universe to z{\ensuremath{\sim}}10 from Two Complementary Catalogs}}, \href{https://doi.org/10.3847/1538-4365/ac3078}{\emph{\apjs} {\bfseries 258} (2022) 11} [\href{https://arxiv.org/abs/2110.13923}{{\ttfamily 2110.13923}}].

\bibitem{Mueller2021fnl}
E.-M.~{Mueller}, M.~{Rezaie}, W.J.~{Percival}, A.J.~{Ross}, R.~{Ruggeri}, H.-J.~{Seo} et~al., \emph{{The clustering of galaxies in the completed SDSS-IV extended Baryon Oscillation Spectroscopic Survey: Primordial non-Gaussianity in Fourier Space}}, \href{https://doi.org/10.48550/arXiv.2106.13725}{\emph{arXiv e-prints} (2021) arXiv:2106.13725} [\href{https://arxiv.org/abs/2106.13725}{{\ttfamily 2106.13725}}].

\bibitem{Wright2010Wise}
E.L.~{Wright}, P.R.M.~{Eisenhardt}, A.K.~{Mainzer}, M.E.~{Ressler}, R.M.~{Cutri}, T.~{Jarrett} et~al., \emph{{The Wide-field Infrared Survey Explorer (WISE): Mission Description and Initial On-orbit Performance}}, \href{https://doi.org/10.1088/0004-6256/140/6/1868}{\emph{\aj} {\bfseries 140} (2010) 1868} [\href{https://arxiv.org/abs/1008.0031}{{\ttfamily 1008.0031}}].

\bibitem{Busca2018quasarnet}
N.~{Busca} and C.~{Balland}, \emph{{QuasarNET: Human-level spectral classification and redshifting with Deep Neural Networks}}, \href{https://doi.org/10.48550/arXiv.1808.09955}{\emph{arXiv e-prints} (2018) arXiv:1808.09955} [\href{https://arxiv.org/abs/1808.09955}{{\ttfamily 1808.09955}}].

\bibitem{Guy2023RROCK}
J.~{Guy}, S.~{Bailey}, A.~{Kremin}, S.~{Alam}, D.M.~{Alexander}, C.~{Allende Prieto} et~al., \emph{{The Spectroscopic Data Processing Pipeline for the Dark Energy Spectroscopic Instrument}}, \href{https://doi.org/10.3847/1538-3881/acb212}{\emph{\aj} {\bfseries 165} (2023) 144} [\href{https://arxiv.org/abs/2209.14482}{{\ttfamily 2209.14482}}].

\bibitem{Schlafly2023effectivetime}
E.F.~{Schlafly}, D.~{Kirkby}, D.J.~{Schlegel}, A.D.~{Myers}, A.~{Raichoor}, K.~{Dawson} et~al., \emph{{Survey Operations for the Dark Energy Spectroscopic Instrument}}, \href{https://doi.org/10.3847/1538-3881/ad0832}{\emph{\aj} {\bfseries 166} (2023) 259} [\href{https://arxiv.org/abs/2306.06309}{{\ttfamily 2306.06309}}].

\bibitem{McQuinn_White_Ly_forest}
M.~{McQuinn} and M.~{White}, \emph{{On estimating Ly{\ensuremath{\alpha}} forest correlations between multiple sightlines}}, \href{https://doi.org/10.1111/j.1365-2966.2011.18855.x}{\emph{\mnras} {\bfseries 415} (2011) 2257} [\href{https://arxiv.org/abs/1102.1752}{{\ttfamily 1102.1752}}].

\bibitem{LBG_Ly_forest_1st_obs}
K.-G.~{Lee}, J.F.~{Hennawi}, C.~{Stark}, J.X.~{Prochaska}, M.~{White}, D.J.~{Schlegel} et~al., \emph{{Ly{\ensuremath{\alpha}} Forest Tomography from Background Galaxies: The First Megaparsec-resolution Large-scale Structure Map at z > 2}}, \href{https://doi.org/10.1088/2041-8205/795/1/L12}{\emph{\apjl} {\bfseries 795} (2014) L12} [\href{https://arxiv.org/abs/1409.5632}{{\ttfamily 1409.5632}}].

\bibitem{CLAMATO_I}
K.-G.~{Lee}, A.~{Krolewski}, M.~{White}, D.~{Schlegel}, P.E.~{Nugent}, J.F.~{Hennawi} et~al., \emph{{First Data Release of the COSMOS Ly{\ensuremath{\alpha}} Mapping and Tomography Observations: 3D Ly{\ensuremath{\alpha}} Forest Tomography at 2.05 < z < 2.55}}, \href{https://doi.org/10.3847/1538-4365/aace58}{\emph{\apjs} {\bfseries 237} (2018) 31} [\href{https://arxiv.org/abs/1710.02894}{{\ttfamily 1710.02894}}].

\bibitem{CLAMATO_II}
B.~{Horowitz}, K.-G.~{Lee}, M.~{Ata}, T.~{M{\"u}ller}, A.~{Krolewski}, J.X.~{Prochaska} et~al., \emph{{Second Data Release of the COSMOS Ly{\ensuremath{\alpha}} Mapping and Tomography Observations: The First 3D Maps of the Detailed Cosmic Web at 2.05 < z < 2.55}}, \href{https://doi.org/10.3847/1538-4365/ac982d}{\emph{\apjs} {\bfseries 263} (2022) 27} [\href{https://arxiv.org/abs/2109.09660}{{\ttfamily 2109.09660}}].

\bibitem{FishLSS_NSailer}
N.~{Sailer}, E.~{Castorina}, S.~{Ferraro} and M.~{White}, \emph{{Cosmology at high redshift - a probe of fundamental physics}}, \href{https://doi.org/10.1088/1475-7516/2021/12/049}{\emph{\jcap} {\bfseries 2021} (2021) 049} [\href{https://arxiv.org/abs/2106.09713}{{\ttfamily 2106.09713}}].

\bibitem{2023MNRAS.521.3648D}
W.~{d'Assignies D}, C.~{Zhao}, J.~{Yu} and J.-P.~{Kneib}, \emph{{Cosmological Fisher forecasts for next-generation spectroscopic surveys}}, \href{https://doi.org/10.1093/mnras/stad611}{\emph{\mnras} {\bfseries 521} (2023) 3648} [\href{https://arxiv.org/abs/2301.02289}{{\ttfamily 2301.02289}}].

\bibitem{DESI2023_science-validation}
{DESI Collaboration}, A.G.~{Adame}, J.~{Aguilar}, S.~{Ahlen}, S.~{Alam}, G.~{Aldering} et~al., \emph{{Validation of the Scientific Program for the Dark Energy Spectroscopic Instrument}}, \href{https://doi.org/10.3847/1538-3881/ad0b08}{\emph{\aj} {\bfseries 167} (2024) 62} [\href{https://arxiv.org/abs/2306.06307}{{\ttfamily 2306.06307}}].

\bibitem{DESI_BAO_forecast2014}
A.~{Font-Ribera}, P.~{McDonald}, N.~{Mostek}, B.A.~{Reid}, H.-J.~{Seo} and A.~{Slosar}, \emph{{DESI and other Dark Energy experiments in the era of neutrino mass measurements}}, \href{https://doi.org/10.1088/1475-7516/2014/05/023}{\emph{\jcap} {\bfseries 2014} (2014) 023} [\href{https://arxiv.org/abs/1308.4164}{{\ttfamily 1308.4164}}].

\bibitem{DESI2024_BAO_gal}
{DESI Collaboration}, A.G.~{Adame}, J.~{Aguilar}, S.~{Ahlen}, S.~{Alam}, D.M.~{Alexander} et~al., \emph{{DESI 2024 III: Baryon Acoustic Oscillations from Galaxies and Quasars}}, \href{https://doi.org/10.48550/arXiv.2404.03000}{\emph{arXiv e-prints} (2024) arXiv:2404.03000} [\href{https://arxiv.org/abs/2404.03000}{{\ttfamily 2404.03000}}].

\bibitem{DESI_2024_bao_ly}
{DESI Collaboration}, A.G.~{Adame}, J.~{Aguilar}, S.~{Ahlen}, S.~{Alam}, D.M.~{Alexander} et~al., \emph{{DESI 2024 IV: Baryon Acoustic Oscillations from the Lyman Alpha Forest}}, \href{https://doi.org/10.48550/arXiv.2404.03001}{\emph{arXiv e-prints} (2024) arXiv:2404.03001} [\href{https://arxiv.org/abs/2404.03001}{{\ttfamily 2404.03001}}].

\bibitem{2024_BAO_cosmo}
{DESI Collaboration}, A.G.~{Adame}, J.~{Aguilar}, S.~{Ahlen}, S.~{Alam}, D.M.~{Alexander} et~al., \emph{{DESI 2024 VI: Cosmological Constraints from the Measurements of Baryon Acoustic Oscillations}}, \href{https://doi.org/10.48550/arXiv.2404.03002}{\emph{arXiv e-prints} (2024) arXiv:2404.03002} [\href{https://arxiv.org/abs/2404.03002}{{\ttfamily 2404.03002}}].

\bibitem{DDE_forecast_Linder}
E.V.~{Linder}, \emph{{The Rise of Dark Energy}}, \href{https://doi.org/10.48550/arXiv.2106.09581}{\emph{arXiv e-prints} (2021) arXiv:2106.09581} [\href{https://arxiv.org/abs/2106.09581}{{\ttfamily 2106.09581}}].

\bibitem{Planck2020}
{Planck Collaboration}, N.~{Aghanim}, Y.~{Akrami}, M.~{Ashdown}, J.~{Aumont}, C.~{Baccigalupi} et~al., \emph{{Planck 2018 results. VI. Cosmological parameters}}, \href{https://doi.org/10.1051/0004-6361/201833910}{\emph{\aap} {\bfseries 641} (2020) A6} [\href{https://arxiv.org/abs/1807.06209}{{\ttfamily 1807.06209}}].

\bibitem{NG_Dalal2008}
N.~{Dalal}, O.~{Dor{\'e}}, D.~{Huterer} and A.~{Shirokov}, \emph{{Imprints of primordial non-Gaussianities on large-scale structure: Scale-dependent bias and abundance of virialized objects}}, \href{https://doi.org/10.1103/PhysRevD.77.123514}{\emph{\prd} {\bfseries 77} (2008) 123514} [\href{https://arxiv.org/abs/0710.4560}{{\ttfamily 0710.4560}}].

\bibitem{Matarrese_Verde_2008}
S.~{Matarrese} and L.~{Verde}, \emph{{The Effect of Primordial Non-Gaussianity on Halo Bias}}, \href{https://doi.org/10.1086/587840}{\emph{\apjl} {\bfseries 677} (2008) L77} [\href{https://arxiv.org/abs/0801.4826}{{\ttfamily 0801.4826}}].

\bibitem{fnl_DESI_CMB}
A.~{Krolewski}, W.J.~{Percival}, S.~{Ferraro}, E.~{Chaussidon}, M.~{Rezaie}, J.N.~{Aguilar} et~al., \emph{{Constraining primordial non-Gaussianity from DESI quasar targets and Planck CMB lensing}}, \href{https://doi.org/10.1088/1475-7516/2024/03/021}{\emph{\jcap} {\bfseries 2024} (2024) 021} [\href{https://arxiv.org/abs/2305.07650}{{\ttfamily 2305.07650}}].

\bibitem{fnl_chaussidon}
E.~{Chaussidon}, C.~{Y{\`e}che}, A.~{de Mattia}, C.~{Payerne}, P.~{McDonald}, A.J.~{Ross} et~al., \emph{{Constraining primordial non-Gaussianity with DESI 2024 LRG and QSO samples}}, \href{https://doi.org/10.48550/arXiv.2411.17623}{\emph{arXiv e-prints} (2024) arXiv:2411.17623} [\href{https://arxiv.org/abs/2411.17623}{{\ttfamily 2411.17623}}].

\end{thebibliography}\endgroup
\newpage
\appendix
\section{Impact of photometric redshifts on the Random Forest performance metrics}
\label{app:diff_z}
\begin{figure}
    \centering
\includegraphics[width=1\linewidth]{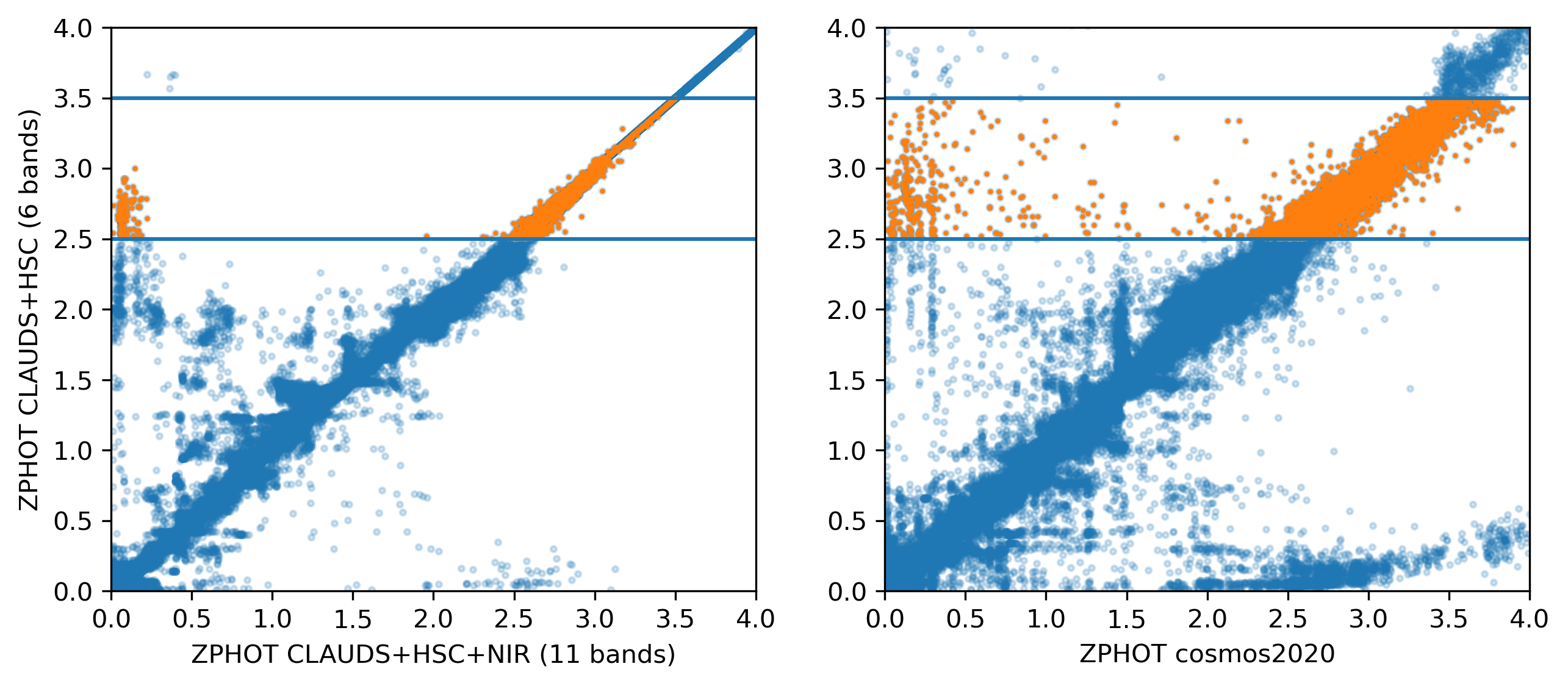}
\caption{\textit{Left}: CLAUDS+HSC photometric redshifts as a function of CLAUDS+HSC+NIR photometric redshifts on COSMOS. \textit{Right}: CLAUDS+HSC photometric redshifts as a function of COSMOS2020 photometric redshifts.}
    \label{fig:diff_zphot}
\end{figure}

Additional photometric redshift datasets are available in the COSMOS field, including the 11-bands CLAUDS+HSC+NIR and COSMOS2020 datasets, both covering a restricted area of $\sim 1.26$ deg$^2$. It is important to note that the footprints of the 6-bands, 11-bands, and COSMOS2020 datasets differ slightly due to varying survey masks applied in COSMOS2020. To compare these datasets, we first match all three catalogs (including the 6-band catalog) for sources with $r < 25$. For a sub-sample defined by $23.75 < r < 24.5$, inspired by the BXU box selection in \citep{RuhlmannKleider2024LBGCLAUDS} we estimate the target densities as $n(2.5 < z_{\rm 6bands} < 3.5) = 2548$ deg$^{-2}$, $n(2.5 < z_{\rm 11bands} < 3.5) = 2655$ deg$^{-2}$, and $n(2.5 < z_{\rm cosmos2020} < 3.5) = 3072$ deg$^{-2}$ after applying a rough correction factor of 0.8 to account for the survey masks on the $1.26$ deg$^2$ field. \\
\indent In \cref{fig:diff_zphot}, we present the 6-bands CLAUDS+HSC photometric redshifts (for $r < 25$) compared to the 11-bands CLAUDS+HSC+NIR photometric redshifts (left panel) and COSMOS2020 photometric redshifts (right panel) obtained for the COSMOS field. These plots reveal that the CLAUDS+HSC redshift catalog contains both high-redshift and low-redshift outliers. Specifically, we observe that \textit{true} low-redshift galaxies are sometimes misclassified as high-redshift galaxies when relying solely on CLAUDS+HSC imaging, and vice versa. This misclassification is particularly pronounced in the redshift range $z \in [2.5, 3.5]$ (orange region), which is crucial for the Random Forest training.

To assess the impact of using improved photometric redshift catalogs in Random Forest training, we consider the CLAUDS+HSC colors $u-g$, $g-r$, $r-i$, and $i-z$, with the target class defined as $1$ if $z_{\rm phot} \in [2.5, 3.5]$ and $0$ otherwise. Here, $z_{\rm phot}$ is taken from either the CLAUDS+HSC, CLAUDS+HSC+NIR, or COSMOS2020 catalog. We keep using the BXU cut in \citep{RuhlmannKleider2024LBGCLAUDS}, defined by $23.75 < r < 24.5$, which probes higher redshift than the complementary TMG selection used in \cref{sec:selection_method} ($22.75 < r < 23.75$). We consider the restricted area of 1.26 deg$^2$ on the COSMOS field where the three different photometric datasets are available. From all these changes compared to \cref{sec:selection_method}, the results obtained in this appendix may not be one-to-one comparable with what is obtained in the \cref{sec:selection_method} (we consider a specific COSMOS field with different depth that the overall footprint, as well as fewer statistics, deeper magnitudes), however, the framework we use here ensures self-consistency when comparing the impact of using different photometric redshifts. Imposing a target density of $1,100$ deg$^{-2}$, we find that $(\epsilon_{\rm 6bands}, p_{\rm 6bands}) = (0.42, 0.98)$, $(\epsilon_{\rm 11bands},p_{\rm 11bands}) = (0.43, 0.96)$ and $(\epsilon_{\rm cosmos2020}, p_{\rm cosmos2020}) = (0.34, 0.97)$. Using COSMOS2020 redshifts leads to a reduction in efficiency, however, the purity is stable over the three photometric datasets.

\begin{figure}[t]
    \centering
    \includegraphics[width=0.95\linewidth]{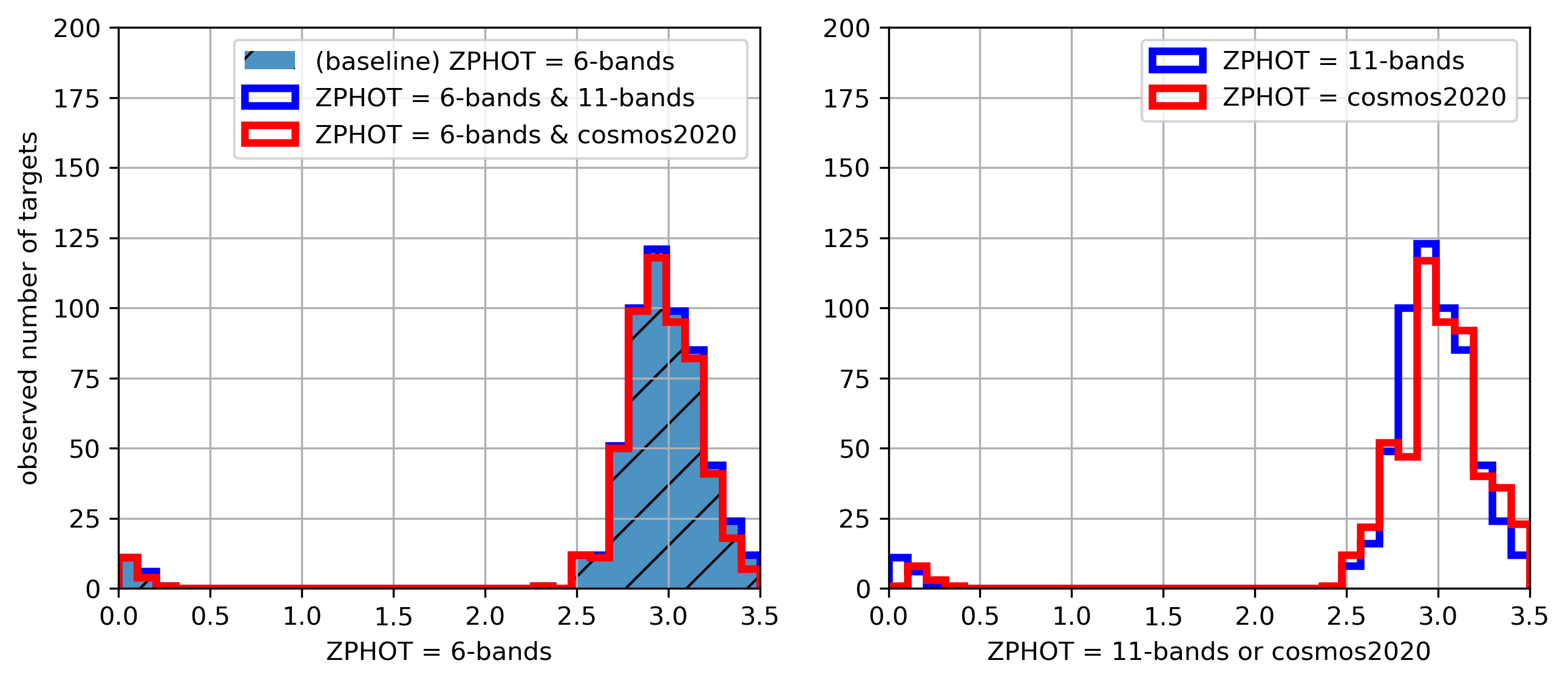}
    \label{fig:nz_diff_z}
    \caption{\textit{Left}: 6-bands {\sc LePhare} photometric redhsift distribution of RF-selected targets (in shaded blue), when cross-matching with the 11-bands {\sc LePhare} catalog (in blue) or with the COSMOS2020 catalog (in red). \textit{Right}: 11-bands {\sc LePhare} (in blue) of COSMOS2020 (in red) photometric redshift distribution of targets.}
\end{figure}

\indent Using a list of targets selected with an RF model trained on $ugriz$ colors and 6-bands {\sc LePhare} photometric redshifts, we compare the shape of the corresponding 6-bands {\sc LePhare} photometric redshift distribution to both the 11-bands {\sc LePhare} and COSMOS2020 photometric redshift distributions of targets. Specifically, (i) we follow our baseline analysis, selecting targets from an RF model trained with $ugriz$ colors and 6-bands {\sc LePhare} photometric redshifts, (ii) we apply an RF quality threshold to achieve a target density of 1,100 deg$^{-2}$, with the observed 6-bands photometric redshift distribution in the 1.26 deg$^{2}$ COSMOS field shown as the shaded blue histogram in \cref{fig:nz_diff_z} (left panel). (iii) We then match these targets to their 11-bands photometric redshifts. The remaining matched distribution is shown in unfilled blue versus 6-bands redshifts. Besides, we match the targets with the COSMOS2020 photometric redshifts, resulting in the final 6-band distribution represented in red. Finally, the 11-band and COSMOS2020 photometric redshift distributions of the selected targets are shown in \cref{fig:nz_diff_z} (right panel), where we find that the recovered photometric redshift distributions are not significantly degraded and remain comparable to the 6-band photometric redshift distributions in the $z>2$ range.

\end{document}